\documentclass[preprint,aps,prd,floatfix,nofootinbib,superscriptaddress,eqsecnum,tightenlines]{revtex4-1}

\usepackage[mathscr]{euscript}
\usepackage{amsmath,amsfonts,amssymb}
\usepackage{bm}
\usepackage{graphicx}% Include figure files
\usepackage[pdftex]{color}
\usepackage[sort&compress]{natbib}
\usepackage[colorlinks=true,linkcolor=blue,filecolor=blue,urlcolor=blue,citecolor=blue,pdftex,plainpages=false]{hyperref}

\allowdisplaybreaks

\newcommand{\eq}[1]{Eq.~\eqref{eq:#1}}

\newcommand{\eqs}[2]{Eqs.~\eqref{eq:#1} and \eqref{eq:#2}}

\def\eqn#1{\label{eq:#1}}

\def\figref#1{Fig.~\ref{fig:#1}}
\def\Figref#1{Figure~\ref{fig:#1}}

\def\secref#1{Sec.~\ref{sec:#1}}

\def\tabref#1{Table~\ref{tab:#1}}
\def\tabrefs#1#2{Tables~\ref{tab:#1} and \ref{tab:#2}}
\def\rcite#1{Ref.~\cite{#1}}

\def\rcites#1{Refs.~\cite{#1}}

\newcommand{\ie}{\emph{i.e.},\ }

\newcommand{\chpt}{$\chi$PT}

\newcommand{\schpt}{S$\chi$PT}

\newcommand{\aschpt}{HMrAS$\chi$PT}

\newcommand{\Aslash}{\ensuremath{A\kern-0.6em/\kern0.2em}}
\newcommand{\dslash}{\ensuremath{\partial\kern-0.5em/\kern0.1em}}
\newcommand{\Dslash}{\ensuremath{D\kern-0.65em/\kern0.2em}}
\newcommand{\pslash}{\ensuremath{p\kern-0.5em/\kern0.1em}}
\newcommand{\vslash}{\ensuremath{v\kern-0.5em/\kern0.1em}}

\newcommand{\MHsinv}{\frac{\Lambda_\text{HQET}}{M_{H_s}}}
\newcommand{\MHsinvDiff}{\frac{\Lambda_\text{HQET}}{M_{H_s}} - \frac{\Lambda_\text{HQET}}{M_{D_s}} }

\newcommand{\Minv}{\frac{\Lambda_\text{HQET}}{M_0}}

\newcommand{\cF}{\ensuremath{\mathcal{F}}}

\newcommand{\cM}{\ensuremath{\mathcal{M}}}

\newcommand{\st}{\text{s}}

\newcommand{\order}{\ensuremath{\text{O}}} %use consistent notation for "order", whatever we decide

\newcommand{\param}[1]{{\textcolor{blue}{#1}}} % make parameters stick out

\newcommand{\half}{\ensuremath{{\textstyle\frac{1}{2}}}}

\newcommand{\sixth}{\ensuremath{{\textstyle\frac{1}{6}}}}
\newcommand{\Sh}{\ensuremath{\widetilde{\mathcal{S}\kern-0.15em\mathit{h}}}}
\newcommand{\Ch}{\ensuremath{\widetilde{\mathcal{C}\kern-0.15em\mathit{h}}}}

\newcommand{\leftvec}{{\raise1.5ex\hbox{$\leftarrow$}\kern-1.00em}}
\newcommand{\rightvec}{{\raise1.5ex\hbox{$\rightarrow$}\kern-1.00em}}

\newcommand{\fm}{\text{fm}} 
\newcommand{\GeV}{\text{GeV}}
\newcommand{\MeV}{\text{MeV}} 
\newcommand{\fpiPDG}{\ensuremath{f_{\pi,\text{PDG}}}}

\newcommand{\LamQCD}{\ensuremath{\Lambda_\text{QCD}}}
\newcommand{\LamHQET}{{\ensuremath{\Lambda_\text{HQET}}}}
\newcommand{\pole}{{\text{pole}}}
\newcommand{\MSbar}{{\ensuremath{\overline{\rm MS}}}}
\newcommand{\Naik}{\ensuremath{\mathbb{N}}}

\newcommand{\EMone}{$K^+$-$K^0$ splitting}
\newcommand{\EMtwo}{$K^0$~mass}
\newcommand{\EMtwos}{$K$-mass scheme}
\newcommand{\EMthree}{$H_x$~mass}
\newcommand{\EMthrees}{$H_s$-mass scheme}

\newcommand{\BR}{\ensuremath{\mathcal{B}}}            % branching ratio 
\newcommand{\oBR}{\ensuremath{\overline{\mathcal{B}}}}            % modified branching ratio 

\def\rSYMANZIK{Symanzik:1980rcnt,*Symanzik:1983dc}
\def\rSYMANZIKGAUGE{Weisz:1982zw,*Weisz:1983bn,*Curci:1983an,*Luscher:1985zq,*Luscher:1984xn,*Alford:1995hw,*Hart:2008sq}
\def\rTADPOLE{Lepage:1992xa}
\def\rHPQCDrHISQ{Follana:2006rc}
\def\rHISQrCONFIGS{Bazavov:2012xda}
\def\rFD2014{Bazavov:2014wgs,*Bazavov:2014lja}
\def\rPDG2016{Olive:2016xmw,Rosner:2015wva}

\def\rTasteSingletTopo{Aubin:2003mg,Billeter:2004wx}

\def\rRHMX{Duane:1985ym,Duane:1986iw,Gottlieb:1987mq,Sexton:1992nu,Kennedy:1998cu,Hasenbusch:2001ne,Omelyan:2002E1,%
Clark:2006fx,Takaishi:2005tz}
\def\rRHMC{Duane:1987de}
\def\rRHMD{Bazavov:2009bb,\rHISQrCONFIGS}

\begin{document}

\title{\boldmath \texorpdfstring{$B$}{B}- and \texorpdfstring{$D$}{D}-meson leptonic decay constants from \\
four-flavor lattice QCD}

\author{A.~Bazavov} 
\affiliation{Department of Computational Mathematics, Science and Engineering,
and Department of Physics and Astronomy, Michigan State University, East Lansing, Michigan 48824, USA}

\author{C.~Bernard}
\email[]{cb@lump.wustl.edu}
\affiliation{Department of Physics, Washington University, St. Louis, Missouri 63130, USA}

\author{N.~Brown}
\affiliation{Department of Physics, Washington University, St. Louis, Missouri 63130, USA}

\author{C.~DeTar}
\affiliation{Department of Physics and Astronomy, University of Utah, Salt Lake City, Utah 84112, USA}

\author{A.X.~El-Khadra}
\affiliation{Department of Physics, University of Illinois, Urbana,  Illinois 61801, USA}
\affiliation{Fermi National Accelerator Laboratory, Batavia, Illinois 60510, USA}

\author{E.~G\'amiz} 
\affiliation{CAFPE and Departamento de F\'isica Te\'orica y del Cosmos, Universidad de Granada, E-18071 Granada, Spain}

\author{Steven~Gottlieb}
\affiliation{Department of Physics, Indiana University, Bloomington, Indiana 47405, USA}

\author{U.M.~Heller} 
\affiliation{American Physical Society, One Research Road, Ridge, New York 11961, USA}

\author{J.~Komijani}
%\altaffiliation[Present address:]{~School of Physics and Astronomy, University of Glasgow, Glasgow G12 8QQ, United~Kingdom}
\email[]{javad.komijani@glasgow.ac.uk}
\affiliation{Physik-Department, Technische Universit\"at M\"unchen, 85748 Garching, Germany}
\affiliation{Institute for Advanced Study, Technische Universit\"at M\"unchen, 85748 Garching, Germany}
\affiliation{School of Physics and Astronomy, University of Glasgow, Glasgow G12 8QQ, United~Kingdom}

\author{A.S.~Kronfeld} 
\email[]{ask@fnal.gov}
\affiliation{Fermi National Accelerator Laboratory, Batavia, Illinois 60510, USA}
\affiliation{Institute for Advanced Study, Technische Universit\"at M\"unchen, 85748 Garching, Germany}

\author{J.~Laiho}  
\affiliation{Department of Physics, Syracuse University, Syracuse, New York 13244, USA}

\author{P.B.~Mackenzie}
\affiliation{Fermi National Accelerator Laboratory, Batavia, Illinois 60510, USA}

\author{E.T.~Neil}
\affiliation{Department of Physics, University of Colorado, Boulder, Colorado 80309, USA}
\affiliation{RIKEN-BNL Research Center, Brookhaven National Laboratory, \\ Upton, New York 11973, USA}

\author{J.N.~Simone}
\affiliation{Fermi National Accelerator Laboratory, Batavia, Illinois 60510, USA}

\author{R.L.~Sugar}
\affiliation{Department of Physics, University of California, Santa Barbara, California 93106, USA}

\author{D.~Toussaint}
\email[]{doug@physics.arizona.edu}
\affiliation{Physics Department, University of Arizona, Tucson, Arizona 85721, USA}

\author{R.S.~Van~de~Water}
\email[]{ruthv@fnal.gov}
\affiliation{Fermi National Accelerator Laboratory, Batavia, Illinois 60510, USA}

\collaboration{Fermilab Lattice and MILC Collaborations}
\noaffiliation

\date{\today}

\begin{abstract}
We calculate the leptonic decay constants of heavy-light pseudoscalar mesons with charm and bottom quarks in lattice quantum
chromodynamics on four-flavor QCD gauge-field configurations with dynamical $u$, $d$, $s$, and $c$ quarks.
We analyze over twenty isospin-symmetric ensembles with six lattice spacings down to $a\approx 0.03$~fm and several values of the
light-quark mass down to the physical value $\frac{1}{2}(m_u+m_d)$.
We employ the highly-improved staggered-quark (HISQ) action for the sea and valence quarks; on the finest lattice spacings,
discretization errors are sufficiently small that we can calculate the $B$-meson decay constants with the HISQ action for the first
time directly at the physical $b$-quark mass.
We obtain the most precise determinations to-date of the $D$- and $B$-meson decay constants and their ratios,
  $f_{D^+}  = 212.7(0.6)$~MeV,
  $f_{D_s}  = 249.9(0.4)$~MeV,
  $f_{D_s}/f_{D^+} = 1.1749(16)$, 
  $f_{B^+}  = 189.4 (1.4)$~MeV,
  $f_{B_s}  = 230.7(1.3)$~MeV, 
  $f_{B_s}/f_{B^+}  = 1.2180(47)$,
where the errors include statistical and all systematic uncertainties.
Our results for the $B$-meson decay constants are three times more precise than the previous best lattice-QCD calculations, and
bring the QCD errors in the Standard-Model predictions for the rare leptonic decays
$\overline{\mathcal{B}}(B_s \to \mu^+\mu^-) =  3.64(11) \times 10^{-9}$,
$\overline{\mathcal{B}}(B^0 \to \mu^+\mu^-) =  1.00(3) \times 10^{-10}$, and
$\overline{\mathcal{B}}(B^0 \to \mu^+\mu^-)/\overline{\mathcal{B}}(B_s \to \mu^+\mu^-) =  0.0273(9)$
to well below other sources of uncertainty.
As a byproduct of our analysis, we also update our previously published results for the light-quark-mass ratios and the
scale-setting quantities $f_{p4s}$, $M_{p4s}$, and $R_{p4s}$.
We obtain the most precise lattice-QCD determination to date of the ratio $f_{K^+}/f_{\pi^+} = 1.1950(^{+16}_{-23})$~MeV.
\end{abstract}

\preprint{FERMILAB-PUB-17/491-T}

\maketitle

\section{Introduction}

Leptonic decays of $B$ and $D$ mesons are important probes of heavy-to-light quark flavor-changing interactions.
The charged-current decays $H^+ \to \ell^+ \nu_\ell$ ($H=D^+,D_s,B^+$; $\ell=e,\mu,\tau$) proceed at tree level in the Standard
Model via the axial-vector current $\mathcal{A}_\mu \equiv \overline{Q}\gamma_5 \gamma_\mu q$, where $Q$ is the heavy charm or
bottom quark and $q$ is the light quark in the pseudoscalar meson.
When combined with a nonperturbative lattice-QCD calculation of the decay constant $f_{H^+}$, an experimental measurement of the
leptonic decay width allows the determination of the corresponding Cabibbo-Kobayashi-Maskawa (CKM) quark-mixing matrix element
$|V_{Qq}|$.
Because the decays $H^0 \to \ell^+\ell^-$ ($H=D^0,B^0,B_s$) proceed via a flavor-changing-neutral-current interaction, and are
forbidden at tree level in the Standard Model, these processes may be especially sensitive to (tree-level) contributions of new
heavy particles.
Both the Standard-Model and new-physics predictions for the rare-decay branching ratios depend upon the decay constants $f_{H^0}$.

Leptonic $B$-meson decays, in particular, make possible several interesting tests of the Standard Model and promising new-physics
searches.
The determination of $|V_{ub}|$ from $B^+ \to \tau^+\nu_\tau$ decay can play an important role in resolving the 2--3$\sigma$ tension
between the values of $|V_{ub}|$ obtained from inclusive and exclusive semileptonic $B$-meson decays (see the recent
reviews~\cite{Amhis:2016xyh,Hamilton:2017nla} and references therein).
Alternatively, the decay $B^+ \to \tau^+\nu_\tau$, because of the large $\tau$-lepton mass, may receive observable contributions
from new heavy particles such as charged Higgs bosons or leptoquarks~\cite{Hou:1992sy,Akeroyd:2003zr}.
The branching ratios for $B^0 \to \ell^+\ell^-$ and $B_s \to \ell^+\ell^-$ can be enhanced with respect to the Standard-Model rates
in new-physics scenarios with tree-level flavor-changing-neutral currents, such as in fourth-generation
models~\cite{Buras:2013uqa,Buras:2010wr}.

Lattice-QCD calculations of the $B$-meson decay constants are especially timely given the wealth of leptonic $B$-decay measurements
from the $B$-factories and, more recently, by hadron-collider experiments at the LHC.
The branching ratio for the charged-current decay $B^+ \to \tau^+\nu_\tau$ has been measured by the BaBar and Belle experiments to
about 20\% precision~\cite{Aubert:2009wt,Lees:2012ju,Adachi:2012mm,Kronenbitter:2015kls}.
The rare decay $B_{s}\to\mu^+\mu^-$ has now been independently observed by the ATLAS, CMS, and LHCb experiments with errors on the
measured branching ratio ranging from around 20\%--100\%~\cite{CMS:2014xfa,Aaboud:2016ire,Aaij:2017vad}; these works have also set
limits on the process $B^0\to\mu^+\mu^-$.
Precise determinations of $f_{B^+}$, $f_{B^0}$, and $f_{B_s}$ are needed to interpret these results.
Such determinations are also necessary to fully exploit coming measurements by
Belle~II~\cite{Bennett:2016qgs}, which will begin running at the Super-KEKb facility next year, as well as future measurements by
ATLAS, CMS, and LHCb after the LHC luminosity and detector upgrades~\cite{Schmidt:2016jra}, which are planned for 2023--2025.
%
% Hack to get a group in the middle of a list
\nocite{Davies:2010ip,McNeile:2011ng,Bazavov:2011aa,Na:2012kp,Na:2012iu,Dowdall:2013tga,Christ:2014uea}
\phantom{\cite{Bazavov:2014wgs,*Bazavov:2014lja}}
\nocite{Yang:2014sea,Carrasco:2014poa}

Several independent three- and four-flavor calculations of heavy-light-meson decay constants using different lattice actions are
available~\cite{Davies:2010ip,McNeile:2011ng,Bazavov:2011aa,Na:2012kp,Na:2012iu,Dowdall:2013tga,Christ:2014uea,%
Bazavov:2014wgs,Yang:2014sea,Carrasco:2014poa,Bussone:2016iua,Boyle:2017jwu,Hughes:2017spc}, with uncertainties ranging from
$\sim0.5\%$--5\% and $\sim2\%$--8\% for the $D_{(s)}$ and $B_{(s)}$ systems, respectively.
The most precise results for $f_D$ and $f_{D_s}$ have been obtained by us~\cite{\rFD2014}, and for $f_{B_s}$ by the HPQCD
Collaboration~\cite{McNeile:2011ng}, in both cases using improved staggered sea quarks and the ``highly-improved staggered quark''
(HISQ) action~\cite{\rHPQCDrHISQ} for the valence light and heavy quarks.
The HISQ action makes possible this high precision because it has both small discretization errors, even at relatively large lattice
spacings, and an absolutely-normalized axial current.
Our previous calculation~\cite{\rFD2014} of the $D_{(s)}$-meson decay constants employed physical-mass light and charm quarks and
gauge-field configurations with lattice spacings down to $a\approx 0.06$~fm; the dominant contribution to the errors on $f_D$ and
$f_{D_s}$ came from the continuum extrapolation.
HPQCD's calculation of $f_{B_s}$ with the HISQ action for the $b$ quark employed five three-flavor ensembles of gauge-field
configurations from the MILC Collaboration~\cite{Bernard:2001av,Aubin:2004wf,Bazavov:2009bb} with lattice spacings as fine as
$a\approx0.045$~fm, enabling them to simulate with heavy-quark masses close to the physical bottom-quark mass.
The statistical errors dominate in their calculation due to the comparatively small number of configurations per ensemble (roughly
200 on their finest up to 600 on their coarsest).
Other important sources of uncertainty are from the extrapolation in heavy-quark mass up to $m_b$ and from the extrapolation to zero
lattice spacing.

In this paper, we present a new calculation of the leptonic decay constants of heavy-light mesons containing bottom and charm quarks
that improves upon prior works in several ways.
As in our previous calculation of $f_D$ and $f_{D_s}$~\cite{\rFD2014}, we employ the four-flavor QCD gauge-field configurations
generated by the MILC Collaboration with HISQ up, down, strange, and charm quarks~\cite{\rHISQrCONFIGS}; we also use the HISQ action
for the light and heavy valence quarks.
We now employ three new ensembles with finer lattice spacings of $a\approx0.042$ and $a\approx0.03$~fm, and also increase 
statistics on the $a\approx 0.06$~fm ensemble with physical-mass light quarks.
Altogether, we analyze 24 ensembles, most of which have approximately 1000 configurations.
We also calculate the $B^+$- and $B^0$-meson decay constant with HISQ $b$ quarks on the HISQ ensembles for the first time.

We fit our lattice data for the heavy-light meson decay constants to a functional form that combines information on the heavy-quark
mass dependence from heavy-quark effective theory, on the light-quark mass dependence from chiral perturbation theory, and on
discretization effects from Symanzik effective theory.
This allows us to exploit our wide range of simulation parameters by including multiple lattice spacings and heavy- and light-quark
mass values in a single effective-field-theory (EFT) fit.
We present results for all charged and neutral heavy-light pseudoscalar-meson decay constants, as well as the SU(3)-breaking
decay-constant ratios and the differences between the charged decay constants and the decay constants in the isospin-symmetric
($m_u=m_d$) limit.
In addition, we provide the correlations between our decay-constant results to facilitate their use in other phenomenological
studies beyond this work.
Preliminary reports of this analysis have been presented in Refs.~\cite{Lattice:2015nee,Komijani:2016jrh}.

This paper is organized as follows.
First, Sec.~\ref{sec:Lattice-Simulations} presents relevant details of the lattice actions, simulation parameters, and
methodology of our calculation, including a discussion of how we deal with nonequilibrated topological charge.
Next, we describe our two-point correlator fits used to obtain the heavy-light-meson decay amplitudes in
Sec.~\ref{sec:Correlator-Fits}.
In Sec.~\ref{sec:physical-mass-analysis}, we determine the lattice spacings and light-quark masses on the ensembles employed in this
calculation, which are parametric inputs to the decay-constant analysis, and also to a determination of heavy-quark masses in a
companion paper~\cite{Bazavov:2018omf}.
Physical quark-mass ratios and the light decay constant ratio $f_{K^+}/f_{\pi^+}$ are obtained as a byproduct.
We then calculate the physical $B$- and $D$-meson decay-constant values in Sec.~\ref{sec:chiral-analysis} by fitting our lattice
decay-amplitude data at multiple values of the light- and heavy-quark masses and lattice spacing to a function based on effective
field theories, and interpolating to the physical light-, charm-, and bottom-quark masses and extrapolating to the continuum limit.
In Sec.~\ref{sec:Errors}, we estimate the systematic uncertainties in the decay constants not included in the EFT fit, and provide
complete error budgets.
We present our final results for the $B$- and $D$-meson leptonic decay constants with total errors and discuss the impact of our
results for determinations of CKM matrix elements and tests of the Standard Model in Sec.~\ref{sec:res_pheno}.
Final results for light-quark mass ratios, $f_{K^+}/f_{\pi^+}$, and the scale-setting quantities $f_{p4s}$ and $M_{p4s}$ are also
presented.
Finally, in Sec.~\ref{sec:conc}, we conclude with an outlook to future work.
Two appendices provide useful information about (improved) staggered fermions when the bare lattice quark mass $am_0\not\ll1$.
Appendix~\ref{Appendix:Tree-level-HISQ} discusses the radius of convergence of the expansion in $am_0$, while
Appendix~\ref{app:normalization} derives the normalization factor for staggered bilinears.
Appendix~\ref{Appendix:Covariance-Matrix} provides the correlation and covariance matrices between our $B$- and
$D$-meson decay constant results.

\section{Simulation Parameters and Methods}
\label{sec:Lattice-Simulations}

In this paper, we use the MILC Collaboration's ensembles of QCD gauge-field configurations with four flavors of dynamical quarks.
This simulation program is described in detail in Ref.~\cite{\rHISQrCONFIGS}, and since then it has been extended to smaller lattice
spacings.
Here we provide information on our current calculation, and also document the new ensembles.
First, in Sec.~\ref{subsec:simsim}, we summarize the parameters of the actions and two-point correlation functions used in the
analysis presented below.
Three ensembles with approximate lattice spacings 0.042 and 0.03~fm are new since Ref.~\cite{\rHISQrCONFIGS}, while some of the older
ensembles have been extended.
In Sec.~\ref{subsec:RHMDvsRHMC}, we update the discussion in Ref.~\cite{\rHISQrCONFIGS} on possible effects from using different
algorithms in different parts of the simulation.
Finally, in Sec.~\ref{subsec:topology}, we discuss effects of poor sampling of the distribution of topological charge and how to
compensate for these effects.

\subsection{Simulation parameters}
\label{subsec:simsim}

The gauge action~\cite{\rSYMANZIKGAUGE} is one-loop Symanzik~\cite{\rSYMANZIK} and tadpole~\cite{\rTADPOLE} improved, using the
plaquette to determine the tadpole quantity~$u_0$.
The fermion action is the HISQ action introduced by the HPQCD collaboration~\cite{\rHPQCDrHISQ}.
The ensembles all have an isospin-symmetric sea.
A single staggered-fermion field yields four species, known as tastes, in the continuum limit~\cite{Karsten:1980wd}.
To adjust the number of species in the sea, we take the fourth (square) root of the quark determinant for the strange and charm 
(up and down) sea~\cite{Marinari:1981qf}.
In addition to the perturbative arguments~\cite{Karsten:1980wd,Giedt:2006ib}, this procedure passes several nonperturbative tests~%
\cite{Follana:2004sz,Durr:2004as,Durr:2004ta,Wong:2004nk,Shamir:2004zc,Prelovsek:2005rf,Bernard:2006zw,Durr:2006ze,Bernard:2006ee,%
Shamir:2006nj,Bernard:2007qf,Kronfeld:2007ek,Donald:2011if}, providing confidence that continuum QCD is obtained as $a\to0$.

Table~\ref{tab:ensembles} summarizes the ensembles used in this work.
\begin{table}
\newcommand{\h}{\phantom{2}}
\caption{Ensembles used in this calculation.
The notation and symbols are discussed in the text.
In the first column the approximate lattice spacings are mnemonic only;
the precise values are tabulated in Table~\ref{tab:lattice-spacing:p4s-method}.
The second column is used as a key to identify the ensembles at a given approximate lattice spacing. 
A dagger ($\dagger$) on $am'_s$ flags ensembles for which the simulation strange-quark mass is deliberately chosen far from the 
physical value. The $M_\pi$ and $L$ values are different from those listed in Table~I of Ref.~\cite{Bazavov:2014wgs},
because those values assumed a mass-dependent scale setting
scheme.}
\label{tab:ensembles}
\begin{tabular*}{\textwidth}{@{\extracolsep{\fill}}llclllr@{$\times$\kern-0.35em}lcccc}
\hline\hline
$\approx a$ & Key & $\beta$ & $am'_l$ & $am'_s$ & $am'_c$ & $(L/a)^3$&$(T/a)$ & $L$  & $M_\pi$ & $M_\pi L$ & $N_\text{conf}$ \\
 (fm) &           &         &         &         &         \multicolumn{3}{c}{}& (fm) &  (MeV)  &           &  \\
\hline
0.15 & $m_s/5$    & 5.80    & 0.013   & 0.065   & 0.838 & $16^3$&$48$   & 2.45 & 305 & 3.8  & 1020\\
0.15 & $m_s/10$   & 5.80    & 0.0064  & 0.064   & 0.828 & $24^3$&$48$   & 3.67 & 214 & 4.0  & 1000\\
0.15 & physical   & 5.80    & 0.00235 & 0.0647  & 0.831 & $32^3$&$48$   & 4.89 & 131 & 3.3  & 1000\\
\hline
0.12 & $m_s/5$    & 6.00    & 0.0102  & 0.0509  & 0.635 & $24^3$&$64$   & 2.93 & 305 & 4.5  & 1040 \\
%unphysical strange quark mass
0.12 & unphysA    & 6.00    & 0.0102  & 0.03054$^\dagger$  & 0.635 & $24^3$&$64$ & 2.93 & 304 & 4.5  & 1020\\
0.12 & small      & 6.00    & 0.00507 & 0.0507  & 0.628 & $24^3$&$64$   & 2.93 & 218 & 3.2  & 1020\\
0.12 & $m_s/10$   & 6.00    & 0.00507 & 0.0507  & 0.628 & $32^3$&$64$   & 3.91 & 217 & 4.3  & 1000\\
0.12 & large      & 6.00    & 0.00507 & 0.0507  & 0.628 & $40^3$&$64$   & 4.89 & 216 & 5.4  & 1028\\
%unphysical strange quark masses
0.12 & unphysB    & 6.00    & 0.01275 & 0.01275$^\dagger$     & 0.640 & $24^3$&$64$   & 2.93 & 337 & 5.0  & 1020\\
0.12 & unphysC    & 6.00    & 0.00507 & 0.0304$^\dagger$      & 0.628 & $32^3$&$64$   & 3.91 & 215 & 4.3  & 1020\\
0.12 & unphysD    & 6.00    & 0.00507 & 0.022815$^\dagger$    & 0.628 & $32^3$&$64$   & 3.91 & 214 & 4.2  & 1020\\
0.12 & unphysE    & 6.00    & 0.00507 & 0.012675$^\dagger$    & 0.628 & $32^3$&$64$   & 3.91 & 214 & 4.2  & 1020 \\
0.12 & unphysF    & 6.00    & 0.00507 & 0.00507$^\dagger$     & 0.628 & $32^3$&$64$   & 3.91 & 213 & 4.2  & 1020 \\
0.12 & unphysG    & 6.00    & 0.0088725 & 0.022815$^\dagger$  & 0.628 & $32^3$&$64$   & 3.91 & 282 & 5.6  & 1020\\
0.12 & physical   & 6.00    & 0.00184 & 0.0507  & 0.628 & $48^3$&$64$  & 5.87 & 132 & 3.9  &\h 999 \\
\hline
0.09 & $m_s/5$    & 6.30    & 0.0074  & 0.037   & 0.440 & $32^3$&$96$  & 2.81 & 316 & 4.5  & 1005 \\
0.09 & $m_s/10$   & 6.30    & 0.00363 & 0.0363  & 0.430 & $48^3$&$96$  & 4.22 & 221 & 4.7  &\h 999 \\ 
0.09 & physical   & 6.30    & 0.0012  & 0.0363  & 0.432 & $64^3$&$96$  & 5.62 & 129 & 3.7  &\h 484 \\
\hline
0.06 & $m_s/5$    & 6.72    & 0.0048  & 0.024   & 0.286 & $48^3$&$144$ & 2.72 & 329 & 4.5  & 1016 \\
0.06 & $m_s/10$   & 6.72    & 0.0024  & 0.024   & 0.286 & $64^3$&$144$ & 3.62 & 234 & 4.3  &\h 572 \\
0.06 & physical   & 6.72    & 0.0008  & 0.022   & 0.260 & $96^3$&$192$ & 5.44 & 135 & 3.7  &\h 842 \\
\hline
0.042 & $m_s/5$   & 7.00    & 0.00316  & 0.0158   & 0.188 & $64^3$&$192$  & 2.73 & 315 & 4.3  & 1167 \\
0.042 & physical  & 7.00    & 0.000569 & 0.01555 & 0.1827 & $144^3$&$288$ & 6.13 & 134 & 4.2  &\h 420 \\
\hline
0.03 & $m_s/5$    & 7.28    & 0.00223  & 0.01115  & 0.1316 & $96^3$&$288$ & 3.09 & 309 & 4.8  &\h 724 \\
\hline\hline
\end{tabular*} 
\end{table}
In this table, we identify the ensembles by the approximate lattice spacing $a$ and the ratio of light sea-quark ($m'_l$) to strange
sea-quark mass ($m'_s$).
The exact lattice spacing and physical strange-quark mass ($m_s$) are outputs of our decay-constant analysis and can be found in
Table~\ref{tab:lattice-spacing:p4s-method} in Sec.~\ref{sec:chiral-analysis}.
The six lattice spacings range from approximately $0.15$~fm to $0.03$~fm, and the sea has light sea-quark masses $0.2m'_s$,
$0.1m'_s$, and approximately physical.
In most ensembles, $m'_s$ is chosen close to the physical strange-quark mass, but sometimes it is deliberately chosen far from
physical to provide useful information about the sea-quark-mass dependence.
In all ensembles, the charm-quark mass is chosen close to its physical value.
In Table~\ref{tab:ensembles}, $\beta = 10/g^2$ is the gauge coupling, $T$ and $L$ are the lattice temporal and spatial extents, and
$M_\pi$ is the mass of the taste-Goldstone sea pion.

% valence quark masses
% 4 or 6 sources
% precessing sources
For each ensemble, the light, strange, and charm sea-quark masses are estimated either from short tuning runs or from tuned masses
on nearby ensembles.
These values are always found to be slightly in error once higher statistics become available, so it is necessary to adjust for this
small sea-quark-mass mistuning \emph{a posteriori}, as we do in the fitting procedure described in Sec.~\ref{sec:chiral-analysis}.

\begin{table}
\newcommand{\h}{\phantom{7}}
\newcommand{\Oo}{0.1\phantom{n7}}
\newcommand{\To}{0.2\phantom{57}}
\caption{Valence-quark masses used in each ensemble.
The first two columns identify the ensemble.
The third column gives the lightest valence-quark mass in units of the sea strange-quark mass.
(The full set of light valence-quark masses is listed in the text.) The fourth column shows the heavy valence-quark masses in units
of the sea charm-quark mass.
The last column shows the number of configurations and the number of source time slices used on each.
}
\label{tab:valencemasses}
\begin{tabular}{l@{\quad}l@{\quad}c@{\hspace{1em}}lr@{$\,\times\,$}l}
\hline\hline
$\approx a$~(fm) & Key & $m_\text{min}/m'_s$ & ~ $m_h/m'_c$  & $N_\text{conf}$&$N_\text{src}$\\
\hline
0.15 & $m_s/5$   & \Oo    & \{0.9, 1.0\} &   1020&4\\
0.15 & $m_s/10$  & \Oo    & \{0.9, 1.0\} &   1000&4 \\
0.15 & physical  & 0.037  & \{0.9, 1.0\} &   1000&4 \\
\hline
0.12 & $m_s/5$   & \Oo    & \{0.9, 1.0\} &   1040&4 \\
%unphysical strange quark mass
0.12 & unphysA   & \Oo    & \{0.9, 1.0\} &   1020&4 \\
0.12 & small     & \Oo    & \{0.9, 1.0\} &   1020&4 \\
0.12 & $m_s/10$  & \Oo    & \{0.9, 1.0\} &   1000&4 \\
0.12 & large     & \Oo    & \{0.9, 1.0\} &   1028&4 \\
%unphysical strange quark masses
0.12 & unphysB   & \Oo    & \{0.9, 1.0\} &   1020&4 \\
0.12 & unphysC   & \Oo    & \{0.9, 1.0\} &   1020&4 \\
0.12 & unphysD   & \Oo    & \{0.9, 1.0\} &   1020&4 \\
0.12 & unphysE   & \Oo    & \{0.9, 1.0\} &   1020&4 \\
0.12 & unphysF   & \Oo    & \{0.9, 1.0\} &   1020&4 \\
0.12 & unphysG   & \Oo    & \{0.9, 1.0\} &   1020&4 \\
0.12 & physical  & 0.037  & \{0.9, 1.0\} &   999&4 \\
\hline
0.09 & $m_s/5$   & \Oo    & \{0.9, 1.0\} &   1005&4 \\
0.09 & $m_s/10$  & \Oo    & \{0.9, 1.0\} &   999&4 \\
0.09 & physical  & 0.033  & \{0.9, 1.0, 1.5, 2.0, 2.5, 3.0\} &   484&4 \\
\hline
0.06 & $m_s/5$   & 0.05\h & \{0.9, 1.0\} &   1016&4 \\
0.06 & $m_s/10$  & 0.05\h & \{0.9, 1.0, 2.0, 3.0, 4.0\} &   572&4 \\
0.06 & physical  & 0.036  & \{0.9, 1.0, 1.5, 2.0, 2.5, 3.0, 3.5, 4.0, 4.5\} &   842&6 \\
\hline
0.042 & $m_s/5$   & 0.036 & \{0.9, 1.0, 1.5, 2.0, 2.5, 3.0, 3.5, 4.0, 4.5\} &   1167&6 \\
0.042 & physical  & 0.037 & \{0.9, 1.0, 1.5, 2.0, 2.5, 3.0, 3.5, 4.0, 4.5, 5.0\} &   420&6 \\
\hline
0.03 & $m_s/5$   & \To    & \{0.9, 1.0, 1.5, 2.0, 2.5, 3.0, 3.5, 4.0, 4.5, 5.0\} &   724&4 \\
\hline\hline
\end{tabular} 
\end{table}

\pagebreak
We compute pseudoscalar correlators for several valence-quark masses on each ensemble.
In almost all cases, we use light valence-quark masses of $0.1m'_s$, $0.2m'_s$, $0.3m'_s$, $0.4m'_s$, $0.6m'_s$, $0.8m'_s$ and
$1.0m'_s$, where the prime distinguishes the strange sea-quark mass from the post-production, better-tuned mass.
To save computer time, however, for the finest ensemble with $a \approx 0.03$~fm and $m'_l=m'_s/5$, we only use valence-quark masses
greater than or equal to the light sea-quark mass $0.2m'_s$.
For the physical quark-mass ensembles and the ensembles with $a\approx 0.06$ and 0.042~fm, we use lighter valence-quark masses,
usually going down to the estimated physical light-quark mass.
The wide range of valence-quark masses on the ensembles with $a \ge 0.042$~fm are used to determine the light-quark-mass dependence,
while the 0.03~fm ensemble helps guide the continuum limit.
This strategy saves computer time, since light-quark propagators on these lattices are expensive, the cost being approximately
proportional to~$1/am_q$.
In all cases, we compute valence heavy-quark propagators with masses of $1.0m'_c$ and $0.9m'_c$, to allow interpolation or
extrapolation to the physical charm-quark mass.
Finally, on six of the ensembles we use valence-quark masses heavier than charm to allow us to extrapolate, and on the finest
lattices \emph{interpolate}, to the $b$-quark mass.
Table~\ref{tab:valencemasses} shows the lightest valence-quark mass used on each ensemble in units of the strange sea-quark mass,
and also the heavy-quark masses used on each ensemble.

On each configuration, we compute quark propagators from four or six evenly-spaced source time slices.
We change the location of the first source time slice from configuration to configuration, shifting by an amount approximately equal
to half the spacing between source time slices but incommensurate with the lattice size, so that all possible source locations are
used.
Table~\ref{tab:valencemasses} also shows the number of source time slices used on each ensemble.

\subsection{RHMC and RHMD algorithms}
\label{subsec:RHMDvsRHMC}

The coarser ensembles were all generated using the rational hybrid Monte Carlo (RHMC) algorithm~\cite{\rRHMX,\rRHMC}, but some of
the finer ensembles were generated with a mixture of the RHMC and the rational hybrid molecular dynamics (RHMD)~\cite{\rRHMX,\rRHMD}
algorithms.
The two most recently generated ensembles, one with $a\approx 0.042$~fm and physical light-quark mass and another with
$a\approx0.03$~fm and $m'_l=m'_s/5$, were generated entirely with the RHMD algorithm.
The considerations behind these choices, and the effects of using the RHMD algorithm, are discussed in detail in
Ref.~\cite{\rHISQrCONFIGS}.
Since the preparation of Ref.~\cite{\rHISQrCONFIGS}, three of the ensembles have been enlarged, which enables us to update the
comparison of the RHMC and RHMD algorithms in that work.

Table~\ref{tab:plaquette} shows the differences in the plaquettes between the parts of the ensembles generated with RHMC and RHMD
algorithms for the ensembles where both algorithms were used.  The numbers of configurations used in this comparison differ from
those in Table~\ref{tab:ensembles} because heavier-than-charm correlators were only run on parts of the first two ensembles listed,
and the third ensemble was extended slightly after this comparison was done.  In addition, the plaquette was measured after every
trajectory, giving 2--3 times larger statistics than used in our decay-constant calculation. 
Motivated by the expectation that using an approximate integration
procedure amounts to simulating with a slightly different action, we can estimate the importance of these shifts by asking how much
the bare coupling or, equivalently, the lattice spacing would need to be adjusted to change the average plaquette by this amount.
From looking at the plaquette at a couple of lattice spacings, we find $\Delta\ln(a)/\Delta\,\text{plaq}\approx-4.2$, which leads
to the corresponding values of ${\Delta a}/{a}$ given in the final column of Table~\ref{tab:plaquette}.  Clearly, these differences
are quite small.  In fact, they are negligible, because in the analysis reported below we use $f_\pi$ to set the scale, and the
fractional error on the current value for $f_\pi$ from the Particle Data Group (PDG)~\cite{\rPDG2016} is about
$150\times10^{-5}$.

\begin{table}
\newcommand{\h}{\phantom{1}}
\newcommand{\n}{\phantom{2}}
\caption{Results for the plaquette from the RHMC and RHMD algorithms.
The first two columns give the approximate lattice spacing and the ratio of the light- to strange sea-quark masses.
The third and fourth columns give the time-step sizes used with the RHMC and RHMD algorithms, respectively, while the fifth and
sixth columns give the simulation time multiplied by the acceptance rate for the two algorithms; the ``effective time units", which
is the molecular dynamics time multiplied by the acceptance rate, indicates the amount of data used in each measurement.
The seventh column is the difference in the plaquette, $\Delta(\text{plaq})$, from the two algorithms.
and the last column the fractional change in the lattice spacing, $\Delta a/a$, needed to create such a difference in the
plaquette.}
\label{tab:plaquette}
\begin{tabular*}{\textwidth}{@{\extracolsep{\fill}}cccccccccc}
\hline\hline
$\approx a$ (fm) &  $m'_l/m'_s$ & RHMC & RHMD & RHMC & RHMD & $\Delta(\text{plaq})$ & 
${\Delta a}/{a}$ \\
     &     & \multicolumn{2}{c}{time step} & \multicolumn{2}{c}{effective time units} & & \\
\hline
0.09 & 1/27 & 0.0115 & 0.0133 &   1339 & 2962 & $-3.0(5)\times10^{-5}$ &  $13\times10^{-5}$ \\
0.06 & 1/10 & 0.0141 & 0.0143 &   2703 & 2180 & $-1.2(5)\times10^{-5}$ & \h$5\times10^{-5}$ \\
0.06 & 1/27 & 0.0100 & 0.0125 & \n 288 & 3432 & $-1.1(4)\times10^{-5}$ & \h$5\times10^{-5}$ \\
\hline\hline
\end{tabular*}
\end{table}

The new $a\approx0.042$~fm physical-mass ensemble has the largest physical volume of the four-flavor MILC ensembles, with a spatial
size of about 6~fm, while the new $a \approx 0.03$~fm ensemble with $m'_l/m'_s=1/5$ has the smallest lattice spacing.
When the physical volume is made larger, more low-momentum (long-distance) modes are added to the system.
Based on these considerations, we do not expect this added physics to be very sensitive to the molecular dynamics step size.
On the other front, the lattice spacing is made smaller by making $\beta$ larger.
If the ultraviolet gauge modes are viewed as free fields, the coefficient of the gauge fields in the molecular-dynamics Hamiltonian
is proportional to $\beta$ while the coefficient of the conjugate momenta added for the molecular-dynamics time evolution is held
fixed.
Thus, the frequency of the modes in molecular dynamics time is proportional to $\beta^{1/2}$.
Strictly speaking, if we wish to keep the fractional error fixed while increasing $\beta$, we should reduce the step size as
$\beta^{-1/2}$.
That dependence is very weak---the square root of $\ln a$.
It turns out that this scaling is more or less what was chosen empirically in going from $a\approx 0.09$~fm to $0.042$~fm.
The step size was decreased from 0.0133 to 0.0125, or by about 6\%, as $\beta$ was increased from 6.3 to 7.0, corresponding to
$\beta^{1/2}$ changing by~5\%.

\subsection{Correction for nonequilibrated topological charge}
\label{subsec:topology}

\begin{figure}
    \includegraphics[trim={0.5in 5.65in 0.5in 1.55in},clip,width=0.7\textwidth]{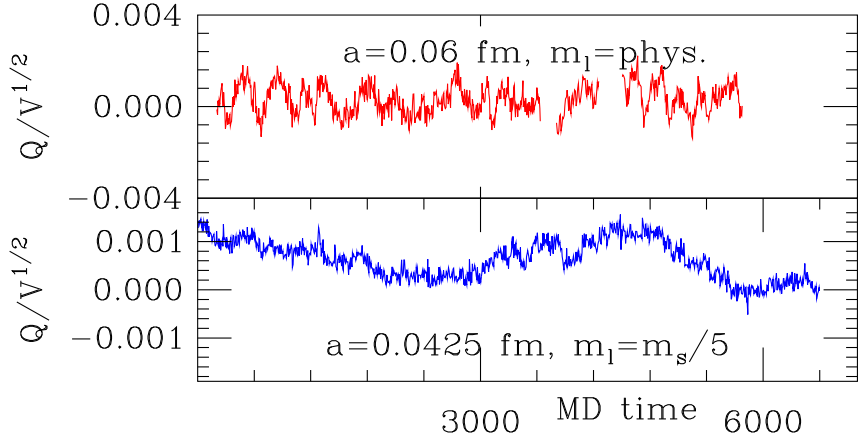}
    \caption{Simulation time history of the topological charge in two cases.
        The upper panel is for the physical quark mass run at $a\approx0.06$~fm, and shows a case where the distribution of $Q$ is
        well sampled.
        The three sections of the trace correspond to three separate runs with the same parameters.
        The lower panel, for the $m'_l=m'_s/5$ run at $a\approx0.042$~fm, shows a case where the time history is not well sampled,
        and where we will apply the correction factors discussed in Ref.~\cite{Bernard:2017npd}.}
    \label{fig:topo_hist}
\end{figure}

Because QCD simulations use approximately continuous update algorithms, the topological charge $Q$ evolves more and more slowly as
the lattice spacing becomes smaller.
In our finest ensembles, the evolution has slowed so much that the distribution of $Q$ has not been sampled properly.
Time histories of the topological charge in many of the HISQ ensembles can be found in Ref.~\cite{Bernard:2017npd}.
In Fig.~\ref{fig:topo_hist}, we show one case, $a \approx 0.06$~fm and physical $m'_l$, where the topological charge is well
equilibrated, and a second case, $a \approx 0.042$~fm and $m'_l=m'_s/5$, where its distribution is clearly not well sampled.

As first discussed in \rcite{Brower:2003yx}, one can study the $Q$-dependence of observables in chiral perturbation theory (\chpt).
Bernard and Toussaint \cite{Bernard:2017npd} recently extended this approach to heavy-light decay constants in the context of
heavy-meson \chpt.
We use their results to adjust the raw decay-constant results to account at lowest order for the incomplete sampling of $Q$ in the
small-$a$ ensembles.
The amount of the adjustment is smaller than our statistical errors, but not negligible in comparison to other systematic effects.

We summarize here the key results that allow us to make this adjustment.
Let $\Phi_{H_x} = f_{H_x}\sqrt{M_{H_x}}$ be the heavy-light decay constant, in the normalization suitable for heavy quarks.
Let $B$ denote either the meson mass $M$, the decay constant $f$, or the combination $\Phi_H$.
In a finite volume $V$ at fixed $Q$, the masses and decay constants obey~\cite{Brower:2003yx,Aoki:2009mx}.
\begin{equation}
    \eqn{topo_dep}
    B \big|_{Q, V} = B + \frac{1}{2\chi_T V} B'' \left( 1-\frac{Q^2}{\chi_T V} \right) + \order\left( (\chi_T V)^{-2} \right),
\end{equation}
where on the right-hand side $B$ is the infinite-volume value, properly averaged over $Q$, $B''$ is its second derivative
with respect to the vacuum angle $\theta$, evaluated at $\theta=0$, and $\chi_T$ is the topological susceptibility
\begin{equation}
    \eqn{chi}
    \chi_T = \frac{\langle Q^2\rangle}{V}
\end{equation}
in a fully-sampled, large-volume ensemble.
For three sea quarks with masses $m_u=m_d=m_l$ and $m_s$, light-meson \chpt\ for the valence-meson mass and decay constant gives
\cite{Aoki:2009mx,Bernard:2017npd}
\begin{align}
    M_{xy}'' &= -M_{xy} \frac{m_l^2 m^2_s}{2(m_l+2m_s)^2} \frac{1}{m_x m_y} , 
    \eqn{mxy-nf3} \\
    f_{xy}'' &= -f_{xy} \frac{m_l^2 m_s^2}{4(m_l+2m_s)^2} \frac{(m_x-m_y)^2}{m^2_x m^2_y} ,
    \eqn{fxy-nf3}
\end{align}
where subscripts $x$ and $y$ denote flavor, and the meson mass and decay constant are at $\theta=0$.
A~similar calculation in heavy-meson \chpt\ gives~\cite{Bernard:2017npd}
\begin{align}
    \Phi''_{H_x} &= -\Phi_{H_x} \frac{m_l^2 m_s^2}{4(m_l+2m_s)^2} \frac{1}{m^2_x},
    \eqn{Phi-result} \\
    M''_x &= -2B_0\lambda_1\frac{m_l^2 m_s^2}{(m_l+2m_s)^2} \frac{1}{m_x} - 2B_0\lambda'_1\frac{m_l m_s}{m_l+2m_s},
    \eqn{MH-result}
\end{align}
where $m_x$ is the mass of the light valence quark, and $B_0$, $\lambda_1$, and $\lambda_1^\prime$ are low energy constants, which
are estimated in a companion paper on heavy-light meson masses~\cite{Bazavov:2018omf}.
These are the appropriate results even with 2+1+1 flavors of sea quark, because the charmed
sea quark decouples from the chiral theory.
Although the dependence of masses and decay constants are usually small compared to our statistical errors, we have been able to
resolve them in some of our well-equilibrated ensembles and confirm, within limited statistics, that our data agree with
these formulas~\cite{Bernard:2016yha, Bernard:2017npd}.

Knowing the dependence of masses and decay constants on the average $Q^2$, one can correct the
simulation results to account for the difference of the simulation average  $\langle Q^2\rangle_\text{sample}$, and the
correct~$\langle Q^2\rangle$.  The lowest order \chpt\ result for the topological susceptibility is~\cite{Leutwyler:1992yt}
\begin{equation}
    \chi_T = \frac{f_\pi^2}{4} \left(\frac{2}{M_{\pi, I}^2} + \frac{1}{M_{ss, I}^2}\right)^{-1}\,,
    \label{eq:Heiri}
\end{equation}
where the effect of staggered taste-violations has been included at leading order by using the taste-singlet meson masses
\cite{\rTasteSingletTopo}, indicated by ``$I$.''
The correction to the decay constants is then given by
\begin{equation}
    f_\text{corrected} = f_\text{sample} -
        \frac{1}{2\chi_T V} F'' \left(1 - \frac{\langle Q^2\rangle_\text{sample}}{\chi_T V}\right)
   \label{eq:topofix}
\end{equation}
with $\chi_T$ from Eq.~(\ref{eq:Heiri}).

MILC has calculated $\langle Q^2\rangle_\text{sample}$ on all ensembles listed in Table~\ref{tab:ensembles}.
For more details, see Ref.~\cite{Bernard:2017npd}.
For three of the finest ensembles, namely those at $a\approx0.042$ and $0.03$~fm, the simulation time histories of $Q^2$ show that
it is not well equilibrated.
In the analysis below, we use Eq.~(\ref{eq:topofix}) with $\langle Q^2\rangle_\text{sample}$ calculated by MILC to adjust the
decay-constant data.
The adjusted data are used in our central fit, and we take 100\% of the difference between fit results with the adjusted data and
with the unadjusted data as the systematic error in our results from incomplete equilibration of the topological charge.

\section{Two-point correlator fits }
\label{sec:Correlator-Fits}

Our procedures for calculating pseudoscalar meson correlators and for finding masses and amplitudes from these correlators are the
same as those used in our earlier computation of charm-meson decay constants in Ref.~\cite{\rFD2014}.
Our analysis includes new and extended ensembles, however, so the fit ranges and the number of states employed have been updated.

We compute quark propagators with both ``Coulomb-wall'' and ``random-wall'' sources, using four source time slices per gauge-field
configuration in most cases, but six source time slices on the $0.042$~fm $m_s/5$ ensemble and the $0.06$ and $0.042$~fm physical
quark mass ensembles.
The pseudoscalar decay constant is obtained from the amplitude of a correlator of a single-point pion operator, $|M^{-1}(x,y)|^2$,
where $M$ is the lattice fermion matrix $\Dslash + m$.
The random-wall source consists of a randomly oriented unit vector in color space at each spatial lattice point at the source time.
When averaged over sources, contributions to the correlator where the quark and antiquark are on different spatial points average to
zero, so the average correlator is just the point-to-point correlator multiplied by the spatial size of the lattice, and the
improved statistics from averaging over all the spatial source points more than makes up for the noise introduced from contributions
with the quark and antiquark at different spatial points.
We use three random source vectors at each source time slice.

\begin{figure}
    \renewcommand{\baselinestretch}{1.1}
    \includegraphics[trim={0.4in 1.75in 0.4in 1.75in},clip,width=0.48\textwidth]{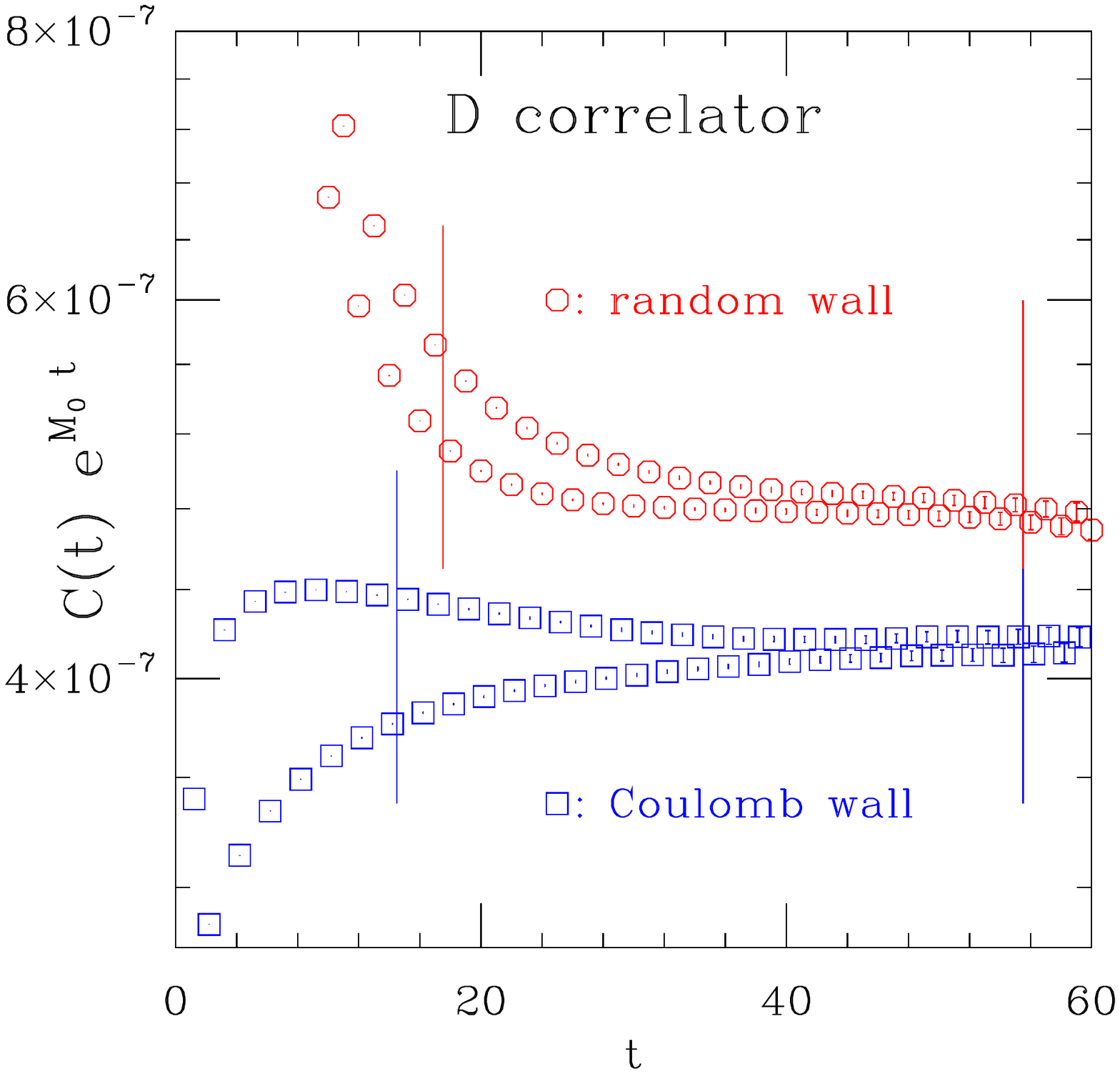} \hfill
    \includegraphics[trim={0.4in 1.75in 0.4in 1.75in},clip,width=0.48\textwidth]{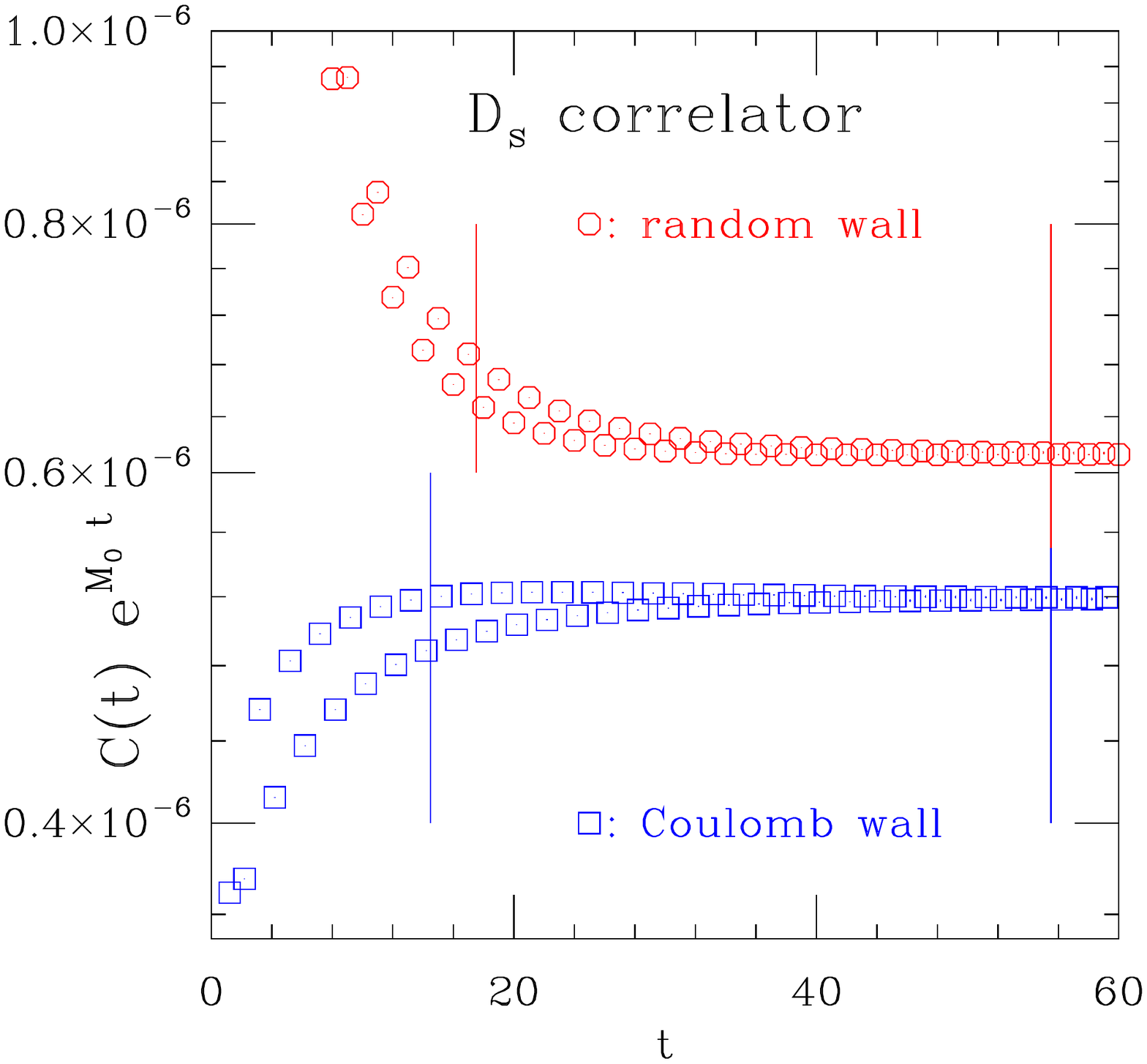}  \\
    \includegraphics[trim={0.4in 1.75in 0.4in 1.75in},clip,width=0.48\textwidth]{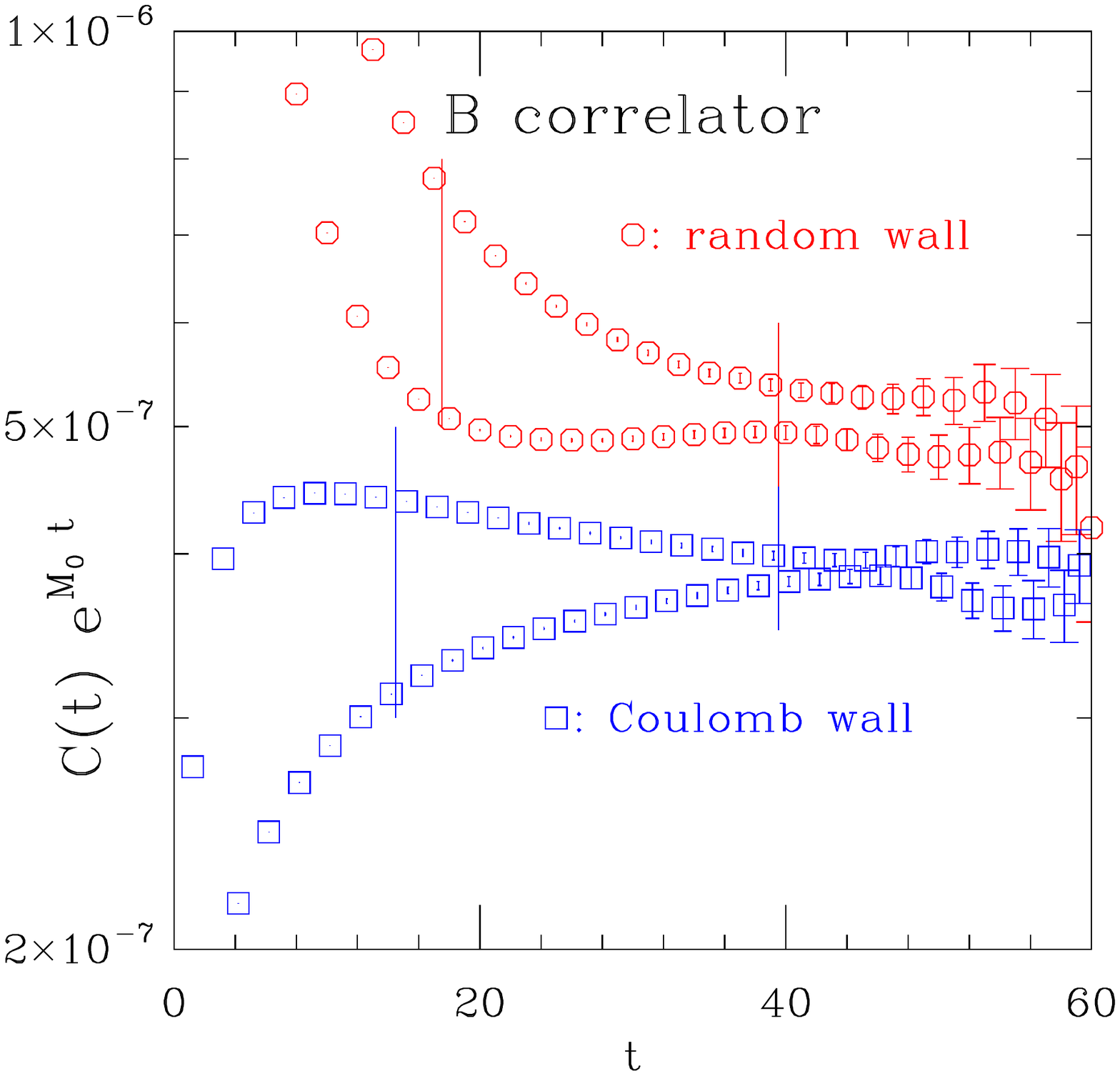}  \hfill
    \includegraphics[trim={0.4in 1.75in 0.4in 1.75in},clip,width=0.48\textwidth]{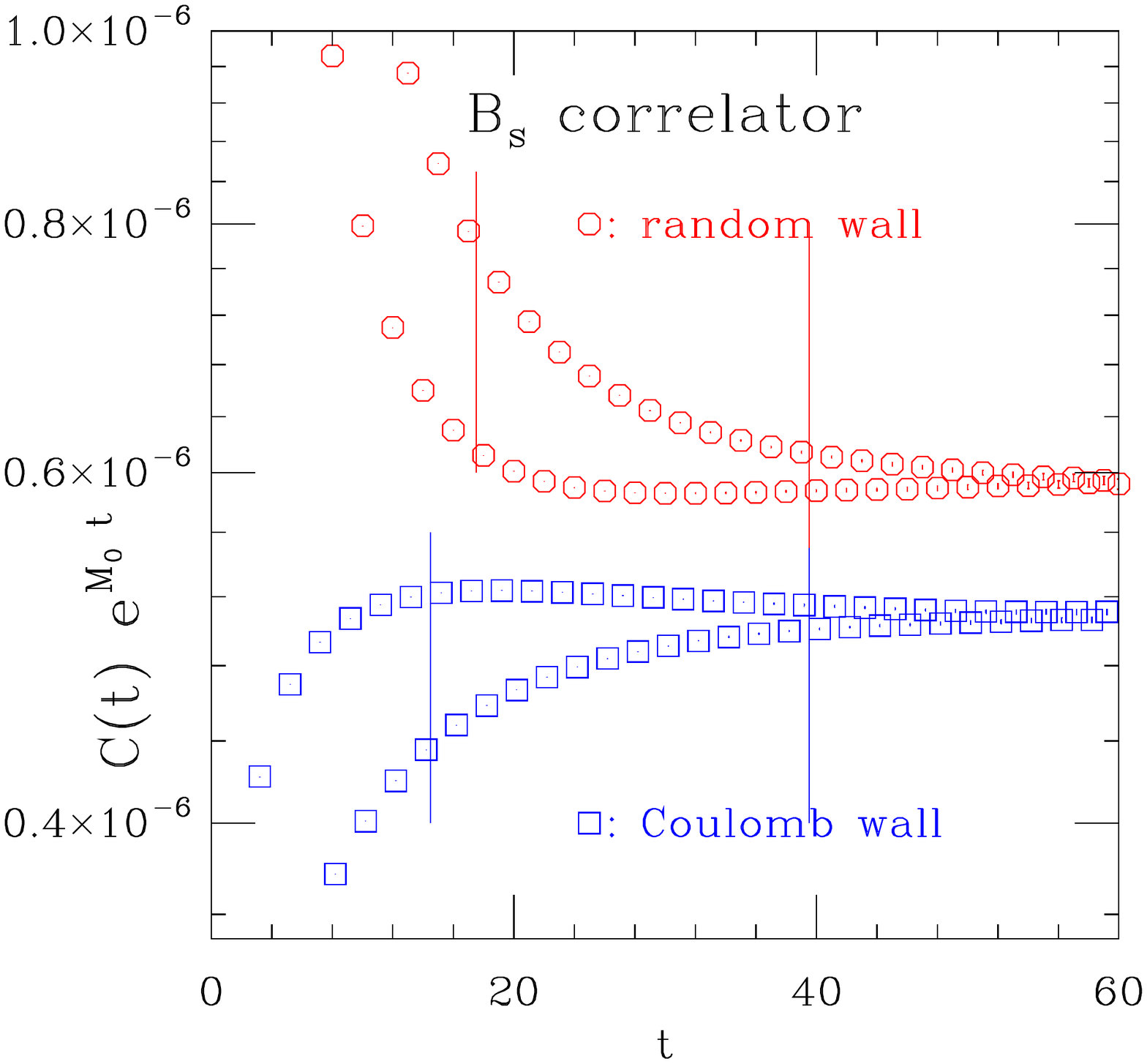}
    \caption{Pseudoscalar correlators for the $D$ (top left), $D_s$ (top right), $B$ (bottom left), and $B_s$ (bottom right) mesons 
        on the $a\approx0.042$~fm physical-quark-mass ensemble.
        Here the valence charm-quark mass is equal to the sea charm-quark mass, and the bottom-quark mass is equal to 4.5 times the
        charm-quark mass.
        The red octagons are the random-wall source correlator and the blue squares the Coulomb-wall correlator.
        Both correlators have been rescaled by $e^{M_0 T}$ where $M_0$ is the ground-state mass; the random-wall correlators have
        also been multiplied by an arbitrary factor to make the vertical scale convenient.
        The vertical lines show the fit ranges used in the 3+2 state fits in our analysis.
        The $D$- and $D_s$-meson fits have $p$-values 0.66 and 0.71 respectively, while the $B$- and $B_s$-meson fits  have 
        $p$-values 0.29 and 0.40 respectively.
        (The oscillatory behavior in $t$ comes from the positive parity states in the correlator.)}
    \label{fig:correlators_both}
\end{figure}

For the Coulomb-wall source we fix to the lattice Coulomb gauge, and then use a source in a fixed direction in color space at each
spatial lattice point.
We use three such vectors, chosen to lie along the three coordinate axes in color space.
The Coulomb-wall source is effectively smeared over the whole spatial slice, which we expect to suppress the overlap with excited
hadrons, allowing us to use smaller distances in our fits.
The Coulomb-wall correlators also have smaller statistical errors.
We fit the correlators from the Coulomb-wall and random-wall sources simultaneously with different amplitudes for each source but
common masses.
The ground-state amplitude from the random-wall source gives the decay constant, but the Coulomb-wall source helps in accurately
fixing the ground state mass, which in turn improves the determination of the random-wall amplitude.
Figure~\ref{fig:correlators_both} shows an example of heavy-light pseudoscalar correlators from the $a\approx0.042$~fm physical
quark-mass ensemble for the light-charm and strange-charm masses, showing the smaller excited state contamination in the
Coulomb-wall correlator.

In all cases the sink operator is point-like, with quark and antiquark propagators contracted at each lattice sites.
We sum the correlators over all spatial slices to project onto zero three-momentum.

The source time slices are equally spaced throughout the lattice.
The location of the first source time slice varies from configuration to configuration by adding an increment close to one half the
source separation, but such that all source slices are eventually used.
For example, on the $a\approx0.042$~fm physical quark-mass ensemble, where we use six source time slices with a separation $t/a=48$,
the location of the first source time slice on the $N^\text{th}$ configuration is $19N$~mod~48, or a shift of 19 slices between
successive configurations.
Meson masses and decay constants are obtained from fitting to these correlators.
For the light-light mesons, we include contributions from the ground state and one opposite parity state in the fit function, taking
a large enough minimum distance to suppress excited states.
This procedure works well for the light-light pseudoscalars, for which broken chiral symmetry makes the ground state mass much
lighter than all the excited state masses.

Because the heavy-light correlators are noisier than the light-light correlators, and the gap in mass between the ground state and
excited states is smaller, we include smaller distances and more states in the two-point correlator fits.
The fits that yield the central values employed in the subsequent EFT analysis include three states with negative parity
(pseudoscalars) and two states with the opposite parity, corresponding to the oscillations in~$t$ seen in
Fig.~\ref{fig:correlators_both}.
We refer to these as ``3+2'' state fits.
For these fits, the minimum distances and fit ranges used vary with the heavy-quark mass.
However, they are kept constant in physical units across all ensembles with different sea-quark masses and lattice spacings, subject
to being truncated to an integer in lattice units.
In these fits, the mass gaps are constrained with Gaussian priors~\cite{Lepage:2001ym}, but the amplitudes are left unconstrained.
Table~\ref{tab:priors} shows the constraints on the mass gaps used in the heavy-light correlator fits.
\begin{table}
    \caption{Bayesian prior constraints on the mass splittings used in our heavy-light correlator fits.
        Here $M_0$ is the ground-state mass, $M_1$ and $M_2$ are the first and second same-parity excited-state masses,
        and $M_0'$ and $M_1'$ are the ground and first excited-state opposite-parity masses.}
    \label{tab:priors}
    \vspace{0.1in}
    \begin{tabular}{ccccc}
        \hline\hline
        $N_\text{states}$ &    $M_0'-M_0$ (MeV)     &     $ M_1-M_0$ (MeV)   &   $M_1'-M_0'$ (MeV)   & $M_2-M_1$ (MeV) \\
        \hline
        3+2               & $400 \pm 200$ & $700\pm 200$ & $700\pm 70$ & $700\pm 60$  \\
        2+1               & $400 \pm 200$ & $700\pm 200$ &      na         &      na \\
        \hline\hline
    \end{tabular}
\end{table}
Although we use loose priors for the lower splittings, tighter priors are needed for the higher splittings to ensure stable fits.

\begin{figure}
    \includegraphics[trim={0.4in 1.75in 0.4in 1.75in},clip,width=0.48\textwidth]
        {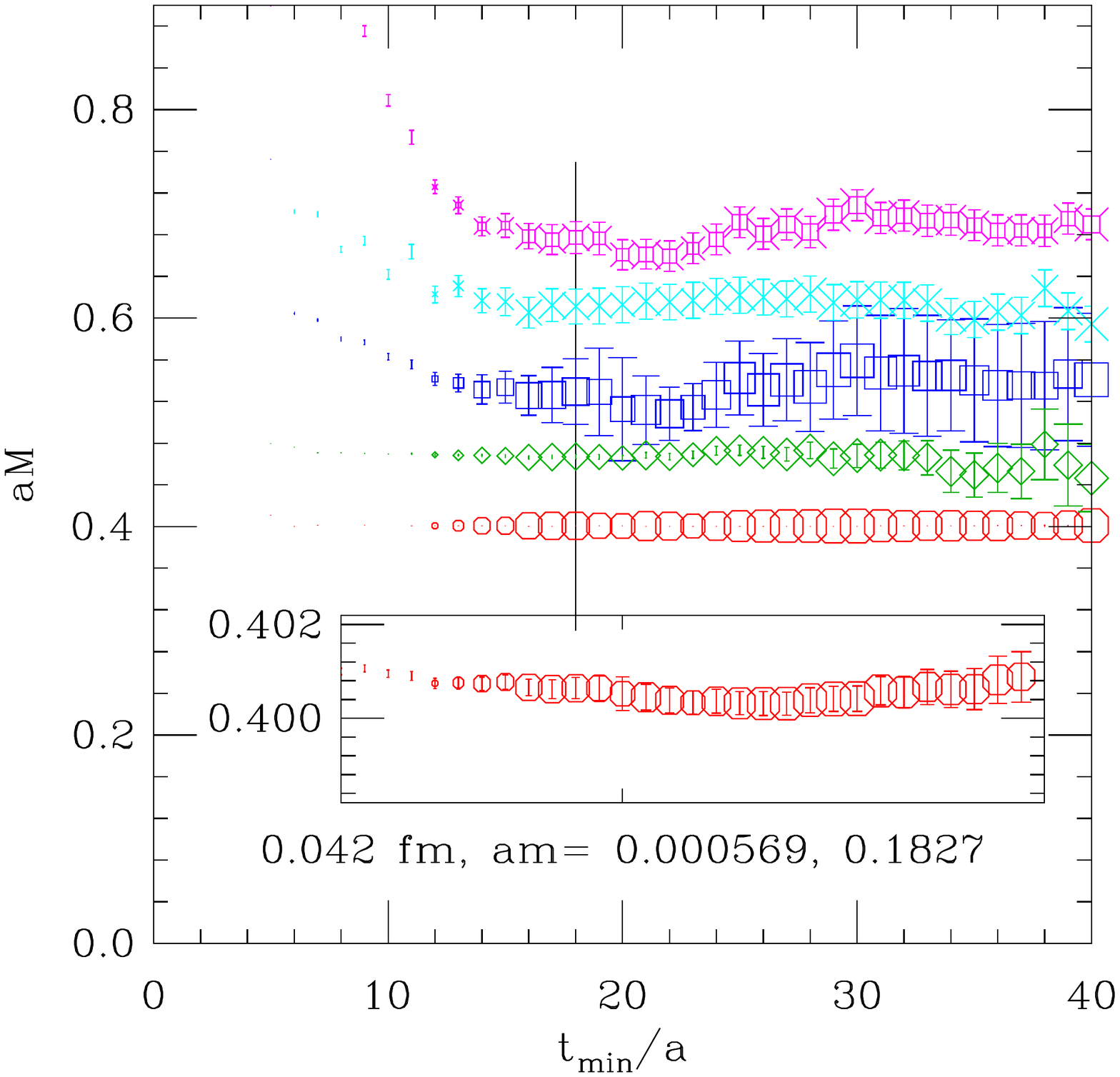} \hfill
    \includegraphics[trim={0.4in 1.75in 0.4in 1.75in},clip,width=0.48\textwidth]
        {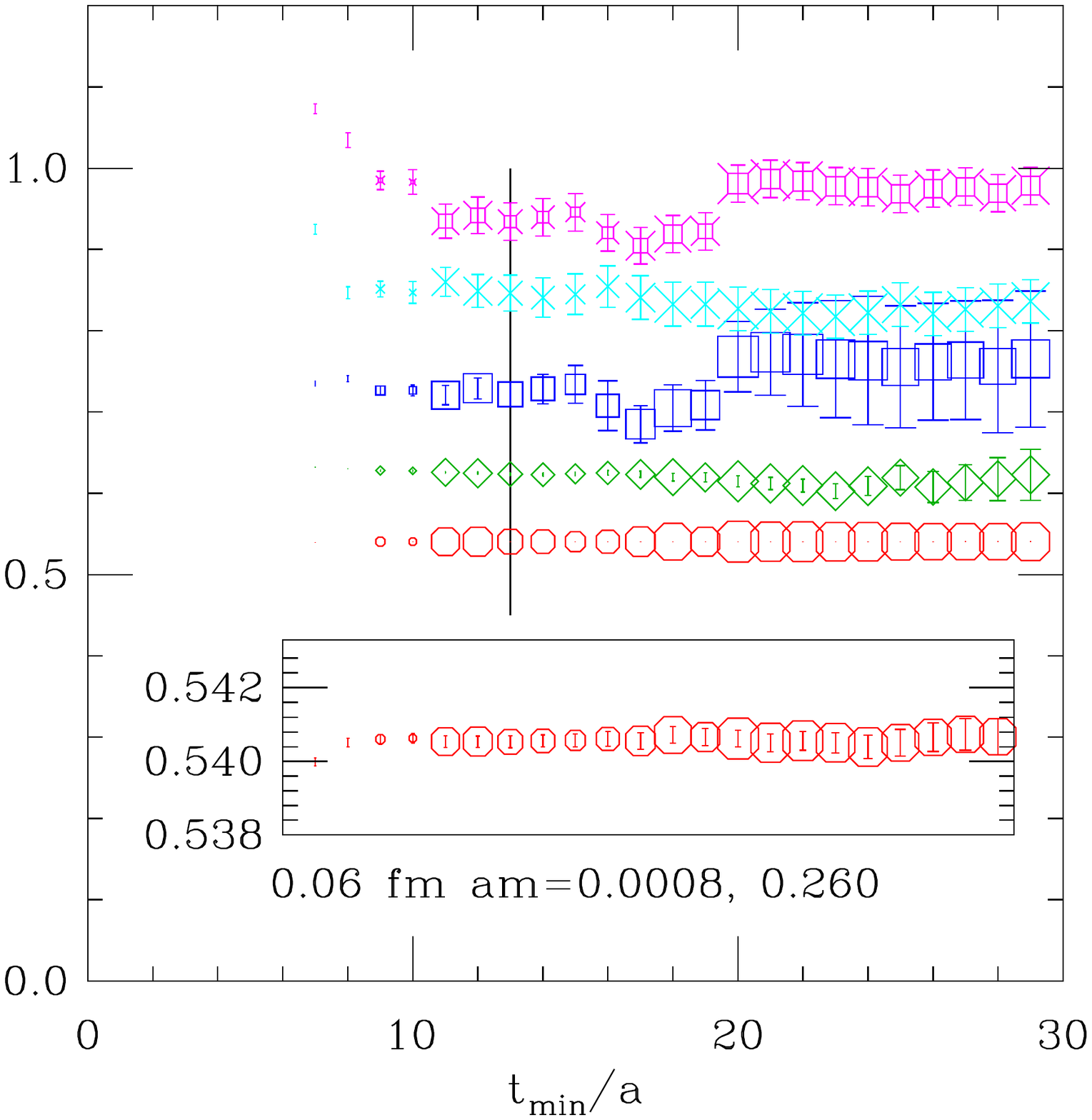} \\ 
    \includegraphics[trim={0.4in 1.75in 0.4in 1.75in},clip,width=0.48\textwidth]
        {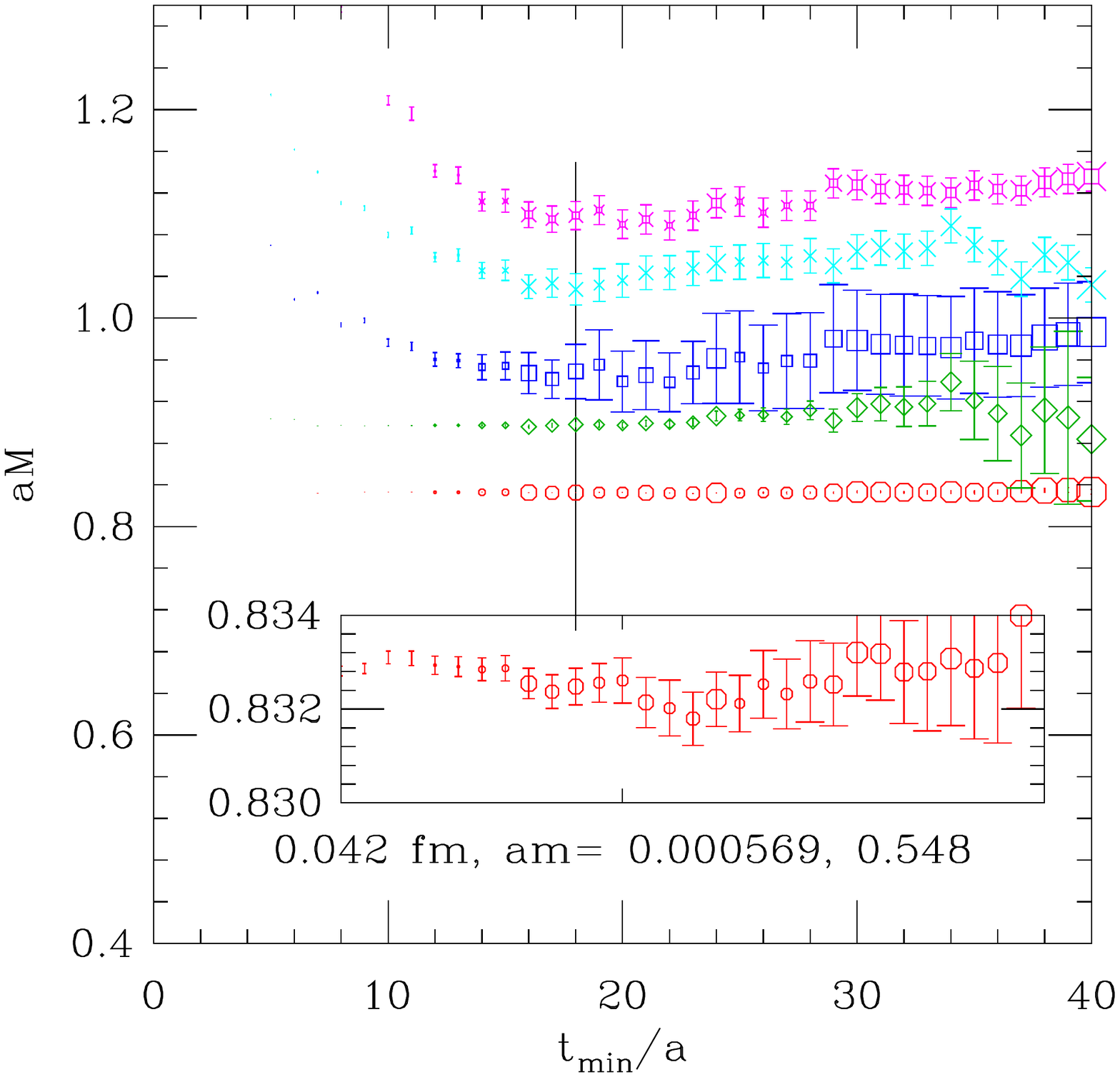}  \hfill 
    \includegraphics[trim={0.4in 1.75in 0.4in 1.75in},clip,width=0.48\textwidth]
        {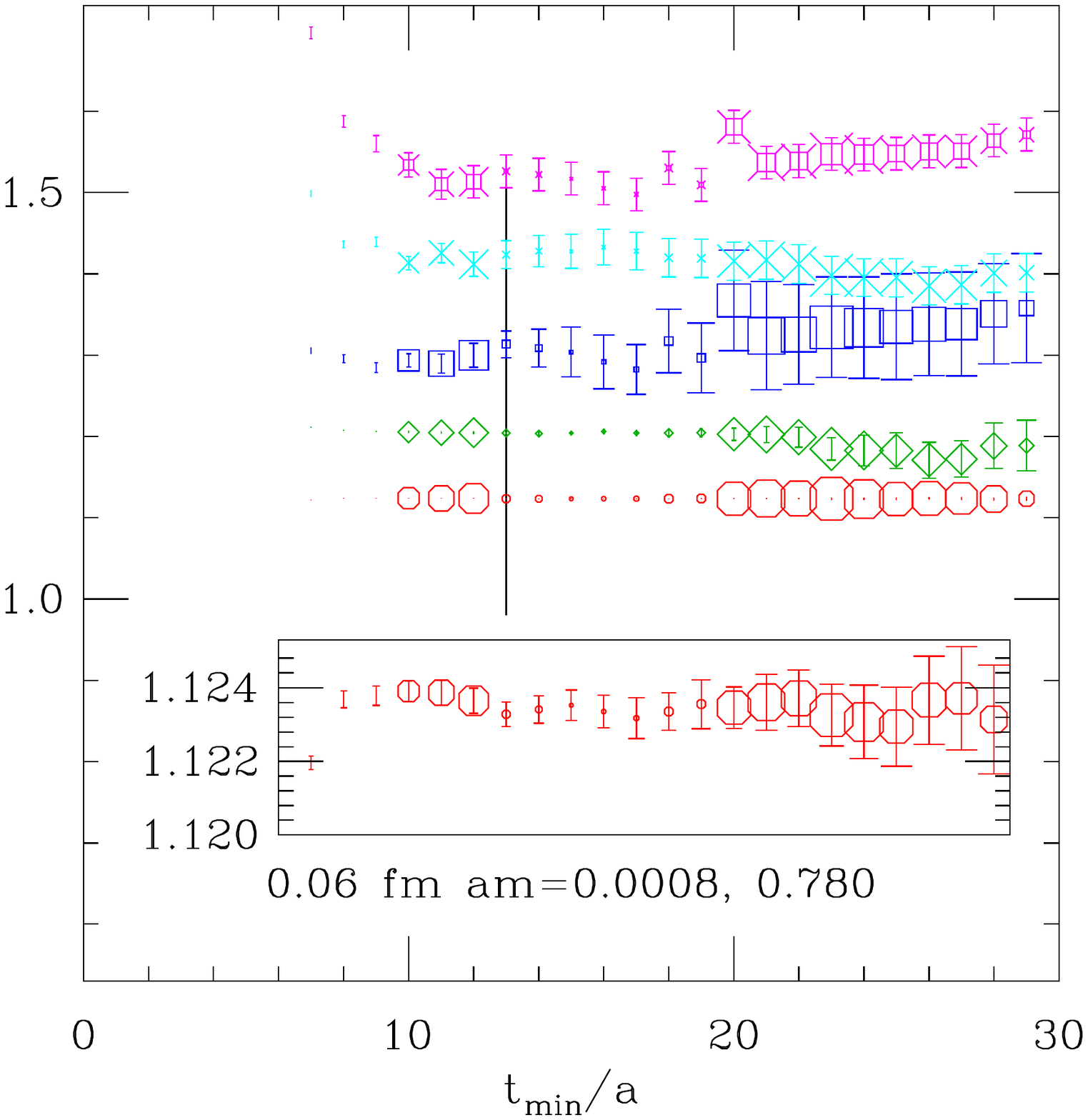}
    \caption{(\emph{top}) Masses for all 3+2 states in light-charm fits on the $a\approx0.042$~fm (left) and 0.06~fm (right) physical 
    quark-mass ensembles versus the minimum distance used for the random-wall correlator.
    (\emph{bottom}) Same as top panels but for light-heavy fits where the heavy-quark mass is three times the charm quark mass.
    The vertical and horizontal ranges in all plots are matched in physical units.
    The vertical lines show the minimum distance in the random wall source correlators used in our analysis.
    The inserts show the ground-state mass with an expanded vertical scale.
    The size of the symbol is proportional to the $p$-value of the fit, with a $p$-value of 0.5 corresponding to the size 
    of the label text.}
    \label{fig:fits_5masses_both}
\end{figure}

Figure~\ref{fig:fits_5masses_both} shows the masses of the five fitted states as a function of the minimum distance included in the
fit on the $a\approx0.042$~fm (left) and 0.06~fm (right) physical quark-mass ensembles.
In this plot, the size of the symbols is proportional to the quality of the fit~$p$.
We compute the $p$ values of our fits using the augmented $\chi^2$ that includes both data and prior contributions, and counting the
degrees of freedom as the number of data points minus the number of unconstrained fit parameters.
Thus it provides a measure of the compatibility of the fit result with both the data and the prior constraints.
At small $t_\text{min}$ the $p$-value is poor, and more states would be required to get a good fit.
At intermediate distances, the masses are mostly determined by the data, while at the largest distances the fit simply returns the
prior central values and errors for excited-state and opposite-parity masses.
We also perform heavy-light fits using 2+1 states with larger minimum distances as a check, and use the difference between results
of the 3+2 state fits and 2+1 state fits to estimate systematic errors coming from excited state contamination.
Based on studies like Fig.~\ref{fig:fits_5masses_both} on every ensemble, we choose the minimum distances $t_\text{min}/a$ so that
$t_\text{min}$ is as close as possible to the minimum distances given in Table~\ref{tab:dmin}.
As seen in this table, we use a slightly smaller $t_\text{min}/a$ for the Coulomb wall source since these correlators have smaller
excited state contamination than the random wall source correlators.

\begin{figure}
    \includegraphics[trim={0.in 1.75in 0.4in 1.875in},clip,width=0.50\textwidth]{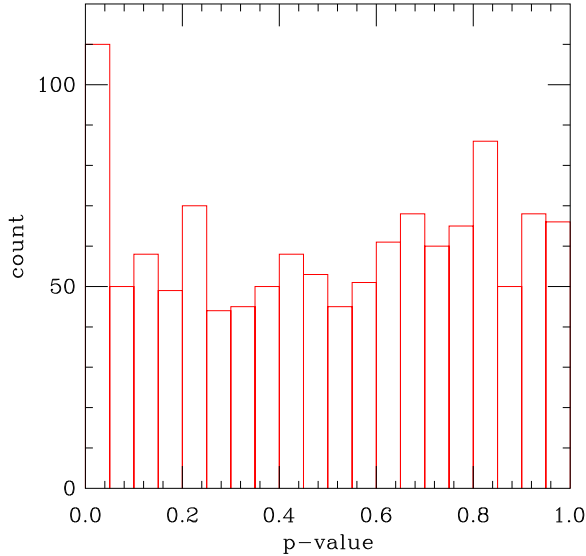}
    \caption{Distribution of $p$ values for our preferred two-point correlator fits in Table~\ref{tab:dmin}.}
    \label{fig:pvalue_hist}
\end{figure}

We expect the $p$ values to be approximately uniformly distributed, with possible systematic deviations from uniformity coming from
artificially loose or tight priors on the mass gaps, and, more importantly, neglecting effects of autocorrelations on the covariance
matrix of the correlator at different distances.
Figure~\ref{fig:pvalue_hist} shows the distribution of $p$ values for our full set of correlator fits using the fit ranges and
number of states in Table~\ref{tab:dmin}.
It is approximately uniform from $0$ to $1$, indicating that we have not introduced any systematic bias in our fits from the choice
of fit ranges or number of states.
Because the $p$-values from correlators with different valence-quark masses in the same ensemble are strongly correlated, the
statistical fluctuations in this histogram are larger than the expectation $1/\sqrt{N}$ for independent data.

\begin{table}
    \newcommand{\h}{\phantom{1}}
    \newcommand{\s}{$^*$\kern0.1em}
    \caption{Minimum distances used in our two-point correlator fits.
    Here ``light'' quarks include masses up to $m_s$ and ``heavy'' quarks masses beginning at $m_c$, and the two numbers in the
    second column are the number of pseudoscalar and opposite-parity states included in the fit.
    The ``$*$'' indicates that this minimum distance is actually taken to depend weakly on the heavy quark mass,
    with the quoted distance the one used for the $D_s$ correlator.}
    \label{tab:dmin}
    \vspace{0.1in}
    \begin{tabular}{lc@{\quad}c@{\quad}c}
        \hline\hline
               meson & $N_\text{states}$ & random wall & Coulomb wall \\
        \hline
        light-light & 1+1 &  2.3\h~fm  & 2.1\h~fm  \\
        heavy-light & 3+2 &  0.77~fm   & 0.68~fm \\
        heavy-light & 2+1 &  1.13\s fm & 1.01~fm \\
        heavy-heavy & 3+2 &  0.80~fm   & 0.68~fm \\
        heavy-heavy & 2+1 &  1.40~fm   & 1.28~fm \\
        \hline\hline
    \end{tabular}
\end{table}

\newcommand{\MF}{M$\Phi$}

In order to subsequently fit the decay constants and masses obtained from these two-point correlator fits to an EFT function of the
quark masses and lattice spacing, we need an estimate of the covariance matrix between these data.
(Here the heavy-light decay constant is to be understood as $\Phi$.) To distinguish this covariance matrix from the matrix of
covariances of the correlators at different distances used in the two-point fits, in this section we refer to matrices of
covariances of masses~$M$ and decay constants~$\Phi$ as ``\MF\ covariance matrices''.
In the \MF\ covariance matrix, all of the amplitudes and decay constants for different sets of valence quark masses are correlated,
while those from different ensembles are uncorrelated.
Thus, the \MF\ covariance matrix is a large block-diagonal matrix, with each block corresponding to a single ensemble.

To obtain each block of the \MF\ covariance matrix, we use a single-elimination jackknife procedure, omitting one configuration at a
time from the two-point fits.
This approach does not account for autocorrelations.
Unfortunately, however, it is not practical to eliminate large enough blocks in the jackknife to suppress the autocorrelations,
since we need a number of jackknife blocks that is large compared to the dimension of the block of the \MF\ covariance matrix for
that ensemble.
We therefore use an approximate procedure.
We first compute the block of the \MF\ covariance matrix from the single-elimination jackknife, and then compute the dimensionless
correlation matrix by rescaling rows and columns so that the diagonal elements are one.
Next we compute the diagonal elements of the \MF\ covariance matrix (that is, the variances of the masses and decay constants) using
a block size large enough to reasonably well suppress the effects of autocorrelation, and rescale the rows and columns of the \MF\
covariance matrix to set its diagonal elements equal to the variances obtained from blocking.
On all ensembles with $a\gtrsim0.09$~fm, we blocked the configurations by four; we used larger block sizes of up to 24
configurations on ensembles with finer lattice spacings to account for the longer autocorrelation times.
This approach uses the single-elimination jackknife to determine the (dimensionless) correlations of all the masses and decay
constants, and the blocked jackknife, which accounts for autocorrelations between gauge-field configurations, to determine the
variances of each mass or decay constant.

The \MF\ covariance matrix used in the EFT fit affects the $p$-value of the fit and the central values obtained for the
decay constants at the physical quark masses and in the continuum limit.
The statistical errors on the masses and decay constants in the \MF\ covariance matrix range from $0.005\%$ to $0.12\%$ and $0.04\%$
to $1.4\%$, respectively.
The statistical errors quoted on the physical, continuum-limit decay constants are, however, obtained by an overall jackknife
procedure, where we repeat the entire fitting chain 20 times, each time omitting $1/20$ of the configurations from each ensemble.

\section{Lattice spacing and quark-mass tuning}
\label{sec:physical-mass-analysis}

Tuning the masses of the light and charm quarks and the determination of the lattice spacings follow the procedure described in
detail in Ref.~\cite{\rFD2014}.
In this procedure, we use the meson masses and decay constants in the physical quark mass ensembles (with a small correction for
mistuned light quark mass), extrapolated to the continuum, to find the $u$, $d$, $s$, and $c$ quark masses used in subsequent steps,
and the lattice spacings of each ensemble.
For setting the overall scale we use the pion decay constant~$f_\pi$.
We also compute an intermediate scale $f_{p4s}$, the decay constant of a fictitious pseudoscalar meson with degenerate valence quark
with mass $m_{p4s}=0.4m_s$.
To obtain $f_{p4s}$ and the associated meson mass $M_{p4s}$, we draw quadratic functions in the valence-quark
mass through the decay-constant and meson-mass data with degenerate valence quarks at 0.3, 0.4 and 0.6 times $m'_s$, and evaluate
these quadratic functions at 0.4 times the tuned strange quark mass $m_s$.
The quantity $f_{p4s}$ is convenient since it has small statistical errors and can be computed without light valence quark mass
correlators.
This feature is essential for the $0.03$~fm ensemble where the lightest valence quark mass is $m'_s/5$, so an extrapolation to
$f_\pi$ on this ensemble would have large errors.

An initial value for the charm quark mass comes from matching the $D_s$ mass.
With this $m_c$ and the light quark masses, we evaluate the masses of the $D^0$ and $D^+$ mesons.
The difference between them, $2.6$~MeV, can be considered to be the part of the $D^+$-$D^0$ mass difference
coming from the difference in the up and down quark masses.
In \secref{Errors}, this quantity is denoted $C(m_d-m_u)$ and used to estimate the electromagnetic contribution to the mass
splitting.

As discussed in Sec.~\ref{sec:Lattice-Simulations}, the main new aspects of this work are the
addition of three new ensembles  and the increased statistics on some of the others.
We also make some minor updates of the input parameters.
The value of $f_\pi$, used to set the scale, has been updated to $130.50\pm 0.13$~MeV following the
PDG~\cite{\rPDG2016}, and the experimental neutral kaon and charmed meson masses have also seen slight changes.

In contrast with Ref.~\cite{\rFD2014}, we now use the strong coupling $\alpha_V$ at scale $q=2.0/a$ obtained from
Ref.~\cite{Chakraborty:2014aca,*Chakraborty:2017aca} in our central fit, and use $\alpha_T$, inferred from taste splittings, in an
alternative fit to estimate systematic errors.

We also update the quantities $(M^2_{K^0})^\gamma$ and $\epsilon'$, which describe electromagnetic
effects, to reflect the most recent results from the MILC Collaboration \cite{Basak:2014vca,Basak:2015lla,Basak:2018yzz}.
The quantity~$(M^2_{K^0})^\gamma$ is the electromagnetic contribution to the squared mass of the neutral kaon.
The quantity~$\epsilon'$ captures
higher-order corrections to Dashen's theorem:
\begin{equation}
    \epsilon' \equiv \frac{(M^2_{K^\pm}-M^2_{K^0})^\gamma - (M^2_{\pi^\pm}-M^2_{\pi^0})^\textrm{expt}}
        {(M^2_{\pi^\pm}-M^2_{\pi^0})^{\textrm{expt}}}.
    \label{eq:epsp-def}
\end{equation}
We use $\epsilon'$ rather than the closely related quantity $\epsilon$ defined in 
\rcite{Aoki:2016frl} as
\begin{equation}
    \epsilon \equiv \frac{(M^2_{K^\pm}-M^2_{K^0})^\gamma - (M^2_{\pi^\pm}-M^2_{\pi^0})^\gamma}
        {(M^2_{\pi^\pm}-M^2_{\pi^0})^{\textrm{expt}}}.
    \label{eq:eps-def}
\end{equation}
Because the experimental pion splitting is largely due to electromagnetism, $\epsilon$ and $\epsilon'$ are close in size.
The difference is estimated in \rcites{Gasser:1984gg,Aoki:2016frl} to be
\begin{equation}
    \label{eq:eps-epsp-diff}
    \epsilon-\epsilon' \equiv \epsilon_m = 0.04(2),
\end{equation}
which is used to find~$\epsilon'$.

In this paper, we use~\cite{Basak:2018yzz}
\begin{align}
    \epsilon^\prime &= 0.74(1)_{\rm stat}({}^{+\phantom{1}8}_{-11})_{\rm syst},
    \label{eq:eps-all-errors} \\
    (M^2_{K^0})^\gamma &= 44(3)_{\rm stat}(25)_{\rm syst}~\MeV^2 .
    \label{eq:K0-result}
\end{align}
Our adjusted kaon masses, or ``QCD masses'', are then found from 
\begin{align}
  ({M_{K^+}^2})^\text{QCD} &=  M_{K^+}^2 - \left( 1+\epsilon' \right)
      \left( M_{\pi^+}^2 - M_{\pi^0}^2 \right) - (M^2_{K^0})^\gamma, \\
  ({M_{K^0}^2})^\text{QCD} &=  M_{K^0}^2 - (M^2_{K^0})^\gamma .
  \label{eq:K_mass_adjust}
\end{align}
These quantities are used to match pure QCD to the more fundamental QCD+QED.
Consequently, any pure QCD calculation will have uncertainties coming from the particular scheme for separating electromagnetic and
isospin effects.
Our scheme is the one introduced for $u$ and $d$ quarks in \rcite{Borsanyi:2013lga} and extended naturally to the $s$ quark using
the fact that mass renormalization for staggered quarks is multiplicative~\cite{Basak:2018yzz}.
As an estimate of the change that would result from the use of a different, but still reasonable, scheme, MILC compares to a scheme
where the EM mass renormalization is calculated perturbatively (at one loop).
While the resulting scheme dependence of $\epsilon'$ is small, $\pm0.038$~\cite{Basak:2018yzz}, that of $(M^2_{K^0})^\gamma$ is
$\sim420~\MeV^2$, much larger than the errors in this quantity in a fixed scheme, although still small compared to $M_{K^0}^2$.%
\footnote{A preliminary value for $(M^2_{K^0})^\gamma$ was reported in \rcite{Basak:2013iw}. That
result did not yet take into account EM quark-mass renormalization and is thus not reliable.}

Table~\ref{tab:QCD_masses} summarizes the experimental masses that we use, and also the ``QCD masses''
where we have made the adjustments for electromagnetic effects described above, and the adjustments
for the heavy meson masses from Eq.~(\ref{eq:EM-model}) in Sec.~\ref{sec:Errors}.

\begin{table}
\caption{Experimental inputs to our tuning procedure (left side) \protect\cite{Olive:2016xmw},
and the meson masses after adjusting for electromagnetic effects (right side).
}
\label{tab:QCD_masses}
\begin{tabular}{r@{\,=\,}l@{\qquad}r@{\,=\,}l}
\hline\hline
\multicolumn{2}{c}{Experimental inputs} & \multicolumn{2}{c}{QCD masses} \\
 \hline
$f_{\pi^+}$ & $130.50(1)_\text{exp.}(3)_\text{$V_{ud}$}(13)_\text{EM}$ MeV & \multicolumn{2}{c}{} \\
$M_{\pi^0}$ & 134.9770 MeV & ($M_\pi)^\text{QCD}$ & 134.977 MeV \\
$M_{\pi^+}$ & 139.5706 MeV & \multicolumn{2}{c}{} \\
$M_{K^0}$ & 497.611(13) MeV & ($M_{K^0})^\text{QCD}$ & 497.567 MeV \\
$M_{K^+}$ & 493.677(16) MeV & ($M_{K^+})^\text{QCD}$ & 491.405 MeV \\
$M_{K^0}-M_{K^+}$ & 3.934(20) MeV & \multicolumn{2}{c}{} \\
$M_{D_s}$ & 1968.28(10) MeV & ($M_{D_s})^\text{QCD}$ & 1967.02 MeV \\
$M_{B_s}$ & 5366.89(19) MeV & ($M_{B_s})^\text{QCD}$ & 5367.11 MeV \\
\hline\hline
\end{tabular}
\end{table}

We extrapolate the scale-setting quantities $f_{p4s}$ and $M_{p4s}$ and the quark-mass ratios $m_u/m_d$, $m_s/m_l$, and $m_c/m_s$ on
the physical quark-mass ensembles to the continuum using a quadratic function in $\alpha_s a^2$.
The fit of $m_c/m_s$ including all lattice spacings is poor, with $p=0.01$, because discretization errors from the charm quark are
large at our coarsest lattice spacing.
The $m_c/m_s$ fit improves substantially to $p=0.8$ when the $a\approx 0.15$~fm data are omitted.
In an analysis of the heavy-light-meson masses in Ref.~\cite{Bazavov:2018omf}, we encounter similar problems when including data
from the $a\approx0.15$~fm ensembles.
We therefore omit the $a\approx0.15$~fm ensembles from our central continuum extrapolations here, in Ref.~\cite{Bazavov:2018omf},
and in the EFT analysis of the heavy-light decay constants in Sec.~\ref{sec:chiral-analysis}.
For estimating systematic errors from our choice of continuum extrapolation of scale-setting quantities, we also consider a fit
quadratic in $\alpha_s a^2$ including all five physical quark-mass ensembles (as was done in Ref.~\cite{\rFD2014}), a fit linear in
$\alpha_s a^2$ omitting the $0.15$~fm ensemble, a fit linear in $\alpha_s a^2$ omitting both the $0.15$~fm and $0.12$~fm ensembles,
and a fit using $\alpha_s$ inferred from taste violations.

\begin{figure}
    \includegraphics[width=0.5\textwidth,trim={1cm 4.5cm 0.5cm 4cm},clip]{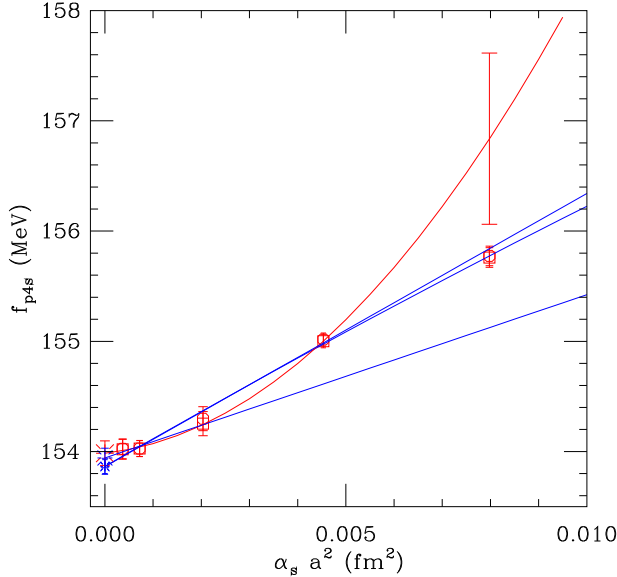}
    \caption{Continuum extrapolations for $f_{p4s}$ on the physical quark mass ensembles.
        Our central fit, shown in red, is quadratic in $\alpha_s a^2$ excluding the 0.15~fm data.
        Alternative fits used for estimating systematic error are shown in blue.
        These include a quadratic fit including all the data, a linear fit including data up to 0.12~fm, and a linear fit including
        data up to 0.09~fm.
        The large error bar on the central fit line shows the statistical error on this fit at $0.15$~fm, the point that is not
        included in this fit.}
    \label{fig:fp4s_extrap}
\end{figure}

Figure \ref{fig:fp4s_extrap} shows these extrapolations for the intermediate scale $f_{p4s}$.
In this fit, as in the other quantities discussed in this section, the central fit, shown
in red, is at one end of the various extrapolations to $a=0$.
We therefore assign a one-sided systematic error from continuum extrapolations equal to the difference between this continuum
extrapolation and the furthest of our alternative fits.

We assign five distinct systematic uncertainties to scale-setting quantities and quark-mass ratios stemming from
electromagnetic effects, and tabulate them in \tabref{lightemerrors}.
\begin{table}
\caption{Electromagnetic errors on, and estimates of scheme dependence of scale-setting parameters, quark mass ratios,
and, for convenience, the phenomenologically interesting ratio $f_{K^+}/f_{\pi^+}$, and
the ratio of the kaon to pion decay constants in the isospin symmetric limit, $f_{K}/f_\pi$.
}
\label{tab:lightemerrors}
\begin{tabular}{lcccccccc}
\hline\hline
Error (\%) & $f_{p4s}$ & $M_{p4s}$ & $f_{p4s}/M_{p4s}$ & $m_u/m_d$ & $m_s/m_l$ &  $m_c/m_s$ & $f_{K^+}/f_{\pi^+}$ & $f_{K}/f_\pi$ \\
\hline
\EMone     & $_{-0.0033}^{+0.0045}$ & $_{-0.011}^{+0.015}$ & $_{-0.011}^{+0.008}$ & $_{-1.44}^{+1.98}$ & $_{-0.021}^{+0.029}$
    & $_{-0.032}^{+0.023}$ & $_{-0.006}^{+0.008}$ & $_{-0.000}^{+0.000}$ \\
\EMtwo     & 0.0014 & 0.006 & 0.003 & 0.000 & 0.011 & 0.012 & 0.001 & 0.007 \\
\EMthree   & na    & na    & na    & na    & na    & 0.109  & na    & na \\
\EMtwos    & 0.027 & 0.093 & 0.065 & 0.691 & 0.188 & 0.205 & 0.025 & 0.025 \\
\EMthrees  & na    & na    & na    & na    & na    & 0.365  & na    & na \\
\hline\hline
\end{tabular}
\end{table}
The first of these, labeled ``\EMone,'' is obtained by shifting $\epsilon'$ by the lower error bar, $-0.11$, in \eq{eps-all-errors},
and the error in the other direction is obtained by scaling by $-8/11$.
Varying the result for $(M^2_{K^0})^\gamma$ in \eq{K0-result} by its total error gives the second error, labeled ``\EMtwo.'' %
The uncertainty labeled ``\EMtwos'' is an estimate of the
variation that would be produced by matching QCD+QED to pure QCD in an alternative reasonable scheme.
This is not taken to be a systematic error in our results, since we work in a fixed, well-defined, scheme.
However, when using our results in a setting that does not take into account the subtleties of the EM scheme, one may wish to
incorporate the estimate of scheme-dependence as an additional uncertainty.
The two remaining electromagnetic uncertainties, which are discussed in more detail in
Sec.~\ref{sec:Errors}, arise from electromagnetic effects on the relevant heavy-light meson masses.
In fact, only the EM effect on the mass of the $D_s$, used to fix the charm quark mass, is needed here.
From the estimates in Sec.~\ref{sec:Errors}, this effect is about 1.3~MeV,
which is subtracted from the experimental $D_s$ mass before tuning the charm-quark mass,
and 100\% of the resulting shift is included in our systematic error estimates in the column labeled
``\EMthree.'' %
Scheme dependence arises again in the EM contribution to the $D_s$
mass, and we estimate it at 4.2~MeV in Sec.~\ref{sec:Errors}.
The resulting uncertainty is listed in the column labeled ``\EMthrees.'' %
The three uncertainties that do not arise from the choice of scheme, namely \EMone, \EMtwo,
and \EMthree, are summed in quadrature to give the error labeled ``Electromagnetic corrections'' in the full error budget, \tabref{lightresults2}.

Another systematic error comes from possible incomplete adjustments for the effects of incorrect sampling of the distribution of the
topological charge.
Using the corrections found in Ref.~\cite{Bernard:2017npd} and described in Sec.~\ref{subsec:topology}, we adjust the meson
masses and decay constants on the 0.042 and 0.03~fm ensembles to compensate for the incorrect average of the squared topological
charge.
We conservatively take 100\% of the effects of this adjustment as a systematic error coming from poor sampling of the topological
charge distribution.

Corrections for finite spatial volume are estimated by the same procedure as in Ref.~\cite{\rFD2014}, where our central fit includes
adjustments calculated in NLO staggered chiral perturbation theory, and an associated systematic error is taken to be the difference
between this adjustment and using nonstaggered finite-volume chiral perturbation theory, at NNLO for $M_\pi$ and $f_\pi$, and NLO
for $M_K$ and~$f_K$.
These estimates are considerably smaller than in Ref.~\cite{\rFD2014} because we have now dropped from the central fit the
coarsest ensembles, with $a\approx0.15$~fm, which dominate the earlier estimate.
The taste-splittings at the next coarsest lattice spacing, $a\approx0.12$~fm, are about a factor of 2 smaller than at $a\approx0.15$
fm \cite{Bazavov:2012xda}, so the difference between staggered and nonstaggered chiral perturbation theory is correspondingly
reduced when the $a\approx0.15$~fm data are dropped.

Finally, we propagate the uncertainty in the PDG value of $f_\pi$.
The main effect is an overall scale error in dimensionful quantities.
Because the decay constants depend on quark masses, an indirect effect also arises, leading to an uncertainty on dimensionless
ratios, and a reduction in the uncertainty on dimensionful quantities, compared to the direct scale error.
For the ratio $m_u/m_d$ the experimental uncertainty in $M_{K^0}-M_{K^+}$ is also included.

Table~\ref{tab:lightresults2} shows the error budgets for the outputs of the scale-setting and quark-mass-ratio
analysis, which are used in the subsequent fitting of the heavy-light results.
\begin{table}
\caption{Error budgets in per cent for scale-setting parameters, quark mass ratios,
$f_{K^+}/f_{\pi^+}$, and $f_{K}/f_\pi$.
}
\label{tab:lightresults2}
\begin{tabular*}{\textwidth}{@{\extracolsep{\fill}}lcccccccc}
\hline\hline
Error (\%) & $f_{p4s}$ & $M_{p4s}$ & $f_{p4s}/M_{p4s}$ & $m_u/m_d$ & $m_s/m_l$ & $m_c/m_s$ & $f_{K^+}/f_{\pi^+}$ & $f_{K}/f_\pi$ \\
\hline
Statistics & 0.072 & 0.033 &  0.080 &  1.20  &  0.17  &  0.12  &  0.13  & 0.10 \\
% \hline
Continuum extrapolation    & $_{-0.078}^{+0}$ & $_{-0}^{+0.036}$ & $_{-0.10}^{+0}$  & $_{-0}^{+1.47}$  & $_{-0}^{+0.24}$
    & $_{-0.47}^{+0}$  & $_{-0.14}^{+0}$ & $_{-0.12}^{+0}$ \\
Electromagnetic corrections  & $_{-0.004}^{+0.005}$ & $_{-0.012}^{+0.016}$ & $_{-0.011}^{+0.008}$ &
    $_{-1.45}^{+1.99}$ & $_{-0.024}^{+0.031}$ & $_{-0.115}^{+0.112}$ & $_{-0.007}^{+0.010}$ & $_{-0.003}^{+0.004}$ \\
Topological-charge distribution  & 0.001 & 0.000  &  0.001  &  0.040  &  0.061  &  0.001  &  0.012  & 0.012 \\
Finite-volume corrections & 0.011  & 0.001  & 0.009  & 0.081  & 0.059  & 0.002  & 0.021  & 0.016 \\
% \hline
$\fpiPDG$ & 0.075   &  0.001   & 0.075   & 0.010   & 0.004   & 0.051   & 0.023   & 0.024 \\
$\Delta M_K$ & 0.000   &  0.000   & 0.000   & 0.283   & 0.000   & 0.000   & 0.001   & 0.000 \\
\hline\hline
\end{tabular*}
\end{table}
The central values for these quantities are listed in \secref{light_results}.

\section{Effective-field-theory analysis}
\label{sec:chiral-analysis}

\begin{figure}%[b]
    \includegraphics[width=0.65\textwidth,trim={0.25in 0.75in 0.125in 0.125in},clip]{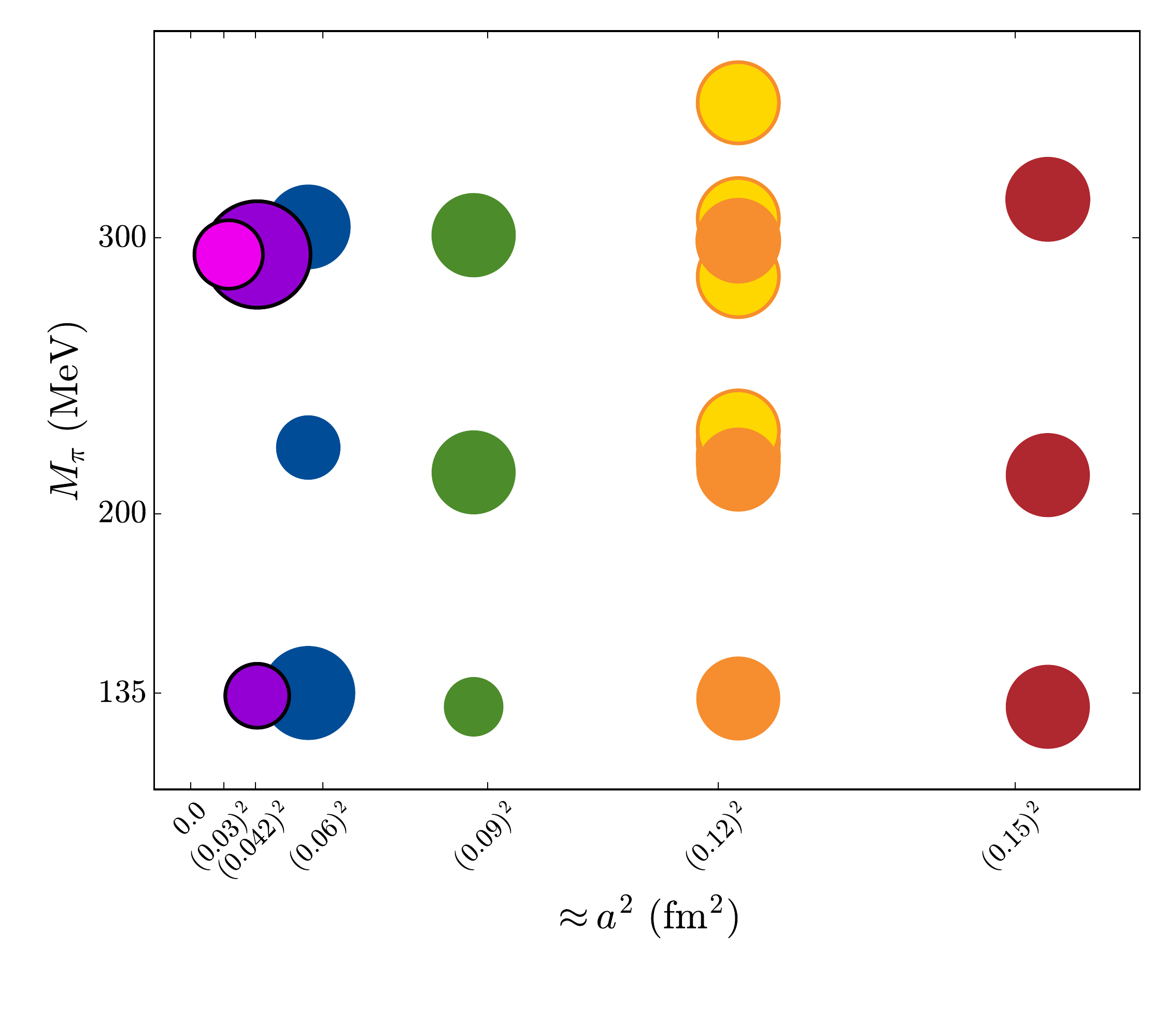}
    \caption{Distribution of four-flavor QCD gauge-field ensembles used in this work.
        Ensembles that are new with respect our previous analysis~\cite{\rFD2014} are indicated with black outlines.
        Ensembles with unphysical strange-quark masses are shown as gold disks with orange outlines.
        The area of each disk is proportional to the statistical sample size $N_\text{conf}\times N_\text{src}$.
        The physical, continuum limit is located at $(a=0, M_\pi \approx 135~\MeV)$.}
    \label{fig:ensembles}
\end{figure}

\begin{figure}
    \includegraphics[width=0.65\textwidth]{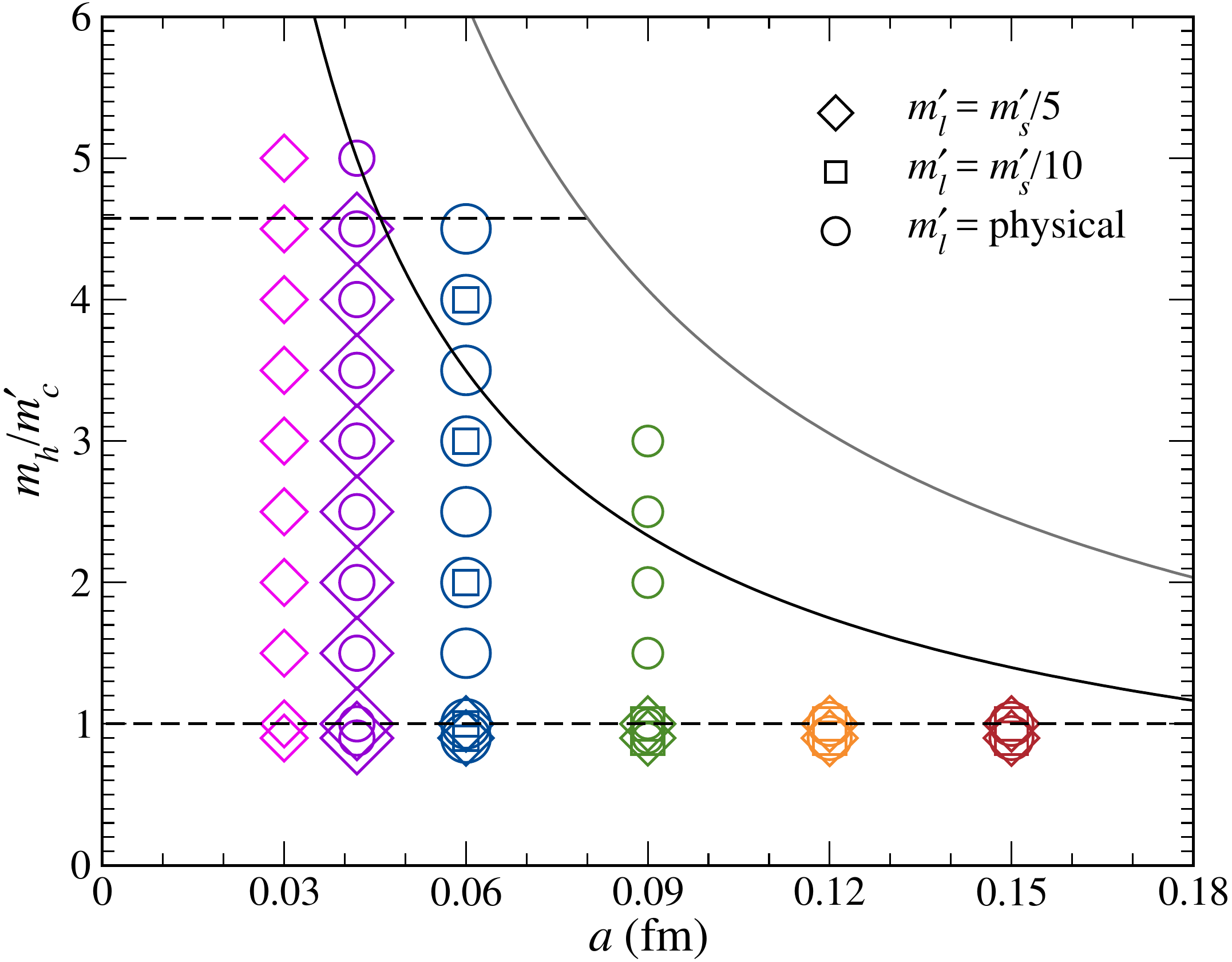}
    \caption{Valence heavy-quark masses vs.\ lattice-spacings of ensembles used in this calculation, in units of the simulation 
        charm sea-quark mass.
        Symbol shapes indicate the value of the light sea-quark masses, with diamonds, squares, and circles corresponding to
        $m'_l=m'_s/5$, $m'_s/10$, and physical, respectively.
        The symbol area is proportional to the statistical sample size. 
        The black (gray) hyperbola shows $am_h=0.9$ ($am_h=\pi/2$).
        The horizontal dashed lines indicate the physical bottom and charm masses.}
    \label{fig:mh_vs_a}
\end{figure}

In this section, we discuss how we combine the lattice data for the meson masses and decay constants described in the previous
sections to obtain continuum-limit, physical-quark-mass results.
There are two crucial features of our data set.
First, as discussed in Sec.~\ref{sec:Lattice-Simulations}, the range of parameters is broader than that commonly encountered in
lattice-QCD calculations.
Figure~\ref{fig:ensembles} shows the lattice spacings and pion masses of the ensembles used in our analysis.
The lattice spacing spans the range $0.03~\fm\lesssim a \lesssim 0.15~\fm$, while the light sea-quark mass lies between
$\half(m_u+m_d)\lesssim m'_l\lesssim 0.2m_s$.
With the HISQ action, it is possible to simulate physical charm and bottom quarks with controlled discretization errors.
Figure~\ref{fig:mh_vs_a} shows the range of valence heavy-quark masses used in our analysis.
On the coarsest $a \approx 0.15$ and 0.12~fm ensembles, we have only two values $m_h = 0.9m'_c$ and $m'_c$; on our finest
$a\approx0.042$ and 0.03~fm ensembles, however, we have several heavy-quark masses between $0.9m'_c\le m_h\le5m'_c$, reaching just
above the physical $b$-quark mass.
Second, as discussed in Sec.~\ref{sec:Correlator-Fits}, we have large statistical sample sizes, with about 4,000 samples on most
ensembles and large lattice volumes; the resulting errors on the decay constants range from $0.04\%$ to~$1.4\%$.

Because of the breadth and precision of the data set, it is a challenge to find a theoretically well-motivated functional form that
is sophisticated enough to describe the whole data set.
We therefore rely on several EFTs to parameterize the dependence of our data on each of the
independent variables just described: Symanzik effective field theory for lattice spacing
dependence~\cite{Symanzik:1983dc}, chiral perturbation theory for light- and strange-quark mass dependence, and heavy-quark
effective theory for the heavy-quark mass dependence.
These EFTs are linked together within heavy-meson rooted all-staggered chiral perturbation theory
(\aschpt)~\cite{Bernard:2013qwa}.  Here we use the one-loop \aschpt\ expression to describe the
nonanalytic behavior of the interaction between pion (and other pseudo-Goldstone bosons) and the heavy-light meson, and
supplement it with higher-order analytic functions in the light- and heavy-quark masses and lattice spacing to enable a good
correlated fit.

Even with these additional terms, however, the extrapolation $a\to0$ and the interpolation $m_h\to m_b$ oblige us
to restrict the range of $am_h$.
In practice, we are able to obtain a good correlated fit of our data with heavy-quark masses $am_h\le0.9$.
Note, however, that our final fit function describes even the data with $am_h>0.9$ quite well.

The rest of this section is organized as follows.
In Sec.~\ref{sec:chiral-function}, we construct an EFT-based fit function with enough parameters (60) to describe the data as a
function of the light- and heavy-quark masses and lattice spacing.
For convenience, the complete final expression is written out in Sec.~\ref{sec:zF}.
Next, Sec.~\ref{sec:p4s-scale-setting} explains how we convert our decay-constant data from
lattice units to ``$p4s$ units'' and, eventually, to MeV.
Finally, we describe how the fit works in practice and present our final fit used to obtain the decay-constant central values and
errors in Sec.~\ref{sec:chiral-fit}.

\subsection{Effective-field-theory fit function for heavy-light decay constants}
\label{sec:chiral-function}

Recall that $H_x$ denotes a generic heavy-light pseudoscalar meson composed of a light valence quark $x$ and a heavy valence
antiquark $\bar h$, with masses $m_x$ and $m_h$, respectively.
The decay constant and mass of $H_x$ are $f_{H_x}$ and $M_{H_x}$, respectively.
In heavy-quark physics, the conventional decay constant is defined and normalized as $\Phi_{H_x} \equiv f_{H_x} \sqrt{M_{H_x}}$.

We start with massless light quarks, with $\Phi_0$ and $M_0$ denoting the 
decay constant and the meson mass in this limit. 
We parametrize $\Phi_0$ as 
\begin{equation}
    \Phi_0 = {C\tilde{\Phi}_0}\left[1 + k_1 \left(\Minv\right) + k_2 \left(\Minv\right)^2 + \cdots \right],
    \label{eq:decay-constant:chiral-limit} 
\end{equation}
where $\tilde{\Phi}_0$ is the matrix element of the HQET current in the infinite-mass limit, 
$\Lambda_\text{HQET}$ is a physical scale for HQET effects that we set to 800~MeV in this analysis, 
and the Wilson coefficient $C$ arises from matching the QCD current and the HQET current~\cite{Ji:1991pr,Manohar:2000dt}
at scale~$m_h$:
\begin{equation} % See Eq. (3.83) of \cite{Manohar:2000dt} and Eq. (10) of \cite{Ji:1991pr}
    C = \left[\alpha_s(m_h)\right]^{{\gamma_0}/{2\beta_0}} \left[1 + \frac{\alpha_s(m_h)}{4\pi}
        \left(-\frac{8}{3} + \frac{\gamma_1}{2\beta_0} - \frac{\gamma_0\beta_1}{2\beta_0^2} \right) + \order(\alpha_s^2) \right] ,
    \label{eq:Wilson-coefficient}
\end{equation}
with $\gamma_0 = -4$, $\gamma_1 = -254/9 - 56\pi^2/27 + 20 n_f/9$, % See Eq. (20) of \cite{Ji:1991pr} 
$\beta_0 = (11 - 2n_f/3)$ and $\beta_1 = (102 - 28n_f/3)$ with $n_f=4$ in our simulations.
The Wilson coefficient is usually defined to depend on the renormalization scale $\mu$ of the HQET current, with the renormalization
scale (and scheme) dependence canceling between the Wilson coefficient and the HQET matrix element.
We have moved this scale dependence%
\footnote{The $\mu$ dependence in the usual Wilson coefficient comes from the exponential of the integral of the anomalous dimension
of the HQET current, and therefore may be factored out.} %
out of $C$ into the matrix element $\tilde{\Phi}_0$, thereby making $\tilde{\Phi}_0$ a renormalization-group invariant quantity.
Consequently, $C$ depends only on the matching scale $m_h$.

As mentioned in Sec.~\ref{sec:Lattice-Simulations}, we use $m'_l$, $m'_s$, and $m'_c$ to denote the simulation masses of the light
(up-down), strange, and charm quarks, respectively; without the primes $m_l = \half(m_u+m_d)$, $m_s$, and $m_c$ denote the correctly
tuned masses of the corresponding quarks.

We now discuss the dependence of $\Phi_{H_x}$ on the deviation of $m'_c$ from $m_c$.
The charm quark can be integrated out for processes that occur at energies well below its mass.
By decoupling~\cite{Appelquist:1974tg}, the effect of a heavy (enough) sea quark on low-energy quantities occurs only through the
change it produces in the effective value of $\LamQCD$ in the low-energy (three-flavor) theory~\cite{Bernreuther:1981sg}.
We use $\LamQCD^{(3)}(m'_c)$ to denote the effective value of $\LamQCD$ 
when the charm quark with mass $m'_c$ is integrated out. 
At leading order in weak-coupling perturbation theory, one obtains \cite[Eq.~(1.114)]{Manohar:2000dt} 
\begin{equation}
    \frac{\LamQCD^{(3)}(m'_c)}{\LamQCD^{(3)}(m_c)} = \left(\frac{m'_c}{m_c}\right)^{2/27}. 
\end{equation}
Noting that $\tilde{\Phi}_0$ has mass-dimension 3/2, we take into account the effects of the
mistuned mass $m'_c$ by assuming $m'_c\approx m_c$ and replacing
\begin{equation}
    \tilde{\Phi}_0 \to \tilde{\Phi}_0  \left(1 + \frac{3}{27}k'_1 \frac{\delta m'_c}{m'_c}\right)
        \left(\frac{m'_c}{m_c}\right)^{3/27},
\end{equation}
where $\delta m'_c = m'_c - m_c$, and $k'_1$ is a new fit parameter to describe higher-order effects.

Within the framework of \aschpt~\cite{Bernard:2013qwa}, \eq{decay-constant:chiral-limit} can be extended to include the light-quark
mass dependence and taste-breaking discretization errors of a generic $H_x$ meson.
This provides a suitable fit function to perform a combined EFT fit to lattice data at multiple lattice spacings and various
valence- and sea-quark masses.
The fit function that we use in this analysis has the following schematic form
\begin{align}
    \Phi_{H_x} &= C\tilde{\Phi}_0
    \left[1 + k_1 \MHsinv + k_2 \left(\MHsinv\right)^2 + k_3\left(\MHsinv\right)^3\right]
    \nonumber \\ & \times
    \left(1 + \frac{3}{27}k'_1\frac{\delta m'_c}{m'_c}\right) \left(\frac{m'_c}{m_c}\right)^{3/27}
%     \nonumber \\ &
    \times
    \big(1+ \delta \Phi_{\text{NLO}} + \delta \Phi_{\text{N$^n$LO,analytic}} \big) ,
    \label{eq:schematic:function}
\end{align}
where $M_{H_s}$ is the mass of a pseudoscalar meson with physical sea-quark masses, physical valence strange-quark mass and
heavy-quark mass $m_h$.
In the last parentheses, $\delta \Phi_{\text{NLO}}$ contains the next-to-leading order (NLO) staggered chiral nonanalytic and
analytic terms, and $\delta\Phi_{\text{N$^n$LO,analytic}}$ contains higher order analytic terms in the valence and sea-quark masses.
For an isospin-symmetric sea with $m_u=m_d\equiv m_l$, we have~\cite{Bernard:2013qwa}
\begin{align}
    \delta \Phi_{\text{NLO}} =& -\frac{1}{16\pi^2f^2} \frac{1}{2} \Biggl\{
	    \frac{1}{16}\sum_{\mathscr{S},\Xi} \ell(m_{\mathscr{S}x_{\Xi}}^2)
            +\frac{1}{3} \sum_{j\in \cM_I^{(2,x)}} \frac{\partial}{\partial m^2_{X_I}} 
             \left[ R^{[2,2]}_{j}( \cM_I^{(2,x)};  \mu^{(2)}_I) \ell(m_{j}^2) \right]
    \nonumber \\ & \qquad\qquad\quad
            +\Bigl( a^2\delta'_V \sum_{j\in \hat{\cM}_V^{(3,x)}} \frac{\partial}{\partial m^2_{X_V}}
            \left[R^{[3,2]}_{j}( \hat{\cM}_V^{(3,x)}; \mu^{(2)}_V) \ell(m_{j}^2)\right] + [V\to A]\Bigr) 
            \Biggr\}
    \nonumber \\ &
        -\frac{1}{16\pi^2f^2} \frac{3g_\pi^2}{2} \Biggl\{
            \frac{1}{16}\sum_{\mathscr{S},\Xi} J(m_{\mathscr{S}x_{\Xi}},\Delta^*+\delta_{\mathscr{S}x})
    \nonumber \\ &\qquad\qquad\quad 
            +\frac{1}{3} \sum_{j\in \cM_I^{(2,x)}} \frac{\partial}{\partial m^2_{X_I}}
            \left[ R^{[2,2]}_{j}(\cM_I^{(2,x)};  \mu^{(2)}_I) J(m_{j},\Delta^*) \right]
    \nonumber \\ &\qquad\qquad\quad 
            +\Bigl( a^2\delta'_V \sum_{j\in \hat{\cM}_V^{(3,x)}} \frac{\partial}{\partial m^2_{X_V}}
            \left[ R^{[3,2]}_{j}( \hat{\cM}_V^{(3,x)}; \mu^{(2)}_V) J(m_{j},\Delta^*)\right] + [V\to A]\Bigr)  
            \Biggr\}\ \nonumber \\
    & { } + L_\text{s}  (2 x_l + x_s) + L_x x_x + \half L_{a} x_{\bar\Delta} ,
    \eqn{chiral-form}
\end{align}
where the indices $\mathscr{S}$ and $\Xi$ run over sea-quark flavors and meson tastes, respectively; $\Delta^*$ is the lowest-order
hyperfine splitting; $\delta_{\mathscr{S}x}$ is the flavor splitting between a heavy-light meson with light quark of flavor
$\mathscr{S}$ and one of flavor $x$; $\delta'_V$ and $\delta'_A$ are taste-breaking hairpin parameters; and $g_\pi$ is
the $H$-$H^*$-$\pi$ coupling.
Definitions of the residue functions $R_j^{[n,k]}$, the sets of masses in the residues, and the chiral functions $\ell$ and $J$ at
infinite and finite volumes are given in \rcite{Bernard:2013qwa} and references therein.
At tree-level in \aschpt, the squared pion mass is linear in the sum of quark masses, $M_\pi^2\approx B_0(m_u+m_d)+a^2\Delta_\Xi$,
where $B_0$ is a low-energy constant (LEC) and the splitting $a^2\Delta_P=0$ for the taste-pseudoscalar pion.
We exploit this relation to define dimensionless quark masses and a measure of the taste-symmetry breaking as
\begin{align}
    x_q &\equiv \frac{2 M_{p4s}^2}{16 \pi^2 f_{\pi}^2} \frac{m_q}{m_{p4s}},
    \label{eq:x-defs} \\
    x_{\bar\Delta} &\equiv \frac{2}{16 \pi^2 f_{\pi}^2} a^2\bar\Delta,
\end{align}
where $q$ denotes the valence or sea light quark%
\footnote{For simplicity, we drop the primes on the simulation $x_q$s in this section.} %
and $a^2\bar\Delta$ is the mean-squared pion taste splitting.
The $x_q$s and $x_{\bar\Delta}$ are natural variables of \aschpt; the LECs $L_\text{s}$, $L_x$, and $L_a$ are therefore
expected to be of order~1.
The taste splittings have been determined to $\sim1$--10\% precision~\cite{Bazavov:2012xda} and are used as input to
\eq{chiral-form}.

Because we have very precise data and approximately 500 data points, NLO \aschpt\ is not adequate to describe fully the quark-mass
dependence, in particular for masses near $m_s$.
We therefore include all mass-dependent analytic terms at next-to-next-to-leading order (NNLO) and next-to-next-to-next-to-leading
order (NNNLO) by defining
\begin{align}
    \delta \Phi_{\text{N$^n$LO,analytic}} &= q_1 x_x^2 + q_2 (2x_l + x_s)x_x + q_3 (2x_l+x_s)^2 + q_4 (2x_l^2+x_s^2)
    \nonumber\\ & \hphantom{=}
        + q_5 x_x^3 + q_6 (2x_l + x_s)x_x^2 + q_7 (2x_l+x_s)^2 x_x + q_8 (2x_l^2+x_s^2) x_x
    \nonumber\\ & \hphantom{=}
        + q_9 (2x_l + x_s)^3 + q_{10} (2x_l+x_s) (2x_l^2+x_s^2) + q_{11} (2x_l^3+x_s^3) + q_{12}x_x^4 . 
  \label{eq:NNLO-analytic}
\end{align}
The terms that depend upon the light valence-quark mass are needed to describe our wide range of correlated data with
$x_l\le x_x\le x_s$.
The terms without $x_x$ are expected to be less important for obtaining a good fit because most of the ensembles
have similar strange sea-quark masses, and because the ensembles are statistically independent,
but we include them to make it a systematic approximation at the level of analytic terms.
We also include a quartic term $q_{12}x_x^4$, again to describe our wide range of valence-quark masses.

The staggered chiral form in \eq{chiral-form} is given at fixed heavy-quark mass $m_h$,
or equivalently at fixed $M_{H_s}$.
As discussed above, the LECs in \eq{chiral-form} encode the effects of short-distance physics,
and the dependence can be parameterized as expansions in inverse powers of the meson mass $M_{H_s}$
and powers of the lattice spacing of each ensemble.
To take the effects at scale $M_{H_s}$ into account, we replace
\begin{equation}
    L_x \to L_x + L'_x \left(\MHsinvDiff\right) + L''_x \left(\MHsinvDiff\right)^2 ,
\end{equation}
and similarly for $L_\text{s}$ and $g_\pi$.
We do not introduce any corrections to $L_a$ because it is suppressed by a factor of $\alpha_s^2 a^2$ at the finest lattice spacings
where the heavy-quark mass dependence could be important.
(At coarsest lattice spacings we only have valence heavy-quark masses near charm and thus the variation due to the valence
heavy-quark masses is less important.) We also add a $1/M_{H_s}$ correction term (but not $1/M^2_{H_s}$) to the four analytic terms
at NNLO:
\begin{equation}
    q_i \to q_i + q'_i \left(\MHsinvDiff\right) ,
\end{equation}
for $i=1,2,3,4$.

Meson-mass dependence also appears implicitly through the hyperfine splitting $\Delta^*$ and the flavor splitting
$\delta_{\mathscr{S}x}$ in \eq{chiral-form}.
To fix the heavy-mass dependence of $\Delta^*$, which first appears at order~$1/m_h$, we use
\begin{equation}
    \Delta^* = A_{\Delta^*} \MHsinv + B_{\Delta^*} \left(\MHsinv\right)^2 ,
\end{equation}
with $A_{\Delta^*}$ and $B_{\Delta^*}$ fixed by demanding that $\Delta^*$ reproduce the experimental values of the hyperfine
splitting in the $D$ and $B$ systems.
Similarly, we determine $\delta_{\mathscr{S}x}$ by writing
\begin{equation}
   \delta_{\mathscr{S}x}=  A_{\delta} + B_{\delta}\MHsinv,
\end{equation}
and fixing $A_{\delta}$ and $B_{\delta}$ from the known flavor splittings in the $D$ and $B$ systems.

To enable a description of our data with a wide range of lattice spacings from $0.03~\text{fm} \lesssim a \lesssim 0.15~\text{fm}$,
we incorporate lattice artifacts into the fit function as follows.
Taste-breaking discretization errors in masses of light mesons, which affect the decay constants of heavy-light mesons at one-loop
in \chpt, are already included in the staggered chiral form in \eq{chiral-form}.
In addition to these NLO effects, various discretization errors in the LECs must be taken into account.
In Appendix~\ref{app:normalization}, we use HQET to study heavy-quark discretization effects at the tree 
level~\cite{ElKhadra:1996mp,Kronfeld:2000ck}.
At the leading order, tree-level heavy-quark discretization errors are eliminated via a normalization
factor, and at the next order in HQET discretization errors start at order $x_h^4$ and $\alpha_sx_h^2$, where
$x_h=2am_h/\pi$.
For these and generic lattice artifacts, we replace in \eq{schematic:function}
\begin{equation}
\label{eq:artifacts:tilde-Phi_0}
    \tilde{\Phi}_0 \to \tilde{\Phi}_0 \left[1 + c_1\alpha_s(a\Lambda)^2 + c_2(a\Lambda)^4 + c_3(a\Lambda)^6 +
        \alpha_s\left(c_4 x_h^2 + c_5 x_h^4 + c_6 x_h^6\right)\right],
\end{equation}
where $\Lambda$ is the scale of generic discretization effects, set to 600~MeV in this analysis.
A factor of $\alpha_s$ is included in the $c_1$ and $c_4$ terms because the HISQ action is tree-level improved to
order~$a^2$~\cite{Naik:1986bn}, so the leading generic discretization errors start at order $\alpha_s(a\Lambda)^2$ or
$\alpha_s(am_h)^2$.
In addition, a factor of $\alpha_s$ is included in the $c_5$ and $c_6$ terms because of the tree-level normalization factor.
For $k_1$ and $k_2$ in \eq{schematic:function}, we likewise replace
\begin{align}
    k_1 &\to k_1 \left[ 1 + c'_1 \alpha_s (a\Lambda)^2 + c'_2(a\Lambda)^4 + c'_3x_h^4 +
        \alpha_s\left(c'_4 x_h^2 + c'_5 x_h^4\right)\right],
    \label{eq:artifacts:k_1} \\
    k_2 &\to k_2 \left[ 1 + c''_1 \alpha_s (a\Lambda)^2 + c''_2 \alpha_s x_h^2 \right].
    \label{eq:artifacts:k_2}
\end{align}
No factor of $\alpha_s$ is included in the $c'_3$ term, because $k_1$ parametrizes effects at NLO in HQET.

Let us return to the parameters $L_x$, $L_\text{s}$, and $g_\pi$ found in $\delta\Phi_\text{NLO}$. 
Owing to the Naik improvement term, it is enough to introduce corrections of order $\alpha_s(a\Lambda)^2$ and $(a\Lambda)^4$.
Similarly, we add $\alpha_s (a\Lambda)^2$ corrections to the NNLO analytic terms in \eq{NNLO-analytic}.
Finally, to incorporate effects of heavy-quark discretization errors, we include
\begin{equation}
    \MHsinv \alpha_s x_h^2
    \label{eq:artifacts:L_v}
\end{equation}
corrections to $L_x$, $L_\text{s}$, and $g_\pi$, as explained in Appendix~\ref{app:normalization}.

Our final EFT fit function has 60 fit parameters.
With reasonable prior constraints on the large number of parameters describing
discretization effects [three parameters at NLO in \schpt\ ($\delta'_V$, $\delta'_A$, $L_a$); 16 parameters for generic
discretization effects in powers of $(a\Lambda)$; 10 parameters for the heavy-quark discretization], the uncertainties from the
continuum extrapolation are propagated to the statistical error reported by
the fit.
We test this expectation in Sec.~\ref{sec:Errors} by looking at the stability of the results to changes
in the widths of the prior constraints, the number of fit parameters, and the data included in the fit.

\subsection{Summary formula}
\label{sec:zF}

In summary, letting $\cF$ be our fit function from Sec.~\ref{sec:chiral-function}, and letting \param{blue} (arXiv) denote fit
parameters, we have
\begin{align}
    \cF &= C \breve{\Phi}_0 \left(1 + \breve{k}_1 w_h + \breve{k}_2 w_h^2 + \param{{k}_3} w_h^3\right)
        \left(1 + \frac{3}{27}\param{k'_1} \frac{\delta m'_c}{m'_c}\right)\left(\frac{m'_c}{m_c}\right)^{3/27}
    \nonumber\\
        &\quad \times \left[1 + \delta \Phi_\text{NLO} + 
	  \sum_{i=1}^4 \left(\param{q_i} + \param{q'_i} \bar{w}_h+\param{\tilde{q}_i}\alpha_s y\right) x_i^2
            + \sum_{j=5}^{11} \param{q_j} x_j^3 + \param{q_{12}}x_x^4 \right]
\end{align}
where $y=(a\Lambda)^2$, $w_h=\LamHQET/M_{H_s}$, $\bar{w}_h=\LamHQET(M_{H_s}^{-1}-M_{D_s}^{-1})$,
and the indices $i$ and $j$ correspond to the labels of the terms in Eq.~(\ref{eq:NNLO-analytic}).
The chiral logarithm term $\delta \Phi_{\text{NLO}}$ is given by~\eq{chiral-form} with the replacements
$L_\text{s}\to\breve{L}_\text{s}$, $L_x\to\breve{L}_x$, and $g_\pi\to\breve{g}_\pi$.
It depends upon the LECs $\param{f}$, $\param{L_{a}}$, $\param{\delta'_V}$, and $\param{\delta'_A}$; the hyperfine
splitting~$\Delta^*$; and the taste-independent flavor splitting~$\delta_{\mathscr{S}x}$.
The breved quantities include terms that allow for the $\chi$PT parameters $\tilde{\Phi}_0$, $k_1$, $k_2$, $L_x$, $L_\st$, and
$g_\pi$ to have heavy-quark mass and lattice-spacing dependence:
\begin{subequations}
\begin{align}
  \breve{\Phi}_0 &= \param{\tilde{\Phi}_0}
    \left[ 1 + \param{c_1}\alpha_s y + \param{c_2} y^2 + \param{c_3} y^3 +
        \alpha_s\left(\param{c_4} x_h^2 + \param{c_5} x_h^4 + \param{c_6} x_h^6\right)\right], \\
    \breve{k}_1 &= \param{k_1} \left[1 + \param{c'_1} \alpha_s y + \param{c'_2} y^2
        + \param{c'_3}x_h^4 + \alpha_s\left(\param{c'_4} x_h^2 + \param{c'_5} x_h^4\right)\right] , \\
    \breve{k}_2 &= \param{k_2}
        \left( 1 + \param{c''_1} \alpha_s y + \param{c''_2} \alpha_s x_h^2 \right), \\
    \breve{L}_x  &= \param{L_x} + \param{L'_x} \bar{w}_h + \param{L''_x} \bar{w}_h^2 
	  + \param{\tilde{L}'_x}\alpha_s y + \param{\tilde{L}''_x} y^2 + \param{L'''_x} w_h \alpha_s x_h^2, \\
    \breve{L}_\st &= \param{L_\st} + \param{L'_\st} \bar{w}_h + \param{L''_\st} \bar{w}_h^2 
	  + \param{\tilde{L}'_\st}\alpha_s y + \param{\tilde{L}''_\st} y^2 + \param{L'''_\st} w_h \alpha_s x_h^2, \\
    \breve{g}_\pi &= \param{g_\pi} + \param{g'_\pi} \bar{w}_h + \param{g''_\pi} \bar{w}_h^2 
	  + \param{\tilde{g}'_\pi}\alpha_s y + \param{\tilde{g}''_\pi} y^2 + \param{g'''_\pi} w_h \alpha_s x_h^2 .
\end{align}
\end{subequations}
Thus, there are a total of 60 fit parameters.
Of these $f$ is constrained by expectations from \chpt, $g_\pi$ is constrained by the results of other lattice-QCD calculations, and
$\delta'_V$ and $\delta'_A$ are constrained by MILC's light-pseudoscalar-meson \chpt\ fits.

\subsection{Setting the lattice scale for the EFT analysis}  
\label{sec:p4s-scale-setting} 

We set the lattice scale with a two-step procedure that combines the pion decay constant with the so-called $p4s$ method, in a way
similar to Ref.~\cite{\rFD2014}.
In the first step of the procedure, we use the PDG value of $f_{\pi}$, $\fpiPDG=130.50(13)~\MeV$~\cite{\rPDG2016},
to set the overall scale and to determine tuned quark masses for each \emph{physical-mass} ensemble.
Then, as described in Sec.~\ref{sec:physical-mass-analysis}, we calculate $M_{p4s}$ and $f_{p4s}$, which are the mass and decay
constant of a pseudoscalar meson with both valence-quark masses equal to $m_{p4s}\equiv0.4m_s$, and with physical sea-quark masses.
The continuum-extrapolated values of $f_{p4s}$, $R_{p4s}\equiv f_{p4s}/M_{p4s}$, and quark mass ratios are then used as inputs to
the second step of the procedure, which we refer to as the $p4s$ method.
In the $p4s$ method, we find $am_{p4s}$ and $af_{p4s}$ on a given physical-mass ensemble by adjusting the valence-quark mass $am_x$
until $(af_x)/(aM_x)$ takes the same value as the continuum-limit ratio $R_{p4s}$ just determined.
In the $p4s$ method, we use a mass-independent scale setting, in which all ensembles at the same $\beta$ as a physical-mass ensemble
have, by definition, the same lattice spacing $a=(af_{p4s})/f_{p4s}$ and~$am_{p4s}$.

%%%%%%%%%%%%%%%%%%%%%%%%%%%
\begin{table}
\newcommand{\h}{\phantom{8}}
\caption{ \label{tab:lattice-spacing:p4s-method}
  Lattice spacing $a$ and $am_s$ (in lattice units) in the $p4s$ mass-independent scale-setting scheme.
  The error associated with $\fpiPDG$ is a multiplicative error for all values of $\beta$;
  the relative error is about $0.15\%$ for lattice spacing $a$
  and about $0.3\%$ for $am_s$. 
  The uncertainty labeled ``EM~scheme'' is an additional uncertainty that can be incorporated when these results are used
  without attention to the EM scheme dependence.
  }
\begin{tabular}{c@{\quad}l@{\quad}l}
  \hline\hline
  $\beta$~{} & \multicolumn{1}{c}{$a$~(fm)} & \multicolumn{1}{c}{$am_s$} \\
  \hline
  $5.8\h$ & 
    $0.15293  (26)_\text{stat}  (19)_\text{syst}  (23)_{\fpiPDG} [07]_\text{EM scheme} $ &  
    $0.06852  (24)_\text{stat}  (22)_\text{syst}  (20)_{\fpiPDG} [05]_\text{EM scheme} $ \\
  $6.0\h$ & 
    $0.12224  (16)_\text{stat}  (15)_\text{syst}  (18)_{\fpiPDG} [05]_\text{EM scheme} $ &
    $0.05296  (15)_\text{stat}  (17)_\text{syst}  (15)_{\fpiPDG} [04]_\text{EM scheme} $ \\
  $6.3\h$ & 
    $0.08785  (17)_\text{stat}  (11)_\text{syst}  (13)_{\fpiPDG} [04]_\text{EM scheme} $ &
    $0.03627  (14)_\text{stat}  (12)_\text{syst}  (10)_{\fpiPDG} [02]_\text{EM scheme} $ \\
  $6.72$ & 
    $0.05662  (13)_\text{stat}  (07)_\text{syst}  (08)_{\fpiPDG} [03]_\text{EM scheme} $ &
    $0.02176  (10)_\text{stat}  (07)_\text{syst}  (06)_{\fpiPDG} [01]_\text{EM scheme} $ \\
  $7.0\h$ & 
    $0.04259  (05)_\text{stat}  (05)_\text{syst}  (06)_{\fpiPDG} [02]_\text{EM scheme} $ &          
    $0.01564  (04)_\text{stat}  (05)_\text{syst}  (04)_{\fpiPDG} [01]_\text{EM scheme} $ \\
  $7.28$ & 
    $0.03215  (14)_\text{stat}  (28)_\text{syst}  (05)_{\fpiPDG} [01]_\text{EM scheme} $ &          
    $0.01129  (10)_\text{stat}  (19)_\text{syst}  (03)_{\fpiPDG} [01]_\text{EM scheme} $ \\
 \hline\hline
\end{tabular}
\end{table}
%%%%%%%%%%%%%%%%%%%%%%%%%%%

To determine $am_{p4s}$ and $af_{p4s}$ accurately, the data must be adjusted for mistunings in the sea-quark masses.
To make these adjustments, we use the derivatives with respect to quark masses, which were calculated in our earlier work and listed
in Table~VII of Ref.~\cite{\rFD2014}.
We then iterate, computing $am_{p4s}$ and $af_{p4s}$, readjusting the data, and repeating the entire process until the values of
$am_{p4s}$ and $af_{p4s}$ converge within their statistical errors.
The results for the lattice spacing $a$ and $am_s = 2.5am_{p4s}$ are listed in Table~\ref{tab:lattice-spacing:p4s-method}.
For the smallest lattice spacing, $a\approx 0.03$~fm, where we do not have an approximately physical-mass ensemble, we rely on the
derivatives to determine $a$ and $am_s$ from data on the $m'_l/m'_s=0.2$ ensemble, leading to larger relative systematic errors
at $\beta=7.28$.

\subsection{Effective-field-theory fit to heavy-light decay constants}
\label{sec:chiral-fit} 

In \secref{chiral-function}, we have constructed an EFT fit function that contains 60 fit parameters.
We use this function to perform a combined, correlated fit to the partially-quenched data at the five lattice spacings, from
$a\approx0.12$~fm to $\approx0.03$~fm, and at several values of the light sea-quark masses.
The sixth lattice spacing, $a\approx0.15$~fm, is used in a check of the estimate of discretization errors, but not included in the
base fit used to obtain our central values and statistical errors.
At the coarsest lattice spacings, we have data with only two different values for the valence heavy-quark mass: $m_h=m'_c$ and
$m_h=0.9m'_c$.
Recall that $m'_c$ is the simulation value of sea charm-quark mass of the ensembles, and is itself not precisely equal to the
physical charm mass $m_c$ because of tuning errors.
At the finest lattice spacings, we have a wide range of valence heavy-quark masses from near charm to bottom.
We include all data with $0.9m'_c \le m_h \le 5m'_c$, subject to condition $am_h<0.9$, which is chosen to avoid large lattice
artifacts.
Note that our analysis includes an $a\approx0.03$~fm, $m'_l/m'_s = 0.2$, ensemble for which $am_b \approx 0.6$, and thus no
extrapolation from lighter heavy-quark masses is needed, although a chiral extrapolation to physical light-quark masses is required.

We use a constrained fitting procedure~\cite{Lepage:2001ym} with priors set as follows.
For the LEC $g_\pi$ of the $D$ system, we use the prior $g_\pi=0.53\pm 0.08$, which is based on lattice-QCD 
calculations~\cite{Becirevic:2012pf,Can:2012tx,Detmold:2012ge}.
For $1/f^2$ in \eq{chiral-form}, our prior is
\begin{equation}
   \frac{1}{f^2} = \frac{1}{2}\left(\frac{1}{f_{\pi}^2} + \frac{1}{f_{K}^2}\right)  
     \pm  \left(\frac{1}{f_{\pi}^2} - \frac{1}{f_{K}^2}\right) , 
\end{equation}
where we set $f_\pi=130.5$~MeV and $f_K=156$~MeV.
For the taste-breaking hairpin parameters, we use priors of $\delta'_A/\bar\Delta= -0.88\pm0.09$ and $\delta'_V/\bar\Delta=
0.46\pm0.23$, which are taken from chiral fits to light pseudoscalar mesons~\cite{Bazavov:2011fh,*Bazavov:2011fh-update}.
The fits of \rcite{Bazavov:2011fh} have been performed at $a\approx0.12$~fm, where ensembles with unphysical strange quark masses
are available (see \tabref{ensembles}).
We take advantage of the fact that both the taste splittings and the hairpin parameters scale like $\alpha^2_s a^2$ at NLO in the
chiral expansion, so their ratio remains constant as $a$ changes.
For $\tilde{\Phi}_0$, we use an extremely wide prior of $0\pm 1000$ in $p4s$ units.
The rest of the fit parameters are normalized to be of order~1, and for them we choose a prior of $0\pm1.5$.
We discuss this choice in \secref{Errors} and argue that it is conservative.
Finally, for $\alpha_s$ we use the coupling $\alpha_V$ at scale $q=2.0/a$, obtained from
Ref.~\cite{Chakraborty:2014aca,*Chakraborty:2017aca}.

\begin{figure}
  \includegraphics[width=0.8\textwidth, trim={0 0.4cm 0 0},clip]{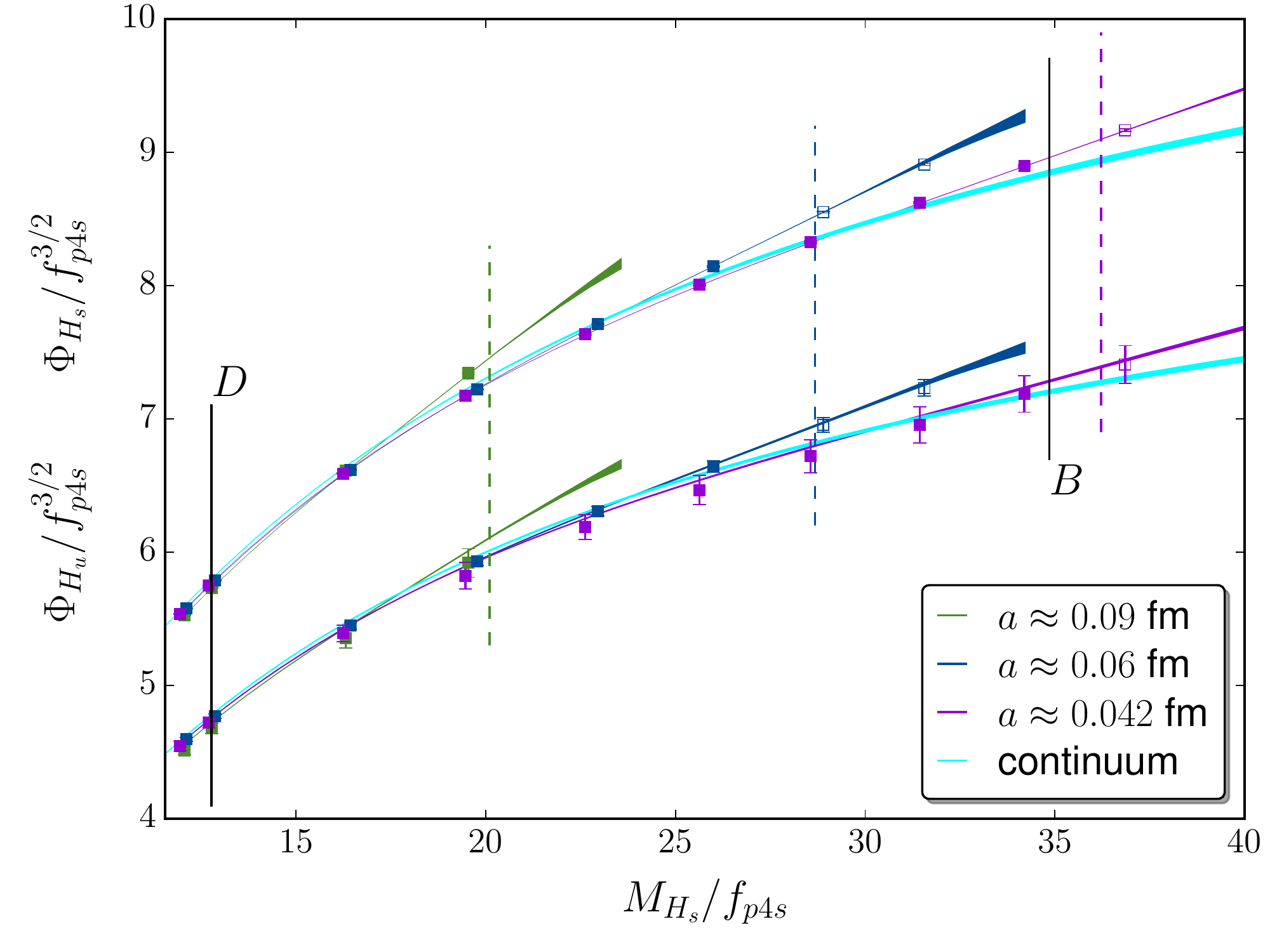}
  \caption{Decay constants plotted in units of $f_{p4s}$ vs the heavy-strange meson mass for
    physical-mass ensembles at three lattice spacings, and continuum extrapolation.
    For each color there are two sets of data and fit lines:
    one with valence light mass $m_x=m_s$ (higher one), and one with $m_x=m_u$.
    The dashed vertical lines indicate the cut $am_h = 0.9$ for each lattice spacing,
    and data points (with open symbols) to the right of the dashed vertical line of the corresponding color are omitted from the fit.
    The width of the fit lines shows the statistical error coming from the fit.
    The solid vertical lines indicate the $D$ and $B$ systems, where $M_{H_s}=M_{D_s}$ and $M_{H_s}=M_{B_s}$, respectively.}
    \label{fig:PhiHu-PhiHs_vs_MHs}
\end{figure}

Altogether we have $492$ lattice data points in the base fit and $60$ parameters in the EFT fit function.
The fit has a correlated $\chi^2_{\text{{data}}}/\text{dof} = 466/432$, giving $p = 0.12$.
\Figref{PhiHu-PhiHs_vs_MHs} shows a snapshot of the decay constants for physical-mass ensembles, plotted versus the corresponding
heavy-strange meson masses $M_{H_s}$ at three lattice spacings.
The continuum extrapolation is also shown.
The valence light mass $m_x$ is tuned either to $m_s$ (upper points) or to $m_u$ (lower points).
Data points with open symbols that are at the right of the dashed vertical line of the corresponding color are omitted from the fit
because they have $am_h>0.9$.
The fact that the fit lines agree well with the omitted points is evidence that we have not overfit the data.
In the continuum extrapolation, the masses of sea quarks are set to the correctly-tuned, physical
quark masses $m_l$, $m_s$, and~$m_c$, while at nonzero lattice spacing the masses of the sea quarks take the simulated values.

The width of the fit lines in \figref{PhiHu-PhiHs_vs_MHs} shows the statistical error coming from the fit,
which is only part of the total statistical error, since it does not include
the statistical errors in the inputs of the quark masses and the lattice scale.
To determine the total statistical error of each output quantity,
we divide the full data set into 20 jackknife resamples. The complete calculation,
including the determination of the inputs, is performed on each resample,
and the error is computed as usual from the variations over the resamples.
(For convenience, we kept the covariance matrix fixed to that from the full data set,
rather than recomputing it for each resample.)
The same procedure is performed to find the total statistical error of $a$ and $am_s$ at each lattice spacing.

The fit function Eq.~(\ref{eq:schematic:function}), evaluated at $a=0$ and physical sea-quark masses, yields a parameterization of
the decay-constant data as a function of the heavy-strange meson mass $M_{H_s}$ and the valence light-quark mass $m_x$.
We ignore isospin violation in the sea, taking the light sea-quark masses to be degenerate with the average $u/d$-quark mass.
Because the \aschpt\ expression for the heavy-light meson decay amplitude is symmetric under the interchange $m_u\leftrightarrow
m_d$, the leading contributions from isospin-breaking in the sea sector are of $\order((m_d-m_u)^2)$, and are expected to be smaller
than the NNLO terms in the chiral expansion.
We can check numerically the effect of sea isospin-breaking using our data by evaluating the fit function with physical up and down
sea-quark masses.
The resulting shifts in the decay constants are less than about 0.02\% for the $B$ system and 0.015\% for the $D$ system, which are
consistent with power-counting expectations and are negligible compared to other uncertainties.
We obtain the physical charged and neutral $B$- and $D$-meson decay constants by setting $m_x$ to either $m_u$, $m_d$ or $m_s$,
and $M_{H_s}$ to the experimental values $M_{B_s}=5366.82(22)$~MeV and $M_{D_s}=1968.27(10)$~MeV~\cite{Olive:2016xmw}, respectively,
along with a prescription to subtract electromagnetic effects from the masses, as discussed below.

\section{Systematic error budgets}
\label{sec:Errors}

\begin{figure}[b]
    \includegraphics[width=1\textwidth]{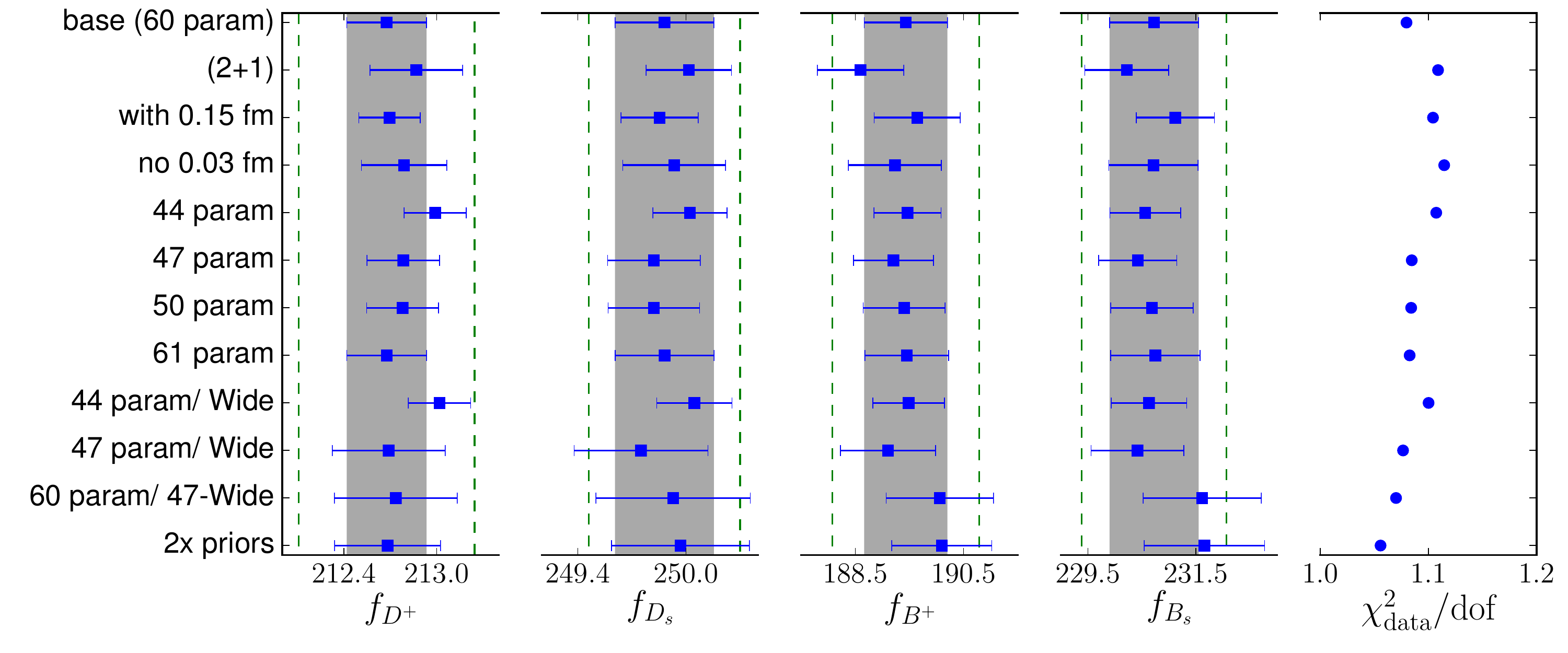}
    \caption{Stability plot showing the sensitivity to different choices of lattice data and fit models.
    (See the text for description.)
    The error bars show only the statistical errors, the gray error bands correspond to the statistical error of the base fit, and
    the green dashed lines correspond to total errors.}
    \label{fig:stability_fB2018}
\end{figure}

\Figref{stability_fB2018} shows the stability of our final results for $f_{D^+}$, $f_{D_s}$, $f_{B^+}$ and $f_{B_s}$
under variations in the data set and the fit models.
In our base fit, we use the decay constants obtained from the (3+2)-state fit to two-point correlators.
To investigate the error arising from excited-state contamination, we perform a fit to the decay-constant data
obtained from the (2+1)-state fit to two-point correlators.
There is some evidence for such contamination, contributing a systematic error that is comparable to the statistical errors for the
$B$ system.
We take the difference between the results from the two types of correlator fits as an estimate of the systematic error due to
excited states.
For consistency, we do so both for the $D$ system as well as the $B$ system, even though there is little evidence
for such contamination for the $D$ system.
It is reasonable that the $B$ correlators suffer from larger excited state effects, because, as seen in \figref{fits_5masses_both},
the fits to correlators with heavier quarks tend to have smaller $p$~values at fixed $T_\text{min}$, as well as larger errors in the
ground state mass.

\Figref{stability_fB2018} also shows a test of the systematic error in the continuum extrapolation from repeating the fit after
either adding in the coarsest ($a\approx 0.15$~fm) ensembles or omitting the finest ($a\approx0.03$~fm) ensemble.
The differences with the base fit are well within the statistical errors, providing support for our earlier assertion that the
continuum-extrapolation errors are already included in our estimate of the statistical uncertainty of our fit.

In our base fit, constrained Bayesian curve fitting~\cite{Lepage:2001ym} is employed to incorporate systematic errors in the
continuum extrapolation.
If the prior values have been chosen in a reasonable way, and if we have sufficiently many parameters in the fit, central values and
error bars of final quantities should not change when more parameters are included in the fit.
The error bars are then expected to capture the systematic errors in the continuum extrapolation.

To test the priors chosen for discretization effects, we repeat the analysis with different numbers of discretization parameters.
The result of this test is shown in \figref{stability_fB2018}.
The base fit has 60 parameters.
We show results from alternative fits with 44, 47, 50, and 61 parameters.
The fit with 50 parameters is constructed from our base EFT fit function by removing 10 terms that describe higher-order
discretization effects in powers of $(a\Lambda)^2$: specifically, the $(a\Lambda)^6$ correction to $\tilde{\Phi}_0$; the
$(a\Lambda)^4$ corrections to $k_1$, $L_\text{s}$, $L_x$ and $g_\pi$; and the $(a\Lambda)^2$ corrections to $k_2$ and the NNLO
analytic terms in \eq{NNLO-analytic}.
In the fit with 47 parameters, three additional terms describing higher-order heavy-quark discretization effects are removed: we set
to zero $c'_3$, $c'_5$ and $c''_2$ in \eqs{artifacts:k_1}{artifacts:k_2}.
The fit with 44 parameters is then obtained by removing, from the 47-parameter fit, the $\alpha_s(a\Lambda)^2$ corrections
to $L_\text{s}$, $L_x$ and $g_\pi$.
Finally, we consider a fit function with 61 parameters, which is constructed from our base EFT fit function 
by adding a term $\alpha_s x_h^8$ to \eq{artifacts:tilde-Phi_0}, which is the most important term at the next order in our 
expansion variables.

The 44-parameter fit shows a significant deviation from the base fit for $f_{D^+}$, but already with 47
parameters the deviations of all quantities are small:   the errors are essentially unchanged from those of the base fit, and the
central values change by no more than half the error bars. Differences between the base fit and the 61-parameter fit 
are not visible at all.
In the context of constrained Bayesian curve fitting~\cite{Lepage:2001ym}, these findings suggests that the posterior
uncertainty captures most or all of the systematic error of the continuum extrapolation.

\begin{figure}
    \includegraphics[width=0.47\textwidth]{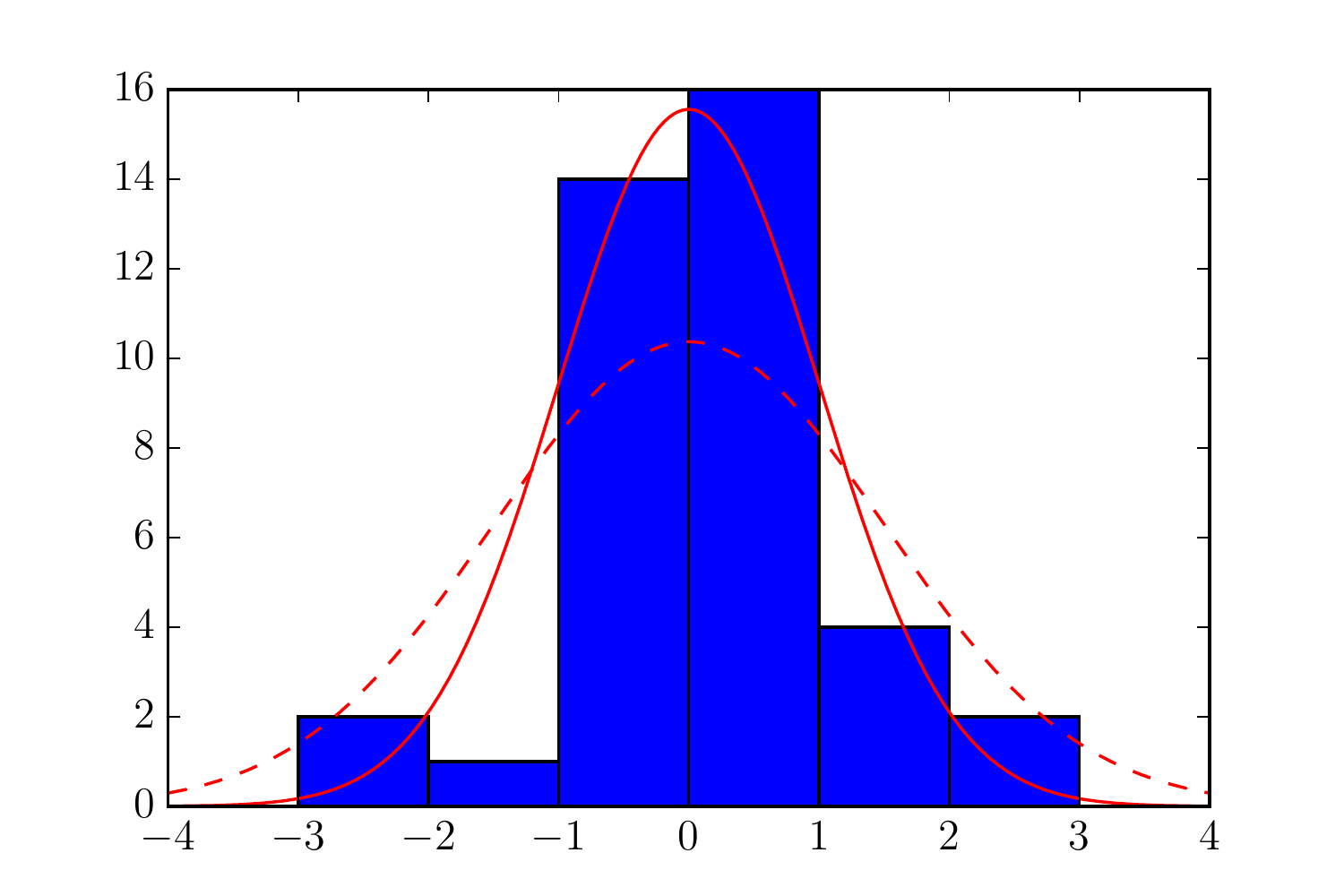}\hfill
    \includegraphics[width=0.47\textwidth]{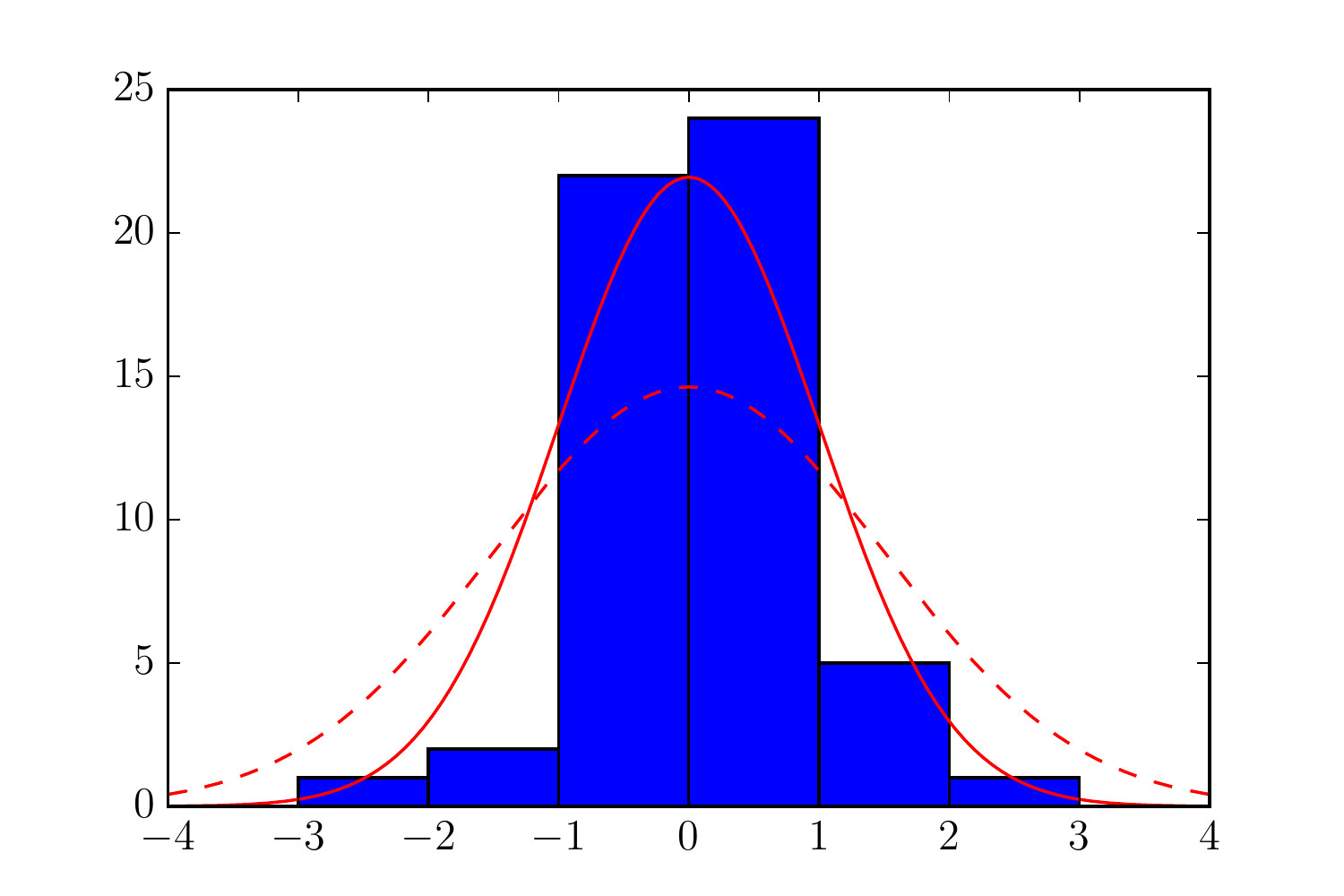}
    \caption{Left: distribution of fit posteriors in a fit with 44 parameters and essentially no prior constraints (prior 
        widths of~100).
        Right: distribution of fit posteriors in the base fit for parameters constrained with priors $0\pm 1.5$.
        In each plot the solid and dashed red curves show Gaussian distributions with width 1 and 1.5, respectively.}
    \label{fig:hist_param}
\end{figure}

The priors may be further tested by monitoring the posteriors in various fits.
\Figref{hist_param} (left) shows the distribution of posterior central values for essentially unconstrained fit parameters (priors
$0\pm100$) in the 44-parameter fit.%
\footnote{The quantities $\delta'_V$, $\delta'_A$, $g_\pi$, $1/f$ and $\tilde{\Phi}_0$, which are set by external 
considerations rather than power counting, have the same priors as in the base fit. \label{foot:not-wide}}
The distribution is compared to Gaussian distributions with widths of 1 and~1.5.
Note that the width-1 Gaussian is already fairly consistent with the distribution,
but there may be some indication of excess in the tails.
On the other hand, the width-1.5 Gaussian clearly encompasses the posterior distribution.
Thus the natural size of these parameters is indeed of order unity, and a prior of $0\pm1.5$ seems to be a conservative assumption
for any additional parameters in other fits that are not well constrained by data.
\Figref{hist_param} (right) shows the corresponding distribution of posterior central values for the 55 parameters in the base fit
that are constrained with priors $0\pm1.5$.
The comparison with the width-1.5 Gaussian indicates that the parameters are not being unnaturally constrained by the Bayesian
priors.

In the Bayesian approach, prior information about fit parameters is explicitly put into the fit.
A non-Bayesian alternative is to limit the number of fit parameters to those constrained
by the data with no external information about what sizes of the parameters are expected.
External information nevertheless enters implicitly by assuming that the parameters omitted from the fit are
all exactly zero.
We apply this alternative approach to test whether there are additional systematic errors in
the continuum extrapolation due to the choice of fit function that are not captured by the Bayesian
analysis.
\Figref{stability_fB2018} shows two fits with fewer parameters than the base fit, which may then be determined by the data, with
essentially no Bayesian constraint.\footnotemark[\value{footnote}]
The fits are labeled ``44 param/ Wide'' and ``47 param/ Wide.'' %
They have the same parameter sets as the 44-parameter and 47-parameter fits discussed above, but now with very wide priors, $0\pm100$.
(The 44 param/ Wide fit yields \Figref{hist_param} (left).) %
We also include a fit, ``60 param/ 47-Wide'' with the same parameters as the base fit, but with the 47 parameters that can be
determined by the data alone now essentially unconstrained by priors and priors of $0\pm1$ for the remaining 13 parameters.
These three new fits have $p$ values larger than 0.05, so we consider them to be acceptable alternatives.
Comparing these fits with the base fit, we find that the central values vary a bit more than we would expect from the Bayesian
analysis.
In particular, $f_{D^+}$ in the 44-param/ Wide fit and $f_{B_s}$ in the 60 param/ 47-Wide fit differ from the base fit by slightly
more than the error bar of the base fit (indicated by the gray band).
We take a conservative approach and take the largest of these differences for each quantity as an additional
systematic error due to the choice of fit model.

A final fit in \figref{stability_fB2018}, labeled ``$2\times$ priors,'' starts with the base fit and doubles, to $0\pm3$, the prior
widths of the 55 parameters constrained by power counting arguments.
The results of this fit are very similar to those from the 60 param/ 47-Wide fit.
In the Bayesian context, it is to be expected that weakening the prior information in the base fit results in an increase in the
resulting errors.
However, the shifts in the central values for the $B$ system are large enough that the inclusion of the fit model error discussed in
the previous paragraph seems prudent.

\tabrefs{results_D_f_details}{results_B_f_details} give representative error budgets for the decay constants and their ratios in the
$D$ and $B$ systems, respectively.
The error listed as ``statistics and EFT fit'' is determined by a jackknife procedure (described at the end of \secref{chiral-fit})
in which we repeat, on data resamples, the EFT fit and its extrapolation to the continuum and interpolation to physical quark masses.
It includes statistical errors in the inputs as well as those from the fit itself.
As explained above, it also includes much of the systematic error associated with the continuum extrapolation.
The small errors from the chiral interpolation are likewise captured by our Bayesian procedure, which includes all analytic chiral
terms at NNLO and NNNLO.

%%%%%%%%%%%%%%%%%%%%%%%%%%%%
\begin{table}
\newcommand{\h}{\phantom{x}}
\caption{\label{tab:results_D_f_details}
Representative error budgets for decay constants of the $D$ system, estimated as described in the text.
Error budgets for $f_{D^0}$ and the isospin-limit value $f_D$ are similar to that for $f_{D^+}$ with one exception.
The uncertainty from the topological-charge correction is larger for lighter valence-quark masses:
0.09\% (0.07\%) for $f_{D^0}$ ($f_D$).}
\begin{tabular}{lccc}
\hline\hline
Error (\%) &  \h$f_{D^+}$\h  &  \h$f_{D_s}$\h  &  \h$f_{D_s}/f_{D^+}$\h  \\
\hline
Statistics and EFT fit & $0.12$ & $0.11$ & $0.05$ \\
Two-point correlator fits & $0.09$ & $0.05$ & $0.04$ \\
Fit model & $0.16$ & $0.07$ & $0.09$ \\
Scale-setting quantities and tuned quark masses & $0.08$ & $0.04$ & $0.05$ \\
Finite-volume corrections & $0.02$ & $0.01$ & $0.01$ \\
Electromagnetic corrections & $0.01$ & $0.01$ & $0.01$ \\
Topological charge distribution & $0.05$ & $0.00$ & $0.05$ \\
$\fpiPDG$ & $0.11$ & $0.08$ & $0.03$ \\
\hline\hline
\end{tabular}
\end{table}

\begin{table}
\newcommand{\h}{\phantom{x}}
\caption{\label{tab:results_B_f_details}
Representative error budgets for decay constants of the $B$ system, estimated as described in the text.
Error budgets for $f_{B^+}$ and the isospin-limit value $f_B$ are similar to that for $f_{B^0}$ with one exception.
The uncertainty from the topological-charge correction is larger for lighter valence-quark masses: 
0.11\% (0.08\%) for $f_{B^+}$ ($f_B$).}
\begin{tabular}{lccc}
\hline\hline
Error (\%) &  \h$f_{B^0}$\h  &  \h$f_{B_s}$\h  &  \h$f_{B_s}/f_{B^0}$\h  \\
\hline
Statistics and EFT fit & $0.39$ & $0.36$ & $0.24$ \\
Two-point correlator fits & $0.39$ & $0.22$ & $0.17$ \\
Fit model & $0.34$ & $0.39$ & $0.08$ \\
Scale-setting quantities and tuned quark masses & $0.10$ & $0.06$ & $0.05$ \\
Finite-volume corrections & $0.03$ & $0.01$ & $0.02$ \\
Electromagnetic corrections & $0.02$ & $0.02$ & $0.01$ \\
Topological charge distribution & $0.07$ & $0.00$ & $0.07$ \\
$\fpiPDG$ & $0.14$ & $0.11$ & $0.04$ \\
\hline\hline
\end{tabular}
\end{table}

%%%%%%%%%%%%%%%%%%%%%%%%%%%%

The error labeled ``two-point correlator fits'' in \tabrefs{results_D_f_details}{results_B_f_details}
is an estimate of the contamination due to excited states. It is determined by comparison of the results from
the base, (3+2)-state, fits and those from (2+1)-state fits.

The error we associate with the choice of fitting function, is labeled ``Fit model'' in each table.
As explained above, it comes from comparing the results of different non-Bayesian (essentially unconstrained) fits to those from the
base fit.
While the differences are not so large that they necessarily invalidate the Bayesian error analysis, they are large enough that we
are inclined to be conservative and include them as a separate source of error.
Since the fit model controls the continuum extrapolation, this error may be interpreted as an estimate of those continuum
extrapolation errors not completely captured by our Bayesian analysis.

The fourth line in each table, labeled ``scale-setting quantities and tuned quark masses," gives the systematic error
associated with the continuum extrapolations of $f_{p4s}$, $R_{p4s}$, and the tuned quark masses.
As described in \secref{physical-mass-analysis}, the central values of these input quantities to the heavy-light analysis come from
a quadratic fit in $\alpha_{{s}} a^2$ to the ensembles with $a \le 0.12$~fm.
We repeat the heavy-light analysis with the inputs instead determined by three alternatives: a quadratic fit including all the data,
a linear fit including data up to 0.12~fm, and a linear fit including data up to 0.09~fm.
The errors shown in \tabrefs{results_D_f_details}{results_B_f_details} are obtained by taking the largest difference between the
base values and the results from each of the three alternatives.

The error labeled ``finite-volume corrections'' gives our estimate of residual finite volume errors, those finite volume effects not
included in our chiral fitting forms.
The errors associated with light-quark and scale-setting inputs are estimated in the same way as those associated with continuum
extrapolation errors of those quantities, using the input finite-volume errors from \tabref{lightresults2}.
To determine the corresponding finite-volume errors arising directly in the heavy-light analysis, we omit the finite-volume
corrections at NLO in \chpt\ from the EFT fits, and then repeat the fits.
We take 0.3 of the differences between the results of the two fits as estimates of the residual finite-volume errors coming from
omitted higher-order terms in \chpt.
We consider the factor 0.3 to be conservative because higher order corrections in SU(3) \chpt\ are typically less than
that; for example, $f_K/f_\pi-1\approx0.2$.
We then add the finite volume errors from the heavy-quark analysis in quadrature with those from the inputs to get the values shown
in \tabrefs{results_D_f_details}{results_B_f_details}.
This is reasonable because we do not know the correlations between the effects of finite volume errors on the light-light and
heavy-light quantities.
For example, the ratios between heavy-light and light-light decay constants, which enter through our scale-setting procedure, are
likely to be less-dependent on volume than either decay constant alone.
In any case, if we instead assumed 100\% correlation between the light-light and heavy-light finite volume errors, it would make
little difference in the total systematic error.

We note that the finite-volume errors in \tabref{lightresults2} are considerably smaller than in
earlier drafts of this paper.
The previous version was inconsistent, in that it took the input estimate of light-quark finite volume errors from a comparison of
fits including the data at $a\approx 0.15$~fm, while our central fit drops that lattice spacing.
As discussed in \secref{physical-mass-analysis}, keeping the $a\approx 0.15$~fm data gives an overestimate of finite-volume effects
due to staggered taste splittings that predominantly affect that lattice spacing.

Despite the fact that the decay constants are by definition pure QCD matrix elements of the axial current, there are
electromagnetic uncertainties in the values that the meson masses (used primarily to fix the physical quark masses) would have in a
pure QCD world.%
\footnote{Electromagnetic effects of course also contribute directly to the leptonic weak decays.
We include an estimate of these effects when we relate the decay constants to experimental decay rates to extract CKM matrix
elements in \secref{res_pheno}.} %
The estimated systematic error labeled ``electromagnetic corrections" in \tabrefs{results_D_f_details}{results_B_f_details} accounts
for the two sources of this uncertainty.
First, there are electromagnetic errors in the tuned values we use for the light-quark masses that arise from errors in the
determinations of the electromagnetic contributions to pion and kaon masses.
These correspond to the ``\EMone'' and ``\EMtwo,'' and errors described in \secref{physical-mass-analysis}.
We vary the values of the tuned light-quark masses by these two EM uncertainties in Table~\ref{tab:lightemerrors} to obtain the
corresponding uncertainties on the decay constants in Table~\ref{tab:results_f_EM}.
In this work, we choose a specific scheme~\cite{Borsanyi:2013lga,Basak:2018yzz} for the electromagnetic contribution to the neutral
kaon masses; other works, for example the FLAG report~\cite{Aoki:2016frl}, choose other schemes.
Changing the scheme so that $\left(M_{K^0}^2\right)^\gamma$ goes from $+44~\MeV^2$ to $+461~\MeV^2$ changes the listed quantities by
the percentages in row ``\EMtwos.''

There are also electromagnetic effects in the heavy-light meson masses, which affect our calculation both directly, in the
meson-mass value we use to convert from a $\Phi$ value to a decay constant $f=\Phi/\sqrt{M}$, and indirectly, through the tuned
values of the heavy-quark masses.
To estimate the resulting electromagnetic errors on the decay constants, we first need to relate the experimental values of the
heavy-light meson masses to QCD-only values.
For this, we use the phenomenological formula~\cite{Davies:2010ip,Goity:2007fu,Rosner:1992qw}
\begin{equation}
   M^\text{expt}_{H_x} = M^\text{QCD}_{H_{l}} + A e_x e_h + B e_x^2\, +\, C(m_x-m_l)\,,
   \label{eq:EM-model}
\end{equation}
where $e_x$ and $e_h$ are charges of the valence light and heavy quarks, respectively, and we have added a term proportional to
$(m_x-m_l)$ to account for the mass difference between $u$ and $d$ quarks.
Physical contributions proportional to $e_h^2$, which come from effects such as the EM correction to the heavy quark's
chromomagnetic moment, are suppressed by $1/m_h$, and are therefore dropped from this simple model.
There are also prescription (scheme) dependent EM contributions to the heavy quark mass renormalization, which are proportional to
$e_h^2m_h$; our choice of scheme is to drop them entirely.
To estimate the parameters $A$ and $B$, we use the experimental $D^0$-, $D^+$-, $B^+$- and $B^0$--meson masses in \eq{EM-model},
which gives
\begin{align} 
    M^\text{expt}_{D^+}-M^\text{expt}_{D^0} &= +4.75~\MeV =  \frac{2}{3} A - \frac{1}{3}B + C(m_d-m_u) ,\\
    M^\text{expt}_{B^+}-M^\text{expt}_{B^0} &= -0.31~\MeV =  \frac{1}{3} A + \frac{1}{3}B - C(m_d-m_u) .
    \label{eq:EM-mass-splitting}
\end{align}
Taking $C(m_d-m_u)=2.6$~MeV as described in Sec.~\ref{sec:physical-mass-analysis},
we then obtain $A=4.44$~MeV and $B=2.4$~MeV.
% \cut{We emphasize that our determination of the parameters $A$ and $B$ omits electromagnetic corrections that depend on quark 
% masses.}

%%%%%%%%%%%%%%%%%%%%%%%%%%%%
\begin{table}
\newcommand{\h}{\phantom{x}}
\caption{Error contributions to, and estimates of scheme dependence of, the decay constants from electromagnetic effects.
The sources of uncertainty are described in the text.}
\label{tab:results_f_EM}
\begin{tabular}{lcccccc}
\hline\hline
Error (\%) &  \h$f_{D^0}$\h  &  \h$f_{D^+}$\h  &  \h$f_{D_s}$\h  &  \h$f_{B^+}$\h  &  \h$f_{B^0}$\h  &  \h$f_{B_s}$\h  \\
\hline
\EMone 	  & $0.02$ & $0.00$ & $0.01$ & $0.02$ & $0.00$ & $0.01$ \\
\EMtwo 	  & $0.00$ & $0.00$ & $0.00$ & $0.00$ & $0.00$ & $0.00$ \\
\EMthree  & $0.05$ & $0.01$ & $0.01$ & $0.04$ & $0.02$ & $0.02$ \\
\EMtwos   & $0.03$ & $0.03$ & $0.05$ & $0.04$ & $0.03$ & $0.06$ \\
\EMthrees & $0.07$ & $0.07$ & $0.07$ & $0.05$ & $0.05$ & $0.04$ \\
\hline\hline
\end{tabular}
\end{table}

%%%%%%%%%%%%%%%%%%%%%%%%%%%%

Using \eq{EM-model}, we estimate that the electromagnetic contribution to the $D_s$-meson mass to be about 1.3~MeV, which is
substantially smaller than the result, $5.5(6)$~MeV, found for this shift in \rcite{Giusti:2017dmp}.
We emphasize that we do not add any terms in \eq{EM-model} proportional to $e_h^2 m_h$.
Such terms, which can explain the difference between results of \rcite{Giusti:2017dmp} and \eq{EM-model}, can be absorbed into the
heavy-quark mass and do not contribute to electromagnetic mass splittings for the heavy-light mesons.
Consequently, these terms only affect the tuned heavy-quark masses, which inevitably depend on the scheme used for matching a pure
QCD calculation onto real-world measurements, which include electromagnetism.

We take the difference between results obtained with and without the electromagnetic
shift from \eq{EM-model} as an estimate of the uncertainty in applying our phenomenological model.
This error includes effects of neglecting mass-dependent corrections to the parameters $A$ and $B$.
We tabulate this error in the row labeled ``\EMthree.''
We also estimate the effect of the scheme dependence of the heavy quark mass,
which we call ``\EMthrees,'' by taking the difference between the electromagnetic contributions to the $D_s$ meson mass obtained
from \eq{EM-model} and the scheme of \rcite{Giusti:2017dmp}, which includes the heavy-quark self-energy.
We do not have corresponding information for the $B_s$ meson, so we take the $D_s$ shift
and simply assume that it is dominated by a mass renormalization term proportional to $e_h^2 m_h$.
Because $m_c e_c^2 \approx m_b e_b^2$, this leads to the same shift, 4.2 MeV,
for both $D_s$ and $B_s$.

The individual electromagnetic EM uncertainties on the decay constants discussed above are tabulated in \tabref{results_f_EM}.
Because we have no information about correlations between the various EM errors, we add the \EMone, \EMtwo, and
\EMthree\ error in quadrature to obtain the total ``electromagnetic corrections" entries given in
\tabrefs{results_D_f_details}{results_B_f_details}.
Even if there were strong correlations between the EM errors, this would make little difference to the total systematic errors of
the heavy-light decay constants, because these errors are subdominant, as can be seen in
\tabrefs{results_D_f_details}{results_B_f_details}.

The error labeled ``topological-charge distribution" accounts for the nonequilibration of topological charge in our finest ensembles.
Before our EFT fit, we adjust the lattice data to compensate for effects of nonequilibration of topological charge as discussed in
Sec.~\ref{subsec:topology}.
We conservatively estimate the uncertainty in our treatment of effects of nonequilibration of topological charge by taking the full
difference between the final results of the analyses with and without adjustments.

The last ``$\fpiPDG$" error included in \tabrefs{results_D_f_details}{results_B_f_details} is the uncertainty due to the
error in the PDG average for the charged-pion decay constant, $f_{\pi^\pm}=130.50(13)$~MeV \cite{Rosner:2015wva}, which is the
physical scale that is used to determine $f_{p4s}$.

All errors in \tabrefs{results_D_f_details}{results_B_f_details} should be added in quadrature to obtain the total uncertainties.
In the following section, when we quote our final results for the physical decay constants, we separate the errors into
``statistical'' errors, which are the ones listed as ``statistics and EFT fit,'' ``systematic'' errors, which are those due to the
systematics of our calculation (rows 2--6 in the tables, added in quadrature), and, finally, the errors due to the PDG value of
$f_\pi$, which is external to our calculation.

As a byproduct of our EFT analysis, we can also obtain the decay amplitudes $\Phi$ for the $D$ and $B$ systems in both the SU(2) and
the SU(3) limits, which are reported in \tabref{results_Phi_chiral}.

%%%%%%%%%%%%%%%%%%%%%%%%%%%%
\begin{table}
\caption{\label{tab:results_Phi_chiral} Results for $\Phi$ in the SU(2) and the SU(3) chiral limits.
  Here $m_x=m'_l=0$ and the strange sea mass is either $m'_s=m_s$ (in the SU(2) case) or $m'_s=0$ (in the SU(3) case).
  The uncertainty labeled ``EM~scheme'' is an additional uncertainty that can be incorporated when these results are used
  without attention to the EM scheme dependence.
}
\begin{tabular}{l l}
\hline\hline
$D$ system \hspace{1.5cm} &
  $\Phi_0^\text{SU(3)} =  8133 (67)_\text{stat} (93)_\text{syst} (12)_{\fpiPDG} [15]_\text{EM scheme}~\MeV^{3/2}$ \\ & 
  $\Phi_0^\text{SU(2)} =  8976 (12)_\text{stat} (24)_\text{syst} (11)_{\fpiPDG} [17]_\text{EM scheme}~\MeV^{3/2}$ \\
\hline
$B$ system &
  $\Phi_0^\text{SU(3)} = 11717 (205)_\text{stat}(181)_\text{syst} (21)_{\fpiPDG} [11]_\text{EM scheme}~\MeV^{3/2}$ \\ &
  $\Phi_0^\text{SU(2)} = 13461  (57)_\text{stat} (73)_\text{syst} (20)_{\fpiPDG} [13]_\text{EM scheme}~\MeV^{3/2}$ \\
\hline\hline
\end{tabular}
\end{table}

%%%%%%%%%%%%%%%%%%%%%%%%%%%%

\section{Results and phenomenological impact} 
\label{sec:res_pheno}

We now present our final results for the heavy-light meson decay constants with total errors and then discuss some of their
phenomenological implications.

\subsection{\boldmath\texorpdfstring{$B$}{B}- and \texorpdfstring{$D$}{D}-meson decay constants}
\label{sec:results}

Our final results for the physical leptonic decay constants of the $D$ and $B$ systems including all sources of systematic
uncertainty discussed in the previous section are
\begin{align}
    f_{D^0}  &= 211.6  (0.3)_\text{stat}  (0.5)_\text{syst}  (0.2)_{\fpiPDG} [0.2]_\text{EM scheme}~\MeV , \label{eq:fD0}\\ 
    f_{D^+}  &= 212.7  (0.3)_\text{stat}  (0.4)_\text{syst}  (0.2)_{\fpiPDG} [0.2]_\text{EM scheme}~\MeV , \label{eq:fD+}\\ 
    f_{D_s}  &= 249.9  (0.3)_\text{stat}  (0.2)_\text{syst}  (0.2)_{\fpiPDG} [0.2]_\text{EM scheme}~\MeV ,
    \label{eq:fDs}\\ 
    f_{B^+}  &= 189.4  (0.8)_\text{stat}  (1.1)_\text{syst}  (0.3)_{\fpiPDG} [0.1]_\text{EM scheme}~\MeV , \label{eq:fB+}\\ 
    f_{B^0}  &= 190.5  (0.8)_\text{stat}  (1.0)_\text{syst}  (0.3)_{\fpiPDG} [0.1]_\text{EM scheme}~\MeV , \label{eq:fB0}\\ 
    f_{B_s}  &= 230.7  (0.8)_\text{stat}  (1.0)_\text{syst}  (0.2)_{\fpiPDG} [0.2]_\text{EM scheme}~\MeV .
    \label{eq:fBs}
\end{align}
These results are obtained in a specific scheme for matching QCD+QED to pure QCD via
the light and heavy meson masses tabulated in Table~\ref{tab:QCD_masses}.
When using our results in a setting that does not take into account the subtleties of the EM scheme, one may wish to also include
the last quantities, in brackets, which are obtained by adding in quadrature the fourth and fifth rows in
Table~\ref{tab:results_f_EM}, as rough estimates of scheme dependence.

Most recent lattice-QCD calculations of heavy-light meson decay constants work, however, in the isospin-symmetric limit.
To enable comparison with these results, we also present results for the $B$- and $D$-meson decay constants evaluated with the light
valence-quark mass fixed to the average $u/d$-quark mass:
\begin{align}
  f_{D}  &= 212.1  (0.3)_\text{stat}  (0.4)_\text{syst}  (0.2)_{\fpiPDG} [0.2]_\text{EM scheme}~\MeV , \label{eq:fD} \\
  f_{B}  &= 190.0  (0.8)_\text{stat}  (1.0)_\text{syst}  (0.3)_{\fpiPDG} [0.1]_\text{EM scheme}~\MeV . \label{eq:fB} 
\end{align}

\begin{figure}
	\centering
    \includegraphics[height=0.5\textwidth]{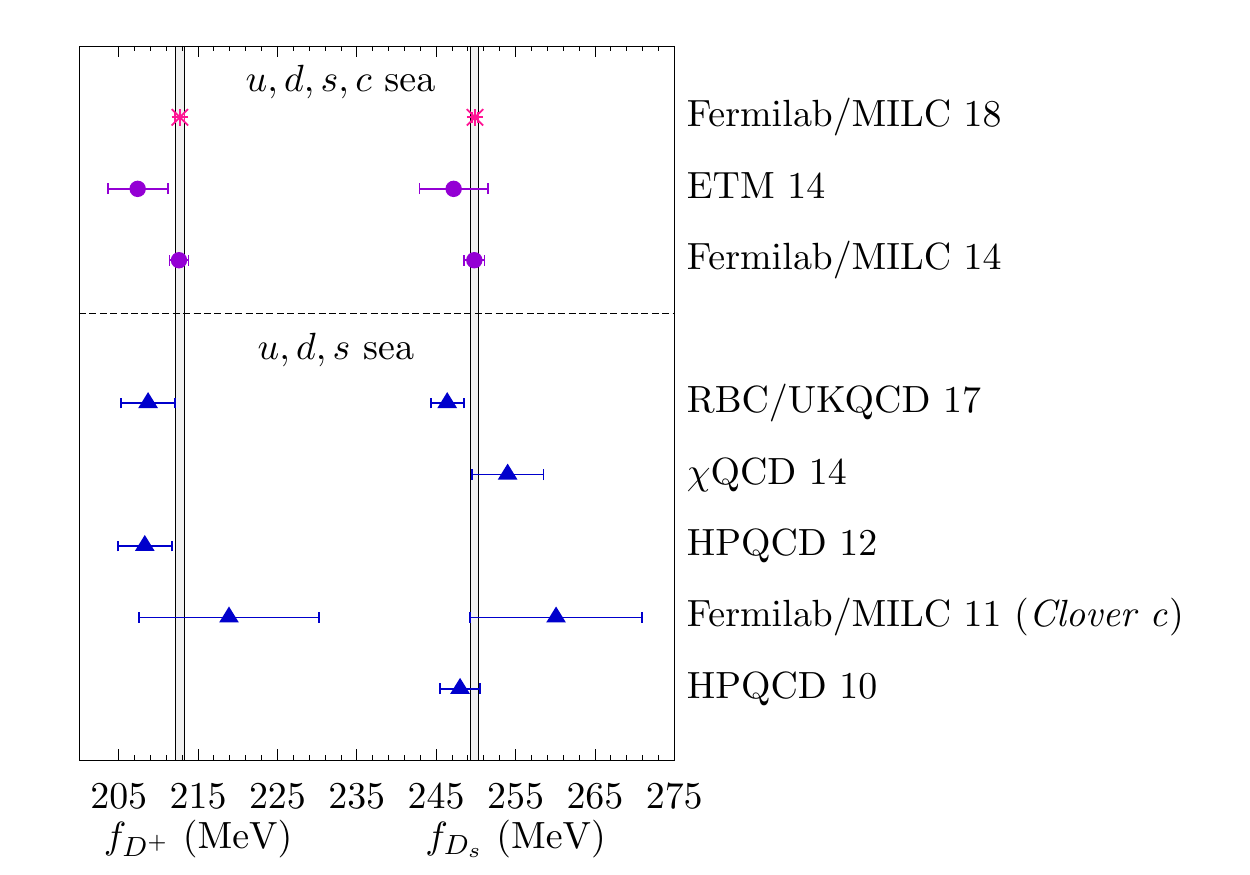}
    \caption{Comparison of our $D$-meson decay-constant results (magenta bursts) with previous three- and four-flavor lattice-QCD
        calculations~\cite{Davies:2010ip,Bazavov:2011aa,Na:2012iu,Bazavov:2014wgs,Yang:2014sea,Carrasco:2014poa,Boyle:2017jwu}.
        The vertical gray bands show the total uncertainties from Eqs.~(\ref{eq:fD+}) and~(\ref{eq:fDs}).
        The asymmetric errors on the RBC/UKQCD 17 results have been symmetrized.}
	\label{fig:fD_summary}
\end{figure}

\begin{figure}
    \centering
    \includegraphics[height=0.5\textwidth]{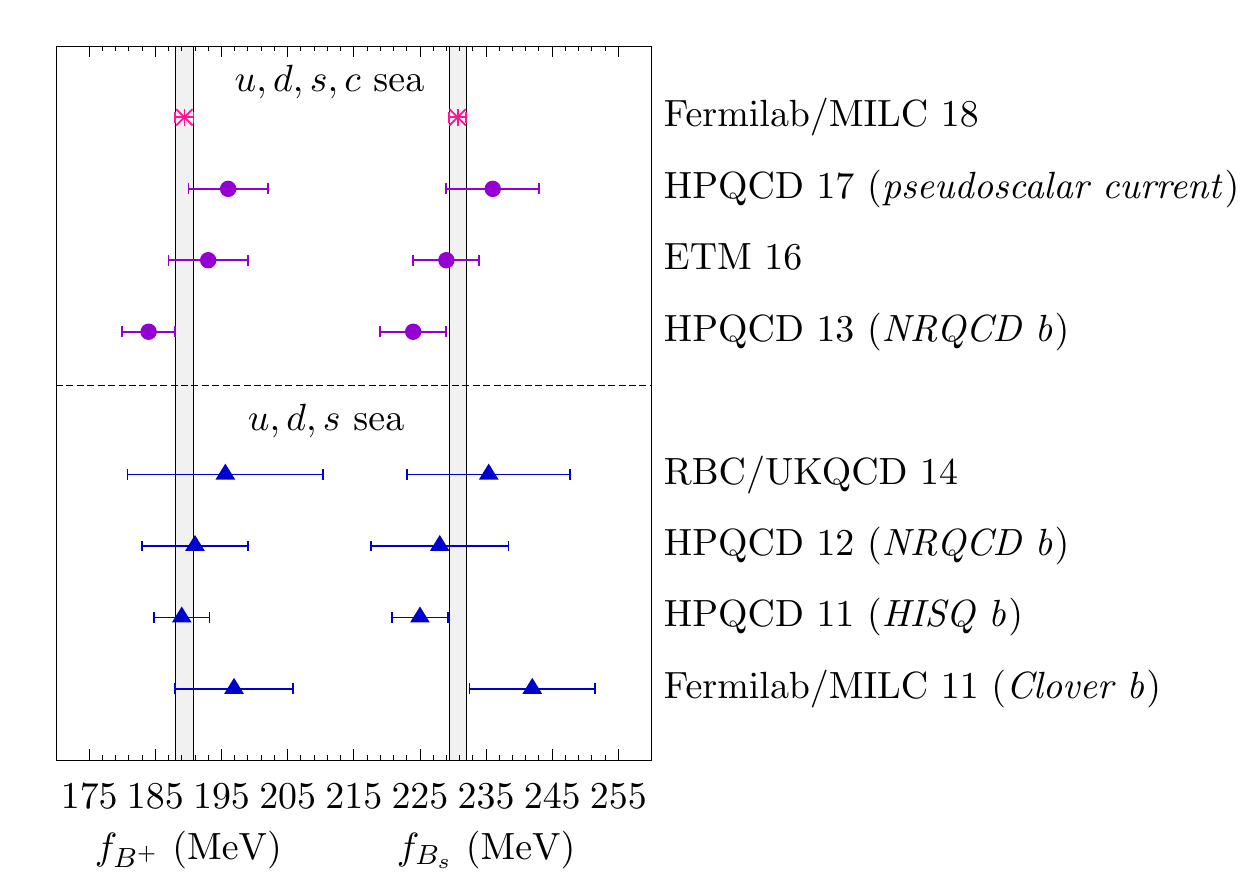}
	\caption{Comparison of $B$-meson decay-constant results (magenta bursts) with previous three- and four-flavor lattice-QCD 
    calculations~\cite{McNeile:2011ng,Bazavov:2011aa,Na:2012kp,Dowdall:2013tga,Christ:2014uea,Bussone:2016iua,Hughes:2017spc}.
    The vertical gray bands show the total uncertainties from Eqs.~(\ref{eq:fB+}) and~(\ref{eq:fBs}).}
	\label{fig:fB_summary}
\end{figure}

Figures~\ref{fig:fD_summary} and~\ref{fig:fB_summary} compare our decay-constant results with previous three- and four-flavor
lattice-QCD calculations~\cite{Davies:2010ip,McNeile:2011ng,Bazavov:2011aa,Na:2012kp,Na:2012iu,Dowdall:2013tga,%
Christ:2014uea,Bazavov:2014wgs,Yang:2014sea,Carrasco:2014poa,Bussone:2016iua,Boyle:2017jwu,Hughes:2017spc}.
They agree with the lattice-QCD averages from the Particle Data Group~\cite{Rosner:2015wva}:
\begin{align}
    f_{D^+, \text{\,PDG}} &= 211.9(1.1)~\MeV, \label{eq:fD_PDG}  \\
    f_{D_s, \text{\,PDG}} &= 249.0(1.2)~\MeV, \label{eq:fDs_PDG} \\
    f_{B^+, \text{\,PDG}} &= 187.1(4.2)~\MeV, \label{eq:fB+_PDG} \\
    f_{B^0, \text{\,PDG}} &= 190.9(4.1)~\MeV, \label{eq:fB0_PDG} \\
    f_{B_s, \text{\,PDG}} &= 227.2(3.4)~\MeV, \label{eq:fBs_PDG}
\end{align}
where we note that the $D_{(s)}$ averages are dominated by our earlier result in Ref.~\cite{\rFD2014}.

For the $D$-meson decay constants, the uncertainties in Eqs.~(\ref{eq:fD+})--(\ref{eq:fDs}) are about 2.5 times smaller than from
our previous analysis.
The improvement stems primarily from the inclusion of finer ensembles with $a\approx 0.042$~fm and 0.03~fm, which reduce the
distance of the continuum extrapolation.

For $B$-meson decay constants, the uncertainties in Eqs.~(\ref{eq:fB+})--(\ref{eq:fBs}) are approximately three times smaller than
from the previous best calculations from HPQCD~\cite{McNeile:2011ng,Dowdall:2013tga}.
For $f_{B_s}$, HPQCD's most precise determination was obtained with the HISQ action for $b$ quarks~\cite{McNeile:2011ng}.
The substantial improvement in our result comes from a combination of higher statistics and the
ensemble with $a \approx 0.03$~fm, which eliminates the need to extrapolate to the bottom-quark mass from lighter quark masses, and
also shortens the continuum extrapolation.
For $f_{B^+}$ and $f_{B^0}$, HPQCD has employed only NRQCD $b$ quarks~\cite{Dowdall:2013tga}.
Thus, our results for these quantities are the first obtained with the HISQ action for the $b$ quarks.
With HISQ, the dominant errors in HPQCD's calculation---from operator matching and relativistic corrections to the
current---simply do not arise.

Because the statistical and several systematic errors are correlated between the decay constants in
Eqs~(\ref{eq:fD+})--(\ref{eq:fBs}), we can obtain combinations of decay constants with even greater precision.
Our results for the decay-constant ratios are
\begin{align}
    f_{D_s}/f_{D^+}  &= 1.1749  (06)_\text{stat}  (14)_\text{syst}  (04)_{\fpiPDG} [03]_\text{EM scheme} , \\
    f_{B_s}/f_{B^+}  &= 1.2180  (33)_\text{stat}  (33)_\text{syst}  (05)_{\fpiPDG} [03]_\text{EM scheme} , \\
    f_{B_s}/f_{B^0}  &= 1.2109  (29)_\text{stat}  (25)_\text{syst}  (04)_{\fpiPDG} [03]_\text{EM scheme} ,
    \label{eq:fBs_fB0} \\
    f_{B_s}/f_{D_s}  &= 0.9233  (25)_\text{stat}  (42)_\text{syst}  (02)_{\fpiPDG} [03]_\text{EM scheme}.
\end{align}
The light quarks in the $D^+$ and $D_s$ mesons have identical charges, so the deviation of $f_{D_s}/f_{D^+}$ from unity
quantifies the degree of $SU(3)$-flavor breaking in the $D$ system.
Similarly, the ratio $f_{B_s}/f_{B_0}$ characterizes the size of $SU(3)$-breaking in the $B$-meson system.
Both yield values of about 20\%, which is consistent with power-counting expectations of $(m_s-m_d)/\Lambda_\text{QCD}$.

For the differences due to strong isospin breaking (\emph{i.e.}, $m_u \neq m_d$) we find
\begin{align}
    f_{D^+}-f_{D}    &= 0.58  (01)_\text{stat}  (07)_\text{syst}  (00)_{\fpiPDG} [01]_\text{EM scheme}~\MeV ,
    \label{eq:IB_fD} \\
    f_{D^+}-f_{D^0}  &= 1.11  (03)_\text{stat}  (15)_\text{syst}  (00)_{\fpiPDG} [01]_\text{EM scheme}~\MeV ,
    \label{eq:IB_fD0}\\
    f_{B}  -f_{B^+}  &= 0.53  (05)_\text{stat}  (07)_\text{syst}  (00)_{\fpiPDG} [00]_\text{EM scheme}~\MeV ,
    \label{eq:IB_fB} \\
    f_{B^0}-f_{B^+}  &= 1.11  (08)_\text{stat}  (13)_\text{syst}  (00)_{\fpiPDG} [01]_\text{EM scheme}~\MeV .
    \label{eq:IB_fB0}
\end{align}
These results can be employed to correct other lattice-QCD results obtained in the isospin limit, which will be essential once other
calculations reach sub-percent precision.
For $f_{D^+}$, the isospin-breaking correction is larger than our total uncertainty in Eq.~(\ref{eq:fD+}), while for $f_{B^+}$ it is
comparable to the total error in Eq.~(\ref{eq:fB+}).
We find a smaller isospin correction to the $B$-meson decay constant than obtained by HPQCD in Ref.~\cite{Dowdall:2013tga},
$(f_{B} -f_{B^+})_\text{HPQCD} = 1.9(5)$~MeV,%
\footnote{The correlated uncertainties were provided by HPQCD (private communication).} %
by more than $2\sigma$.
HPQCD's estimate was obtained, however, by setting both the valence- and sea-quark masses in $f_{B^+}$ to $m_u$ because the analysis
only included unitary data.
Hence their value includes effects both from valence isospin breaking and from reducing the average light sea-quark mass;
when we follow this prescription, we obtain a similarly-large shift of about $1.6(2)$~MeV.
On the other hand, our results for the isospin corrections to both $f_D$ and $f_B$ agree with calculations using Borelized sum
rules~\cite{Lucha:2016nzv,Lucha:2017zng}.

Tables~\ref{tab:correlation_matrix} and~\ref{tab:covariance_matrix} in Appendix~\ref{Appendix:Covariance-Matrix} provide the
correlation and covariance matrices, respectively, between the $B$- and $D$-meson decay constants in
Eqs.~(\ref{eq:fD0})--(\ref{eq:fB}).
They can be used to compute any combination of our results with the correct uncertainties.

\subsection{\boldmath Quark-mass ratios, \texorpdfstring{$f_K/f_\pi$}{fK/fπ}, and scale-setting quantities}
\label{sec:light_results}

In Sec.~\ref{sec:physical-mass-analysis}, we analyze the ensembles with physical light-quark masses to obtain several input
parameters for the EFT fit of heavy-light meson decay constants.
We obtain for the mass and decay constant of a fictitious pseudoscalar-meson with degenerate valence-quark masses $0.4m_s$:
\begin{align}
    f_{p4s} &= 153.98 (11)_\text{stat}(\null_{-12}^{\, +2})_\text{syst}(12)_{\fpiPDG}[4]_\text{EM scheme}~\MeV, \\
    M_{p4s} &= 433.12 (14)_\text{stat}(\null_{-6}^{+17})_\text{syst}(4)_{\fpiPDG}[40]_\text{EM scheme}~\MeV, \\
    f_{p4s}/M_{p4s} &= 0.3555 (3)_\text{stat}(\null_{-4}^{+1})_\text{syst}(3)_{\fpiPDG}[2]_\text{EM scheme} ,
\end{align}
where the last quantity, in brackets, is an additional uncertainty
when these results are used without attention to EM scheme dependence.
These quantities are used to set the scale in our analysis.

\begin{figure}
	\centering
    \includegraphics[height=0.5\textwidth]{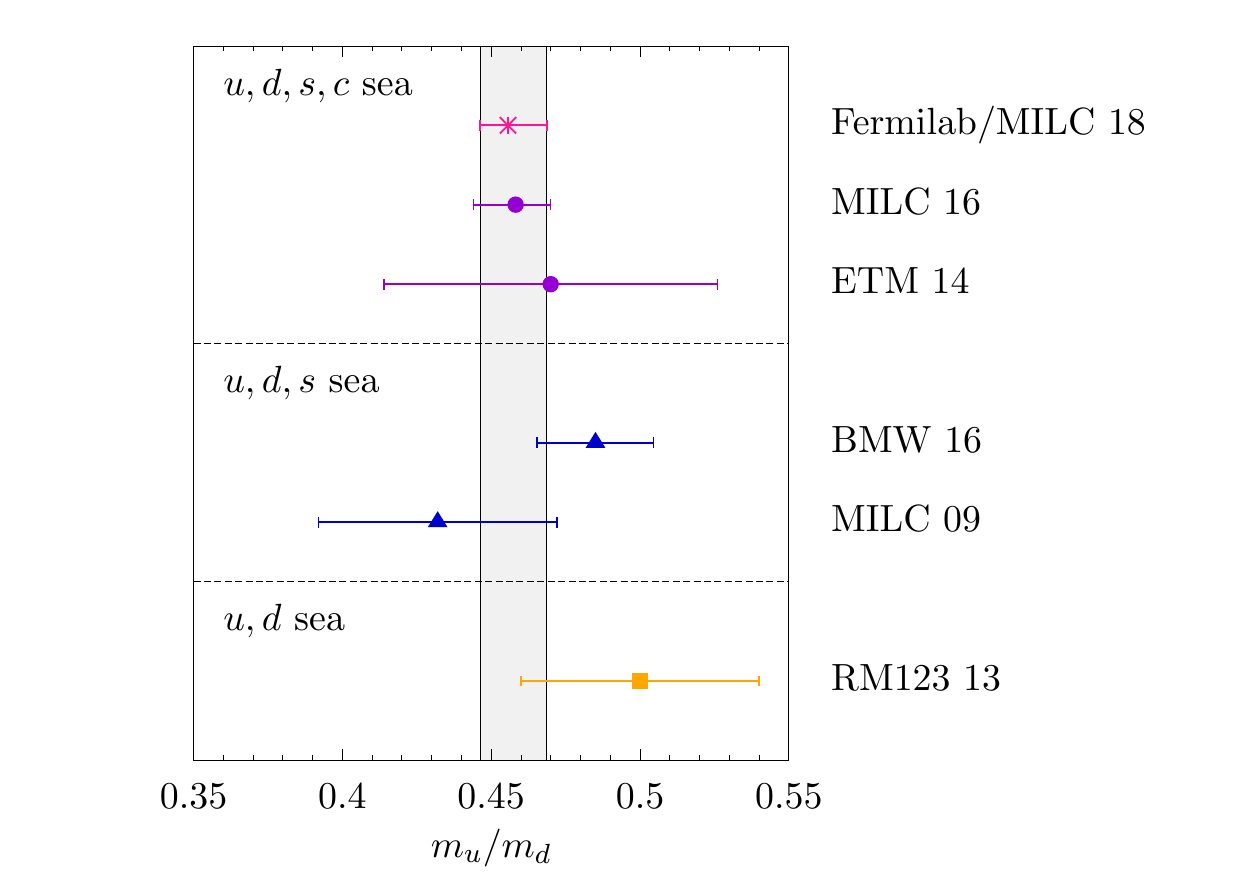}
	\caption{Comparison of $m_u/m_d$ in Eq.~(\ref{eq:mu_md}) (magenta burst) with previous unquenched lattice-QCD
        calculations~\cite{Bazavov:2009fk,deDivitiis:2013xla,Carrasco:2014cwa,Basak:2016jnn,Fodor:2016bgu}.
	\label{fig:mu_md_summary}}
\end{figure}

\begin{figure}[p]
	\centering
    \includegraphics[height=0.5\textwidth]{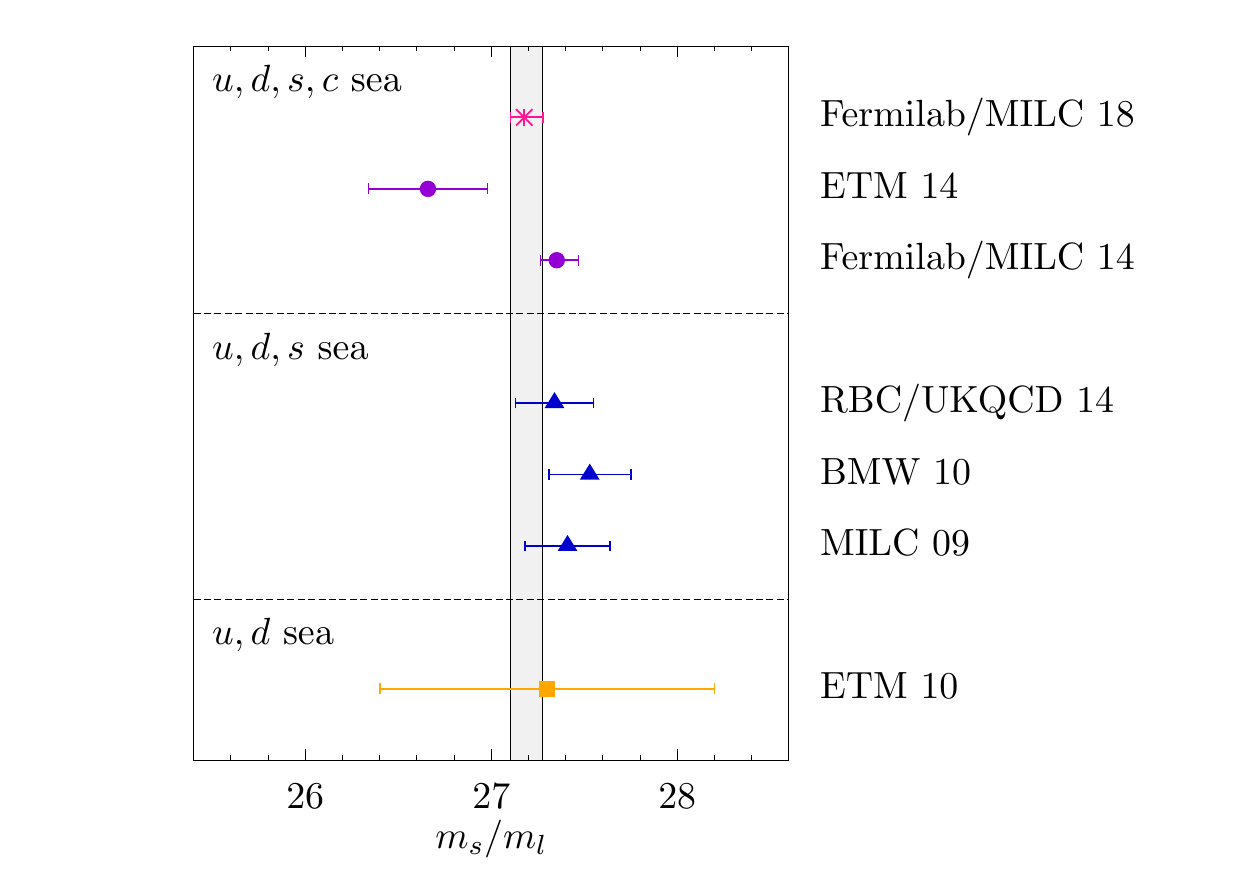}
	\caption{Comparison of $m_s/m_l$ in Eq.~(\ref{eq:ms_ml}) (magenta burst) with previous unquenched lattice-QCD
        calculations~\cite{\rFD2014,Bazavov:2009fk,Baron:2010bv,Durr:2010vn,Carrasco:2014cwa,Blum:2014tka}.}
	\label{fig:ms_ml_summary}
\end{figure}

We obtain for the ratios of quark masses: 
\begin{align}
    m_u/m_d &=  0.4556 (55)_\text{stat}(\null_{-67}^{+114})_\text{syst}(13)_{\Delta M_K}[32]_\text{EM scheme}, 
    \label{eq:mu_md} \\
    m_s/m_l &= 27.178 (47)_\text{stat}( \null_{-26}^{+70})_\text{syst}(1)_{\fpiPDG}[51]_\text{EM scheme}, 
    \label{eq:ms_ml} \\
    m_c/m_s &=  11.773 (14)_\text{stat}( \null_{-57}^{+14})_\text{syst}(6)_{\fpiPDG}[49]_\text{EM scheme} ,
    \label{eq:mqRatios}
\end{align}
where $m_l$ is the average $u/d$-quark mass.
The errors on the quark-mass ratios in Eqs.~(\ref{eq:mu_md})--(\ref{eq:mqRatios}) are smaller than from our previous analysis in
Ref.~\cite{\rFD2014} because the finer lattice spacings employed here reduce the continuum-extrapolation error.
Figures~\ref{fig:mu_md_summary} and~\ref{fig:ms_ml_summary} compare our results for $m_u/m_d$ and $m_s/m_l$, respectively, with
previous unquenched lattice-QCD calculations.
The difference in our value for $m_s/m_l$ relative to Ref.~\cite{\rFD2014} mostly comes from three changes, which all push the value
in the same direction.
In order of size, these are the addition of the 0.042~fm physical-quark-mass ensemble, removing the 0.15~fm ensembles from our
central fits, and adding more data on the 0.06~fm physical-quark-mass ensembles.
An even more precise value for $m_c/m_s$ is reported in a companion paper on the determination of quark masses
from heavy-light meson masses~\cite{Bazavov:2018omf}.

\begin{figure}[tbp]
	\centering
    \includegraphics[height=0.5\textwidth]{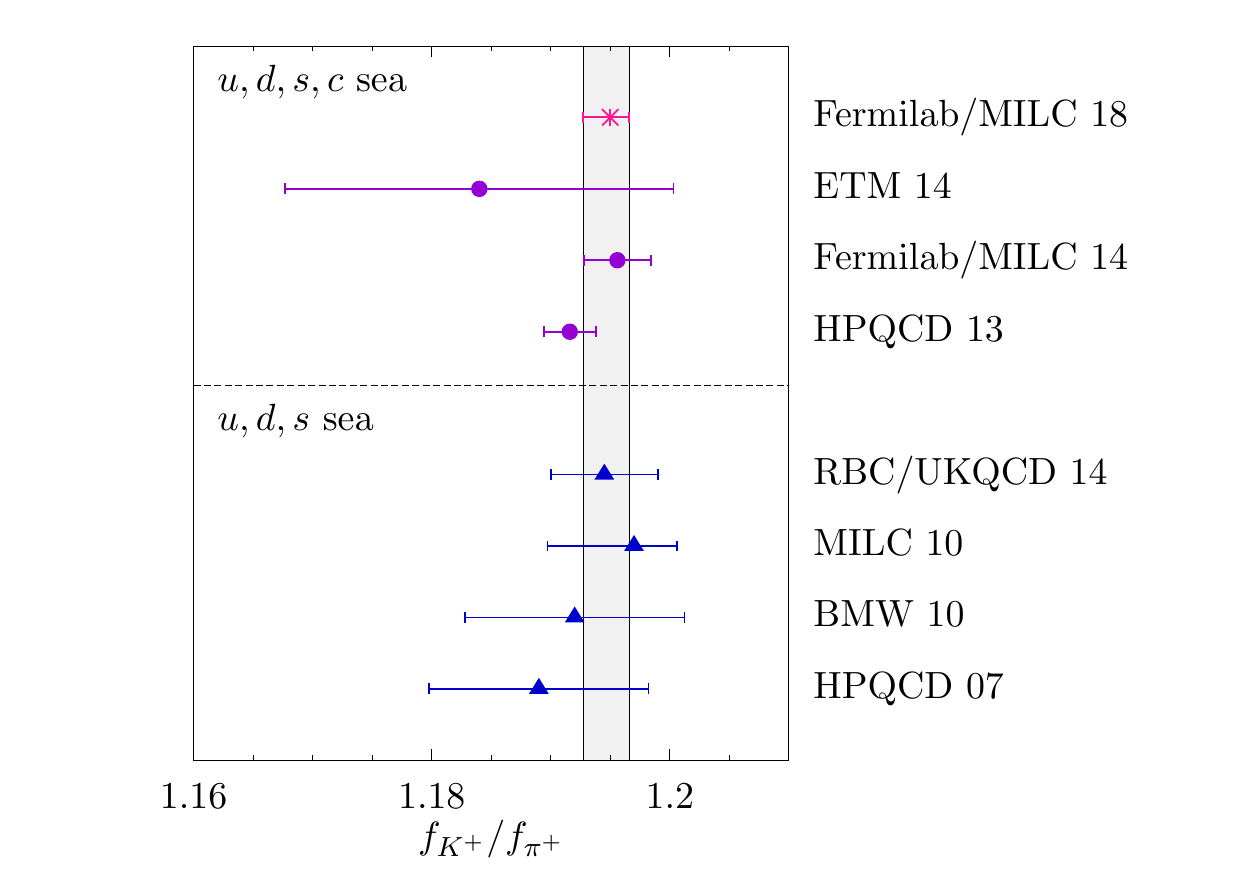}
	\caption{Comparison of $f_{K^+}/f_{\pi^+}$ in Eq.~(\ref{eq:fKfPi}) (magenta burst) with previous three- and four-flavor 
        lattice-QCD calculations~\cite{\rFD2014,Follana:2007uv,Durr:2010hr,Bazavov:2010hj,Dowdall:2013rya,Blum:2014tka,%
        Carrasco:2014poa}.}
	\label{fig:fKfPi_summary}
\end{figure}

Finally, we obtain the ratio of charged pion to kaon decay constants.
We also give the ratio in the isospin symmetric limit, and the difference between the two:
\begin{align}
    f_{K^+}/f_{\pi^+}  &= 1.1950 (15)_\text{stat}(\null_{-17}^{\,+4})_\text{syst}(3)_{\fpiPDG}[3]_\text{EM scheme} ,
    \label{eq:fKfPi} \\
    f_{\bar K}/f_{\pi} &= 1.1980 (12)_\text{stat}(\null_{-14}^{\,+3})_\text{syst}(3)_{\fpiPDG}[3]_\text{EM scheme} , 
    \label{eq:fKfPi-iso} \\
    f_{\bar K}/f_{\pi} - f_{K^+}/f_{\pi^+} &=
        0.00305(50)_\text{stat}(\null_{-12}^{\,+31})_\text{syst}(2)_{\fpiPDG,\Delta M_K}[3]_\text{EM scheme} ,
    \label{eq:fKfPi-diff}
\end{align}
which are again more precise than our previous determination in Ref.~\cite{\rFD2014} because of the shorter 
continuum extrapolation.
Our results agree with previous three- and four-flavor lattice-QCD calculations (see Fig.~\ref{fig:fKfPi_summary}), 
and with the 2016 FLAG averages~\cite{Aoki:2016frl}.

\subsection{CKM matrix elements}
\label{sec:CKM}
We now combine our decay-constant results with experimental measurements of the $D^+_{(s)}$-meson leptonic decay rates to obtain
values for the CKM matrix elements $|V_{cd}|$ and $|V_{cs}|$ within the Standard Model.

The products of decay constants times CKM factors from the Particle Data Group~\cite{Rosner:2015wva},
\begin{align}
    \left(f_{D^+}  |V_{cd}|\right)_\text{expt} &= 45.91(1.05)~\MeV, \label{eq:fD_Vcd_PDG} \\
    \left(f_{D_s^+}|V_{cs}|\right)_\text{expt} &=  250.9(4.0)~\MeV, \label{eq:fD_Vcs_PDG}
\end{align}
are obtained by averaging the experimentally-measured decay rates into muon and tau final states.
The value for $f_{D^+}|V_{cd}|$ in Eq.~(\ref{eq:fD_Vcd_PDG}) includes the correction from structure-dependent bremsstrahlung effects
that lowers the $D^+ \to \mu^+ \nu_{\mu}$ rate by $\sim1\%$~\cite{Burdman:1994ip,Dobrescu:2008er}.
Other electroweak corrections, however, are not accounted for in the PDG averages shown above.
The electroweak contributions to leptonic pion and kaon decays are estimated to be about one or two
percent~\cite{Cirigliano:2007ga,Cirigliano:2011ny}, and the uncertainties in these corrections lead to $\sim 0.1\%$
uncertainties in $|V_{us}|/|V_{ud}|$ and $|V_{us}|$.
Now that the errors on $f_{D}$ and $f_{D_s}$ are well below half a percent, electroweak corrections must also be included when
extracting $|V_{cd}|$ and $|V_{cs}|$ from leptonic $D$-meson decays.

We take the estimate of the electroweak corrections to the leptonic $D^+_{(s)}$-meson decay rates from our earlier
work~\cite{\rFD2014}, which includes all contributions that are included for pion and kaon decays.
We first adjust the experimental decay rates quoted in the PDG by the known long- and short-distance electroweak
corrections~\cite{Kinoshita:1959ha,Sirlin:1981ie}.
The former lowers the $D^+$- and $D_s$-meson leptonic decay rates by about 2.5\%, while the latter increases them by about $1.8\%$,
such that the net effect is a slight decrease in the rates by less than a percent.
We then include a 0.6\% uncertainty to account for unknown electromagnetic corrections that depend upon the mesons' structure.
This estimate is based on calculations of the structure-dependent electromagnetic corrections to pion and kaon
decays~\cite{Knecht:1999ag,Cirigliano:2007ga,DescotesGenon:2005pw}, but allowing for much larger coefficients than for the light
pseudoscalar mesons.

With these assumptions, and taking our $D^+$- and $D_s$-meson decay-constant results from Eqs.~(\ref{eq:fD+}) and~(\ref{eq:fDs}),
we obtain for the CKM matrix elements
\begin{align}
	|V_{cd}|^{\text{SM},\, f_D}  &= 0.2151(6)_{f_D}(49)_{\rm expt}(6)_{\rm EM},
    \label{eq:Vcd_fD} \\
    |V_{cs}|_{\text{SM},\, f_{D_s}} &= 1.000(2)_{f_{D_s}}(16)_{\rm expt}(3)_{\rm EM},
    \label{eq:Vcs_fDs}
\end{align}
where ``EM" denotes the error due to unknown structure-dependent electromagnetic corrections.
In both cases, the lattice-QCD uncertainties from the decay constants are an order of magnitude smaller than those from experiment.
Further, the electromagnetic errors are only a rough estimate, and need to be put on a more robust and quantitative footing by a
direct calculation of the hadronic structure-dependent effects.

The CKM matrix elements $|V_{cd}|$ and $|V_{cs}|$ can also be obtained from semileptonic $D^+\to\pi^0\ell^+\nu$ and $D^+\to
K^0\ell^+\nu$ decays.
Recently the ETM Collaboration published the first four-flavor lattice-QCD determination of the vector and scalar form factors
for these processes~\cite{Lubicz:2017syv}.
Combining their form factors over the full range of momentum transfer with experimental measurements of the decay rates yields for
the CKM elements~\cite{Riggio:2017zwh}
\begin{align}
    |V_{cd}|_{D\to\pi} &= 0.2341(74) , \\
    |V_{cs}|_{D\to K}  &= 0.970(33)   ,
\end{align}
where the errors are primarily from the theoretical uncertainties on the form factors.
Although our result for $|V_{cs}|$ in Eq.~(\ref{eq:Vcs_fDs}) agrees with this
determination, our result for $|V_{cd}|$ in Eq.~(\ref{eq:Vcd_fD}) is about 2.1$\sigma$ lower than the above value from semileptonic
decays.
We note, however, that combining $f_+^{D\pi}(0)|V_{cd}| = 0.1425(19)$ from the Heavy Flavor Averaging Group~\cite{Amhis:2016xyh}
with $f_+^{D\pi}(0)=0.666(29)$ from the most precise three-flavor lattice-QCD calculations by HPQCD~\cite{Na:2011mc} leads to a
lower value of $|V_{cd}|_{D\to\pi} = 0.2140(97)$ that agrees with our result.

Our results for $|V_{cd}|$ and $|V_{cs}|$ make possible a test of the unitarity of the second row of the CKM matrix.
Taking $|V_{cb}|_\text{incl+excl}=41.40(77) \times 10^{-3}$ from a weighted average of determinations from inclusive and exclusive
semileptonic $B$ decays~\cite{Bailey:2014tva,Lattice:2015rga,Na:2015kha,Gambino:2016jkc,Bigi:2016mdz,Bigi:2017njr}, we obtain for
the sum of squares of the CKM elements
\begin{equation}
|V_{cd}|^2 + |V_{cs}|^2 + |V_{cb}|^2 - 1.0 = 0.049(2)_{|V_{cd}|}(32)_{|V_{cs}|}(0)_{|V_{cb}|} ,
\end{equation}
which is compatible with three-generation CKM unitarity within 1.5$\sigma$.
The precision on the above test is only at the few-percent level, and is limited by the experimental error on the leptonic decay
widths for $D_s \to \mu\nu_{\mu}$ and $D_s \to \tau \nu_{\tau}$.

We can also update the determination of the ratio of CKM elements $|V_{us}/V_{ud}|$ from leptonic pion and kaon decays.
Combining our result for $f_{K^+}/f_{\pi^+}$ in Eq.~(\ref{eq:fKfPi}) with the experimental rates and estimated radiative-correction
factor from the Particle Data Group~\cite{Rosner:2015wva}, we obtain
\begin{equation}
    |V_{us}/V_{ud}|_\text{SM} = 0.2310(4)_{f_K/f_\pi}(2)_\text{expt}(2)_\text{EM} ,
\end{equation}
where we have averaged the upper and lower errors from our decay-constant ratio.

\subsection{Branching ratios for \texorpdfstring{\boldmath$B_q\to\mu^+\mu^-$}{Bq to mumu}}

The rare leptonic decays $B^0 \to \mu^+\mu^-$ and $B_{s} \to \mu^+\mu^-$ proceed via flavor-changing-neutral-current interactions
and are therefore promising new-physics search channels.
In the $B_s$-meson system, the difference between decay widths of the light and heavy mass eigenstates is large,
$\Delta\Gamma_s/\Gamma_s\sim0.1$~\cite{Amhis:2016xyh}, and leads to a difference between the CP-averaged and time-averaged branching
ratios.
Because only the heavy $B_s$ eigenstate can decay to $\mu^+\mu^-$ pairs in the Standard Model, to a very good
approximation~\cite{Bobeth:2013uxa}, the two quantities are related simply as
$\oBR(B_s\to\mu^+\mu^-)_\text{SM}=\tau_{H_s}\Gamma(B_s\to\mu^+\mu^-)_\text{SM}$, where $\tau_{H_s}$ is the lifetime of the heavy
mass eigenstate, and the bar denotes time averaging.
The relative width difference $\Delta \Gamma_d/\Gamma_d \sim 0.001$ is 100 times smaller in the $B^0$-meson system, so
$\oBR(B_s\to\mu^+\mu^-) = \BR(B_s \to \mu^+ \mu^-)$.

The LHCb and CMS experiments reported the first observation of $B_s \to \mu^+ \mu^-$ decay in 2014~\cite{CMS:2014xfa}.
This observation was subsequently confirmed by the ATLAS experiment~\cite{Aaboud:2016ire}, and LHCb has since improved upon their
initial measurement using a larger data set~\cite{Aaij:2017vad}.
The most recent results for the $B_s \to \mu^+ \mu^-$ time-integrated branching fraction are marginally compatible:
\begin{align}
    10^9 \times \oBR(B_s \to \mu^+ \mu^-)_\text{ATLAS}   & = 0.9(^{+1.1}_{-0.8}), \\
    10^9 \times \oBR(B_s \to \mu^+ \mu^-)_\text{LHCb 17} & = 3.0(0.6)(^{+0.3}_{-0.2}),     
\end{align}
with the LHCb measurement being about $1.8\sigma$ larger.
The LHCb and CMS experiments also reported 3$\sigma$ evidence for the decay $B^0\to\mu^+\mu^-$, which is suppressed in the
Standard Model relative to $B_s\to\mu^+\mu^-$ by the CKM factor $|V_{td}/V_{ts}|^2 \sim 0.04$.
The significance, however, has subsequently weakened, and ATLAS and LHCb most recently only presented upper limits
of~\cite{Aaboud:2016ire,Aaij:2017vad}
\begin{align}
    \oBR(B^0 \to \mu^+ \mu^-)_\text{ATLAS}   & <  3.4 \times 10^{-10} , \\
    \oBR(B^0 \to \mu^+ \mu^-)_\text{LHCb 17} & <  4.2 \times 10^{-10},
\end{align}
at 95\% confidence level.
 
Here we update the theoretical predictions for the Standard-Model branching ratios using our results for the neutral $B^0$- and
$B_s$-meson decay constants.
We employ the formulae in Eqs.~(6) and~(7) of Ref.~\cite{Bobeth:2013uxa}, which provide the branching ratios in terms of the decay
constants, relevant CKM elements, and a few other parametric inputs.
Using the CKM elements and other inputs listed in Table~\ref{tab:inputs}, and $f_{B^0}$, $f_{B_s}$, and their ratio from
Eqs.~(\ref{eq:fB0})--(\ref{eq:fBs}) and~(\ref{eq:fBs_fB0}), we obtain
\begin{align}
    \overline{\mathcal{B}}(B_s \to \mu^+\mu^-)_\text{SM} &=  3.64(4)_{f_{B_s}}(8)_\text{CKM}(7)_\text{other} \times 10^{-9} ,
    \label{eq:Bstomumu_fBs} \\
    \overline{\mathcal{B}}(B^0 \to \mu^+\mu^-)_\text{SM} &=  1.00(1)_{f_{B^0}}(2)_\text{CKM}(2)_\text{other} \times 10^{-10} ,
    \label{eq:Bdtomumu_fBd} \\
    \left(\frac{\overline{\mathcal{B}}(B^0 \to \mu^+\mu^-)}{\overline{\mathcal{B}}(B_s \to \mu^+\mu^-)}\right)_\text{SM} &=
        0.0273(2)_{f_{B_q}}(5)_\text{CKM}(7)_\text{other} , \label{eq:Bds_fBq}
\end{align}
where the errors are from the decay constants, CKM matrix elements, and the quadrature sum of all other contributions, respectively.
Because $\overline{\mathcal{B}}(B_q \to \mu^+\mu^-)$ is proportional to the square of the decay constant, our three-fold improvement
in the uncertainty on the $B$-meson decay constants reduces the error contributions from the decay constants by almost a
factor of two, such that they are now well below the other sources of uncertainty.

\begin{table}
\centering
\caption{Numerical inputs used to calculate $B_q \to \mu^+\mu^-$ branching ratios.
The strong coupling (in the \MSbar\ scheme) is a weighted average of three- and four-flavor lattice-QCD
results~\cite{Aoki:2009tf,McNeile:2010ji,Blossier:2013ioa,Bazavov:2014soa,Chakraborty:2014aca,Bruno:2017gxd}.
The $B$-meson lifetimes are from the Heavy Flavor Averaging Group's Summer 2017 averages~\cite{Amhis:2016xyh,HFLAV_2017}.
The CKM matrix elements are from the CKMfitter group's global unitarity-triangle analysis including results through ICHEP
2016~\cite{Charles:2004jd}, where we have symmetrized the errors on $|V_{ts}^* V_{tb}^{}|$ and $|V_{td}^* V_{tb}^{}|$, and
used the Wolfenstein parameters $\left\{\lambda = 0.22510(28), A = 0.8341(20), \bar{\rho} = 0.1600(74), \bar{\eta}=0.3500(62) \right\}$ rather than
the simple ratio to obtain $|V_{td}/V_{ts}|$ with a reduced uncertainty.}
\label{tab:inputs}
\begin{tabular}{l@{\hskip 10mm}l}
\hline \hline
%
% \spp
$m_{t,\pole} = 173.1(6)~\GeV$~\cite{Olive:2016xmw} & $\alpha_{s}(m_Z) = 0.1186(4)$ \\ 
%
% \spp
$\tau_{B_d} = {1.518(4)}\; \text{ps}$ & $\tau_{H_s} = {1.619(9)}  \; \text{ps}$ \\
%
% \spp
$|V_{ts}^* V_{tb}^{}| = 40.9(4) \times 10^{-3}$ &  $|V_{td}^* V_{tb}^{}| = 8.56(9) \times 10^{-3} $ \\
%
% \spp
$|V_{td}/V_{ts}| = 0.2085(18)$ &  \\
\hline\hline
\end{tabular}
\end{table}

\section{Summary and outlook}
\label{sec:conc}

In this paper, we have presented the most precise lattice-QCD calculations to-date of the leptonic decay constants of heavy-light
pseudoscalar mesons with charm and bottom quarks.
We use highly improved staggered quarks with finer lattice spacings than ever before, which enables us for the first time to work
with the HISQ action directly at the physical $b$-quark mass.
As shown in Figs.~\ref{fig:fD_summary} and~\ref{fig:fB_summary}, our results agree with previous three- and four-flavor lattice-QCD
determinations using different actions for the light, charm, and bottom quarks.
The errors on our $D$-meson decay constants in Eqs.~(\ref{eq:fD0})--(\ref{eq:fDs}) are about 2.5 times smaller than
those from our earlier analysis~\cite{\rFD2014}.
The error reduction is primarily due to the use of finer lattice spacings, which reduces the continuum-extrapolation uncertainty.
Our $B$-meson decay constants in Eqs.~(\ref{eq:fB+})--(\ref{eq:fBs}) are about three times more precise than the previous best
lattice-QCD calculations by HPQCD~\cite{McNeile:2011ng,Dowdall:2013tga}.
Here the improvement again stems from the use of finer lattice spacings, which enable us to employ the HISQ action directly at the
physical $m_b$ with controlled heavy-quark discretization errors, thereby eliminating the need to extrapolate to the bottom-quark
mass from lighter heavy valence-quark masses or to use an effective action such as NRQCD with its uncertainties from omitted
higher-order corrections in $\alpha_s$ or $1/m_Q$.

Our results for the charged $D^+$- and $D_s$-meson decay constants can be combined with the experimental leptonic decay rates for
$D^+_{(s)}\to l^+\nu_{l}$~\cite{Rosner:2015wva} to yield the CKM matrix elements
\begin{subequations}
    \label{eq:CKMresults}
    \begin{align}
        |V_{cd}| &= 0.2151(6)_{f_D}(49)_{\rm exp.}(6)_{\rm EM}\,, \\
        |V_{cs}| &= 1.000(2)_{f_{D_s}}(16)_{\rm exp.}(3)_{\rm EM}\,.
    \end{align}
\end{subequations}
We note, however, that the uncertainties due to unknown hadronic structure-dependent electromagnetic corrections are only rough
estimates based on the analogous contributions for pion and kaon decay constants (see Sec.~\ref{sec:CKM}), and need to be calculated
directly for the $D$ system.
The determinations of $|V_{cd}|$ and $|V_{cs}|$ from leptonic $D$ decays in Eq.~(\ref{eq:CKMresults}) enable us to test the
unitarity of the second row of the CKM matrix at the few-percent level, and are compatible with three-generation CKM unitarity
within 1.5$\sigma$.
The significance of this test of the Standard Model is presently limited by the experimental errors on the corresponding leptonic
decay widths~\cite{Rosner:2015wva}.
Recently the BES-III Experiment published its first measurements of $\BR(D^+_s \to \mu^+\nu_\mu)$ and
$\BR(D^+_s\to\tau^+\nu_\tau)$~\cite{Ablikim:2016duz}, and presented a preliminary measurement of
$\BR(D^+\to\tau^+\nu_\tau)$~\cite{Ma:2017pgu}; these results are statistics-limited, and will improve with additional running.
The forthcoming Belle~II Experiment will also measure the leptonic $D^+_{(s)}$-meson decay rates, and anticipates obtaining
sufficient precision to determine the CKM element $|V_{cd}|$ with an error below about 2\%~\cite{Schwartz:2017gni}.

The neutral $B_s$- and $B^0$-meson decay constants are parametric inputs to the Standard-Model rates for the rare decays
$B_s\to\mu^+\mu^-$ and $B^0\to\mu^+\mu^-$, respectively.
Using our results for $f_{B_s}$ and $f_{B^0}$, we obtain the predictions
\begin{align}
    \oBR(B_s\to\mu^+\mu^-) &=  3.64(4)_{f_{B_s}}(8)_\text{CKM}(7)_\text{other} \times 10^{-9}\,,
    \label{eq:BsmumuResult}
    \\
    \oBR(B^0\to\mu^+\mu^-) &=  1.00(1)_{f_{B^0}}(2)_\text{CKM}(2)_\text{other} \times 10^{-10}\,,
    \label{eq:BdmumuResult}
\end{align}
where the largest contributions to the errors are from the CKM elements $|V_{ts}|$ and $|V_{td}|$, respectively.
The theoretical uncertainty on $\oBR(B_s\to\mu^+\mu^-)$ in Eq.~(\ref{eq:BsmumuResult}) is more than ten times smaller than recent
experimental measurements~\cite{CMS:2014xfa,Aaboud:2016ire,Aaij:2017vad}, while the prediction for $\oBR(B^0 \to \mu^+\mu^-)$ in
Eq.~(\ref{eq:BdmumuResult}) is half an order of magnitude below present experimental limits~\cite{Aaboud:2016ire,Aaij:2017vad}.

The high-luminosity LHC combined with upgraded ATLAS, CMS, and LHCb detectors should make possible significant
improvements on these measurements in the next decade.
In particular, given Standard-Model expectations, the LHCb Experiment anticipates determining
$\oBR(B_s\to\mu^+\mu^-)$ to about 5\% and the ratio
$\oBR(B^0\to\mu^+\mu^-)/\oBR(B_s \to \mu^+\mu^-)$ to the order of 40\% by the end of the HL-LHC
era~\cite{Schmidt:2016jra}.
Our results for $f_{B_s}$ and $f_{B^0}$ can also be used to improve the Standard-Model predictions for the $B_{(s)}$-meson branching
ratios to electron-positron or $\tau$-lepton pairs, which are of $\order(10^{-6})$ and $\order(10^{-13})$,
respectively~\cite{Bobeth:2013uxa}.
The LHCb experiment recently placed the first direct limit on
$\oBR(B_s\to\tau^+\tau^-)<6.8\times10^{-3}$~\cite{Aaij:2017xqt}, and will continue to improve this measurement
with additional running.
Further, the decay rates $\oBR(B_s\to e^+e^-)$ and $\oBR(B^0\to e^+e^-)$ can be substantially
enhanced in new-physics scenarios in which the Wilson coefficients of the relevant four-fermion operators are independent of the
flavor of the decaying $B_q$ meson and the final-state leptons~\cite{Fleischer:2017ltw}.
In this case, the latter process could be observable by the LHCb and Belle~II Experiments, providing unambiguous evidence for new
physics.

Our result for $f_{B^+}$ can be combined with the experimental average for $\BR(B^+\to\tau^+\nu_{\tau})$%
~\cite{Aubert:2009wt,Lees:2012ju,Adachi:2012mm,Kronenbitter:2015kls,Rosner:2015wva} to yield the CKM matrix element
\begin{equation}
    |V_{ub}| = 4.07(3)_{f_{B^+}}(37)_\text{expt} \times 10^{-3}
    \label{eq:Vub}
\end{equation}
with an about 10\% uncertainty stemming predominantly from the error on the measured decay width.
Within this large uncertainty, Eq.~(\ref{eq:Vub}) agrees with the determinations of $|V_{ub}|$ from both
inclusive~\cite{Bauer:2001rc,Lange:2005yw,Gambino:2007rp,Aglietti:2007ik,Gardi:2008bb} and
exclusive~\cite{delAmoSanchez:2010af,Ha:2010rf,Lees:2012vv,Sibidanov:2013rkk} semileptonic $B$-meson decays.
The Belle~II Experiment expects, however, to collect enough data by 2024 to measure $\BR(B^+ \to \tau^+\nu_\tau)$ with a precision
of 3--5\%~\cite{Bennett:2016qgs}, which will make possible a competitive determination of $|V_{ub}|$ from leptonic decays.
The decay $B^+ \to \tau^+\nu_\tau$ also probes extensions of the Standard Model with particles that couple preferentially to heavy
fermions.
Using $f_{B^+}$ from this work and taking $10^3\, |V_{ub}| = 3.72(16)$ from our recent lattice-QCD calculation of the
$B\to\pi\ell\nu$ form factor~\cite{Lattice:2015tia}, we obtain for the Standard-Model branching ratio
\begin{equation}
    \BR(B^+ \to \tau^+\nu_\tau) =  8.76(13)_{f_{B^+}}(75)_{V_{ub}}(2)_\text{other} \times 10^{-5},
    \label{eq:BtoTaunu}
\end{equation}
in agreement with the experimental average $10^4\BR(B^+\to\tau^+\nu_\tau)=1.06(20)$%
~\cite{Aubert:2009wt,Lees:2012ju,Adachi:2012mm,Kronenbitter:2015kls,Rosner:2015wva}.

Given the current and projected experimental uncertainties on the $D_{(s)}$- and $B_{(s)}$-meson leptonic decay rates, better
lattice-QCD calculations of the decay constants are not needed in the near future.
Nevertheless, there are still opportunities for improvement.
So far, $D$- and $B$-decay constant calculations include neither isospin nor electromagnetic effects from first principles.
Isospin effects can be addressed straightforwardly with $1+1+1+1$ ensembles being generated for problems such as the anomalous
magnetic moment of the muon~\cite{Chakraborty:2017tqp}.
The inclusion of electromagnetism in lattice-QCD simulations is more challenging, but calculations of the light-hadron spectrum and
light-quark masses within quenched QED are available~\cite{Borsanyi:2014jba,Fodor:2016bgu,Basak:2016jnn}, and ensembles with
dynamical photons~\cite{Zhou:2014gga} to be generated for other quantities can again be employed to calculate heavy-light meson
decay constants.
In addition, higher-order electroweak effects are presently ignored when relating experimental measurements of charged leptonic
decays to Standard-Model calculations.
Effective-field-theory techniques can be used to separate effects at the electroweak and QCD scales from long-range radiation from
charged particles.
Further lattice-QCD calculations are needed to fit in with this scale separation.
For leptonic pion and kaon decays, these effects are relevant and being studied~\cite{Carrasco:2015xwa,Tantalo:2016vxk}.
Even if not immediately crucial for leptonic $D$ and $B$ decays, they are relevant for semileptonic $D$ and $B$ (as well as $K$ and
$\pi$) decays; see, for example, the comparison of QED and QCD uncertainties in Ref.~\cite{Bailey:2014tva}.

The next step in our $B$-physics program is to extend the use of HISQ $b$ quarks on the same gauge-field configurations employed in
this work to target other hadronic matrix elements needed for phenomenology.
The analysis of ensembles with physical-mass pions and very fine lattice spacings will address two of the most important sources of
systematic uncertainty in our recent calculations of the $B\to\pi(K)\ell\nu$ and $B\to\pi(K)\ell^+\ell^-$ semileptonic form
factors~\cite{Lattice:2015tia,Bailey:2015nbd,Bailey:2015dka} and of the neutral $B$-mixing matrix elements~\cite{Bazavov:2016nty} by
eliminating the chiral-extrapolation uncertainty and reducing continuum-extrapolation and heavy-quark discretization errors.
When combined with anticipated future measurements, this will enable us to determine more precisely the CKM matrix elements
$|V_{ub}|$ and $|V_{td(s)}|$, which are parametric inputs to Standard-Model and new-physics predictions.
These advances will also make possible more sensitive searches for $b \to d(s)$ flavor-changing neutral currents, charged Higgs
particles, and other extensions of the Standard Model that would give rise to new sources of flavor and $CP$ violation in the
$B$-meson sector.

% \clearpage 
\acknowledgments

We thank Silvano Simula for useful correspondence.

Computations for this work were carried out with resources provided by
the USQCD Collaboration,
the National Energy Research Scientific Computing Center,
the Argonne Leadership Computing Facility, 
the Blue Waters sustained-petascale computing project,
the National Institute for Computational Science,
the National Center for Atmospheric Research,
the Texas Advanced Computing Center,
and Big Red II+ at Indiana University.
USQCD resources are acquired and operated thanks to funding from the Office of Science of the U.S. Department of Energy.
%
%NERSC
The National Energy Research Scientific Computing Center is a DOE Office of Science User Facility supported by the
Office of Science of the U.S. Department of Energy under Contract No.\ DE-AC02-05CH11231.
%
% ALCF
An award of computer time was provided by the Innovative and Novel Computational Impact on Theory and Experiment (INCITE)
program. This research used resources of the Argonne Leadership Computing Facility, which is a DOE Office of Science
User Facility supported under Contract DE-AC02-06CH11357.
%
%BLUEWATERS
The Blue Waters sustained-petascale computing project is supported by the National Science Foundation (awards OCI-0725070 and
ACI-1238993) and the State of Illinois.
Blue Waters is a joint effort of the University of Illinois at Urbana-Champaign and its National Center for Supercomputing
Applications.
This work is also part of the ``Lattice QCD on Blue Waters'' and ``High Energy Physics on Blue Waters'' PRAC allocations supported
by the National Science Foundation (award numbers 0832315 and 1615006).
% Teragrid/XSEDE
% NCAR
% NICS
This work used the Extreme Science and Engineering Discovery Environment (XSEDE), which is supported by National Science Foundation
grant number ACI-1548562~\cite{XSEDE_REF}.
Allocations under the Teragrid and XSEDE programs included resources at the National Institute for Computational Sciences (NICS) at
the Oak Ridge National Laboratory Computer Center, The Texas Advanced Computing Center and the National Center for Atmospheric
Research, all under NSF teragrid allocation TG-MCA93S002.
Computer time at the National Center for Atmospheric Research % Frost
was provided by NSF MRI Grant CNS-0421498, NSF MRI Grant CNS-0420873, NSF MRI Grant CNS-0420985, NSF sponsorship of the National
Center for Atmospheric Research, the University of Colorado, and a grant from the IBM Shared University Research (SUR) program.
% Indiana
Computing at Indiana University is supported by Lilly Endowment, Inc., through its support for the Indiana University Pervasive
Technology Institute.

This work was supported in part by the U.S.\ Department of Energy under grants
No.~DE-FG02-91ER40628 (C.B., N.B.),
No.~DE-FC02-12ER41879 (C.D.),
No.~DE{-}SC0010120 (S.G.),     % braces for arXiv weirdness
No.~DE-FG02-91ER40661 (S.G.),
No.~DE-FG02-13ER42001 (A.X.K.),
No.~DE{-}SC0015655 (A.X.K.), 
No.~DE{-}SC0010005 (E.T.N.),
No.~DE-FG02-13ER41976 (D.T.),
No.~DE{-}SC0009998 (J.L.);
by the U.S.\ National Science Foundation under grants
PHY14-14614 and PHY17-19626 (C.D.),
and PHY13-16748 and PHY16-20625 (R.S.);
by the MINECO (Spain) under grants FPA2013-47836-C-1-P and FPA2016-78220-C3-3-P (E.G.);
by the Junta de Andaluc\'{\i}a (Spain) under grant No.\ FQM-101 (E.G.);
by the UK Science and Technology Facilities Council (J.K.);
by the German Excellence Initiative and the European Union Seventh Framework Program under grant agreement No.~291763 as well as 
the European Union's Marie Curie COFUND program (J.K., A.S.K.).
Brookhaven National Laboratory is supported by the United States Department of Energy, Office of Science, Office of High Energy
Physics, under Contract No.\ DE{-}SC0012704.
This document was prepared by the Fermilab Lattice and MILC Collaborations using the resources of the Fermi National Accelerator
Laboratory (Fermilab), a U.S.\ Department of Energy, Office of Science, HEP User Facility.
Fermilab is managed by Fermi Research Alliance, LLC (FRA), acting under Contract No.\ DE-AC02-07CH11359.
%

% \clearpage
\appendix
% Appendix A

\section{Tree-level calculations of heavy quarks with HISQ action}
\label{Appendix:Tree-level-HISQ}

The HISQ action for one flavor can be written as 
\begin{equation}
    S = \sum_x \overline{\psi}(x)\left\{\sum_\mu \gamma_\mu\left[a\Delta_\mu - \frac{\Naik}{6}a^3\Delta_\mu^3\right] + am_{0}
    \right\} \psi(x),
    \label{eq:HISQ-action}
\end{equation}
where (suppressing the gauge field) $a\Delta_\mu\psi(x)=\half[\psi(x+\hat{\mu}a)-\psi(x-\hat{\mu}a)]$,
$m_{0}$ is the bare mass, and $\Naik=1+\epsilon$ is the coefficient of the Naik improvement term~\cite{Naik:1986bn}.
The correction $\epsilon$ is needed to improve the dispersion relation when $m_0a\not\ll1$~\cite{Follana:2006rc}.
The notation $\epsilon$ is used in Ref.~\cite{Follana:2006rc}; in Appendix~\ref{app:normalization}, however, $1+\epsilon$ appears,
so we use $\Naik$ for brevity.

We are interested in heavy quarks with mass much larger than their typical momentum.
Then, the energy can be expanded as
\begin{equation}
  E(\bm{p}) = m_1 + \frac{\bm{p}^2}{2 m_2} +\cdots, 
  \label{eq:energy:rest-kinetic-mass }
\end{equation}
where $m_1$ and $m_2$ are called the rest and kinetic masses, respectively.
At nonzero lattice spacing, these two masses are no longer identical.
The parameter $\epsilon$ in the HISQ action is supposed to be tuned
such that the kinetic mass of a quark equals its rest mass, \ie
\begin{equation}
  \frac{m_1}{m_2} = \lim_{\bm{p}\to\mathbf{0}} \frac{E^2(\bm{p}) - E(0)^2}{\bm{p}^2} = 1 .
  \label{eq:HISQ:epsilon:tunning}
\end{equation}
This condition yields
\begin{align}
    \epsilon &= \frac{4 - 2\sqrt{1+3X}}{\sinh^2 (am_1)} -1, \label{eq:epsilon:exact} \\
    X &= \frac{2am_1}{\sinh(2am_1)}. \label{eq:epsilon:X}
\end{align}
With this exact expression for $\epsilon$, we have $am_2=am_1$ to all orders in $am_0$, at the tree level.

The Taylor expansion of $\epsilon$, in \eq{epsilon:exact}, about the origin reads
\begin{align}
    \epsilon = -\frac{27}{40} (am_1)^2 &
        + \frac{327}{1120}    (am_1)^4 
        - \frac{5843}{53760}  (am_1)^6
        + \frac{153607}{3942400} (am_1)^8
        - \frac{604604227}{43051008000}   (am_1)^{10} \nonumber \\ &
        + \frac{2175452933}{422682624000} (am_1)^{12}
        - \frac{1398976049}{729966182400} (am_1)^{14} + \cdots.
    \label{eq:epsilon:exact:Taylor} 
\end{align}
The radius of convergence of this series is ${\pi}/{2}$, which is set by the singularities in the complex plane from
the inverse power of $\sinh(2am_1)$ in the exact expression. 
Equation~(\ref{eq:epsilon:exact:Taylor}) can be rewritten as
\begin{equation}
    \epsilon = -1.67\,x_h^2 + 1.78\,x_h^4 - 1.63\,x_h^6 + 1.44\,x_h^8
        -1.28\,x_h^{10} + 1.16\,x_h^{12} - 1.07\,x_h^{14} + \cdots, 
    \label{eq:epsilon:exact:Taylor:Natural}
\end{equation}
where $x_h = 2am_1/\pi$.
(The coefficients have been rounded to two significant figures.) 
This expansion converges inside the unit disc in the complex $x_h$-plane, centered at the origin.
One sees that many of the first several coefficients of this power series are of order~1, 
and in this sense, $x_h$ can be considered to be a natural expansion parameter.

The bare mass $m_{0}$ in the quark action is related to its tree-level pole mass by
\begin{equation}
   am_0 = \sinh(am_1)\,\frac{1+\sqrt{1+3X}}{3},
   \label{eq:am-2-am0}
\end{equation}
with $X$ as in \eq{epsilon:X}. 
As with $\epsilon$, the Taylor expansion of $m_0$ breaks down at $am_1=\pi/2$,
and $m_0$ has a natural series expansion in powers of~$x_h$.

% Appendix B

\section{Normalization of staggered bilinears when \texorpdfstring{\boldmath$am_0\not\ll1$}{am !<< 1}}
\label{app:normalization}

From Ref.~\cite{ElKhadra:1996mp} for massive Wilson fermions, it follows that when $am_0\not\ll1$ a bilinear can lose the
conventional normalization.
In this appendix, we derive the factor needed to restore this normalization for the pseudoscalar density of (improved)
staggered fermions.
To this end, we also need to think of HQET as a theory of cutoff effects, applied directly to lattice gauge 
theory~\cite{Kronfeld:2000ck}.

The starting point is the time evolution of the fermion propagator at zero momentum.
Using the residue theorem ($\delta$ is real, small, and positive),
\begin{align}
    C(\bm{0},x_4) &= \int_{-(\pi-\delta)/a}^{(\pi+\delta)/a}\frac{dp_4}{2\pi}e^{ip_4x_4}
        \frac{-i\gamma_4\tilde{S}_4+m_0}{\tilde{S}_4^2+m_0^2}
    \nonumber \\ &=
        \frac{1}{\Ch} e^{-m_1|x_4|} \left[ \frac{1\pm\gamma_4}{2} + 
            e^{-i\pi|x_4|/a} \frac{1\mp\gamma_4}{2} \right],
    \label{eq:Cpt}
\end{align}
where the upper (lower) sign in front of $\gamma_4$ is for $x_4>0$ ($x_4<0$), and
\begin{align}
    a\tilde{S}_4(p) &= \sin ap_4  \left(1+\sixth\Naik\sin^2  ap_4\right), \\
    a\Sh            &= \sinh am_1 \left(1-\sixth\Naik\sinh^2 am_1\right), \\
    \Ch             &= \cosh am_1 \left(1-\half \Naik\sinh^2 am_1\right).
\end{align}
The rest mass $m_1$ is obtained from the bare mass $m_0$ via
\begin{equation}
    m_0 = \Sh. % (m_1).
\end{equation}
Equation~(\ref{eq:Cpt}) consists of an unwanted normalization factor, the exponential fall-off in Euclidean time, and (correctly 
normalized) Dirac matrices for two species: the one with the factor $e^{-i\pi|x_4|/a}$ is the time doubler.
States with energy near the cutoff are omitted, and one should bear in mind that other doublers with energy $m_1$ can be found in 
other corners of the spatial Brioullin zone.
None of these staggered features is important here.

The first factor implies that the external line factors for zero-momentum fermion and antifermion states are
\begin{align}
    \psi(x)|q(\xi,\bm{0})\rangle &= \Ch^{-1/2} u(\xi,\bm{0}) e^{-m_1x_4}, \\
    \bar{\psi}(x)|\bar{q}(\xi,\bm{0})\rangle &= \Ch^{-1/2} \bar{v}(\xi,\bm{0}) e^{-m_1x_4},
\end{align}
when the fermion states are normalized to
\begin{equation}
    \langle q(\xi',\bm{p}')|q(\xi,\bm{p})\rangle = (2\pi)^3 \delta(\bm{p}'-\bm{p}) \delta^{\xi'\xi},
\end{equation}
and similarly for single-antiquark states.

With naive or staggered fermions, the pseudoscalar density appearing in the Ward identity of the exact remnant of chiral symmetry 
is the local one:
\begin{equation}
    P_{hx}(x) = \bar{\psi}_h(x)i\gamma^5\psi_x(x)
\end{equation}
using the notation of the naive formulation.
Let us consider two matrix elements of the pseudoscalar density, namely when the $x$ quark is nonrelativistic or 
ultrarelativistic.
To the order needed, one finds
\begin{align}
    \langle0|P_{hx}(0)|q_x(\xi_x,\bm{0})\bar{q}_h(\xi_h,\bm{0})\rangle &=
        \Ch^{-1/2}_h \Ch^{-1/2}_x w^\dagger_{\xi_h} w_{\xi_x},
    \label{eq:nonrelP} \\
    \langle0|P_{hx}(0)|q_x(\xi_x,\bm{p}_x)\bar{q}_h(\xi_h,\bm{p}_h)\rangle &= \left(2\,\Ch_h\right)^{-1/2} w^\dagger_{\xi_h}
        \left[1 - \frac{(\bm{\sigma}\cdot\hat{\bm{p}}_x)(\bm{\sigma}\cdot\bm{p}_h)}{2m_{0h}} 
        \right] w_{\xi_x} + \order(\bm{p}^2),
    \label{eq:ultrarelP}
\end{align}
for the nonrelativistic and ultrarelativistic cases, respectively, where $w^\dagger_{\xi_h}$ and $w_{\xi_x}$ are two-component
spinors, and $\hat{\bm{p}}_x=\bm{p}_x/|\bm{p}_x|$.
Similar results hold for other local bilinear currents.

These tree-level calculations reveal two important features about the heavy-quark discretization effects.
First, depending on whether the $x$ quark is a nonrelativistic or ultrarelativistic, matrix elements should be
multiplied by a factor
\begin{align}
    Z_{J_{hx}} &= \Ch^{1/2}_h \Ch^{1/2}_x , \\
    Z_{J_{hx}} &= \Ch^{1/2}_h
    \label{eq:Appendix:Z}
\end{align}
to remove tree-level mass-dependent discretization effects at the leading order in $|\bm{p}_h|/m_{0h}$.%
\footnote{For a light quark ($m_x\lesssim2\Lambda_\text{QCD}$), $\Ch_x$ deviates from 1 by effects as small or smaller than
other discretization effects.
In particular $\Ch_x=1+\order(a^2m_x^2)$ for the unimproved action with $\Naik=0$
and $\Ch_x=1+\order(a^4m_x^4)$ for the improved actions with $\Naik=1$ or $\Naik=1+\epsilon$.} %
Second, the next order in the HQET expansion requires an additional correction (as is the case with Wilson
fermions~\cite{ElKhadra:1996mp,Kronfeld:2000ck}) to ensure the correct normalization of this term.
It is, however, proportional to
\begin{equation}
    \frac{1}{m_{0h}} - \frac{1}{m_{1h}} = \frac{1-m_{0h}/m_{1h}}{m_{0h}}.
\end{equation}
The numerator's leading discretization errors are of order $x_h^4$ and $\alpha_sx_h^2$, owing to the tree-level Naik improvement
term, and the dimensions are balanced, in a heavy-light meson, by $\LamHQET$ or~$m_x$.
As in Appendix~\ref{Appendix:Tree-level-HISQ}, $x_h=2am_{1h}/\pi$ is the natural expansion parameter for organizing
heavy-quark discretization errors.

To arrive at the decay constant, the pseudoscalar density must be multiplied by the sum of the quark masses.
From the axial Ward identity, the combination $m_{0x}+m_{0h}$ is natural.
This quantity would, however, introduce heavy-quark discretization effects that can be avoided by using $m_{1x}+m_{1h}$ instead.
With this choice and Eq.~(\ref{eq:Appendix:Z}) for normalizing $\Phi_{H_x}$, all heavy-quark
discretization errors are suppressed by either $\alpha_s$ or $\LamHQET/M_{H_x}$ or both.

\vfill
\section{Covariance matrix for decay constants}
\label{Appendix:Covariance-Matrix}

Tables~\ref{tab:correlation_matrix} and~\ref{tab:covariance_matrix} provide the correlation and covariance matrices
for our decay-constant results, respectively, to enable future phenomenological studies.

\begin{turnpage}
\begin{table}
\newcommand{\h}{\phantom{x}}
\caption{Correlation matrix between the $D$- and $B$-meson decay constants in Eqs.~(\ref{eq:fD0})--(\ref{eq:fB});
entries are symmetric across the diagonal.} 
\label{tab:correlation_matrix}
\begin{tabular}{l|cccccccc}
\hline\hline
	    & \h$f_{D^0}$\h & $f_{D}$\h & $f_{D^+}$\h & $f_{D_s}$\h & $f_{B^+}$\h & $f_{B}$\h & $f_{B^0}$\h & $f_{B_s}$ \\
\hline
$f_{D^0}$\h & \h1\h          &                &                &                &                &                &                & \\
$f_{D}$\h   & \h0.99034256\h & \h1\h          &                &                &                &                &                & \\
$f_{D^+}$\h & \h0.96489064\h & \h0.99179205\h & \h1\h          &                &                &                &                & \\
$f_{D_s}$\h & \h0.85584800\h & \h0.89529969\h & \h0.91276762\h & \h1\h          &                &                &                & \\
$f_{B^+}$\h & \h0.41698224\h & \h0.42111777\h & \h0.41595657\h & \h0.39194646\h & \h1\h          &                &                & \\
$f_{B}$\h   & \h0.43374664\h & \h0.44096880\h & \h0.43740528\h & \h0.41993616\h & \h0.99827684\h & \h1\h          &                & \\
$f_{B^0}$\h & \h0.45049520\h & \h0.45971393\h & \h0.45703271\h & \h0.44373556\h & \h0.99419014\h & \h0.99877397\h & \h1\h          & \\
$f_{B_s}$\h & \h0.54139865\h & \h0.56564796\h & \h0.57288800\h & \h0.58902865\h & \h0.85069938\h & \h0.87357307\h & \h0.89060925\h & \h1\h \\
\hline\hline
\end{tabular}
\end{table}
\begin{table}[p]
\newcommand{\h}{\phantom{x}}
\caption{Covariance matrix between the $D$- and $B$-meson decay constants in Eqs.~(\ref{eq:fD0})--(\ref{eq:fB});
entries are symmetric across the diagonal and are in MeV$^2$.}
\label{tab:covariance_matrix}
\begin{tabular}{l|cccccccc}
\hline\hline
	    & \h$f_{D^0}$\h & $f_{D}$\h & $f_{D^+}$\h & $f_{D_s}$\h & $f_{B^+}$\h & $f_{B}$\h & $f_{B^0}$\h & $f_{B_s}$ \\
\hline
$f_{D^0}$\h & \h0.34779867\h &                &                &                &                &                &                & \\
$f_{D}$\h   & \h0.33313370\h & \h0.32534065\h &                &                &                &                &                & \\
$f_{D^+}$\h & \h0.32265640\h & \h0.32076578\h & \h0.32151147\h &                &                &                &                & \\
$f_{D_s}$\h & \h0.21136370\h & \h0.21384909\h & \h0.21673461\h & \h0.17536366\h &                &                &                & \\
$f_{B^+}$\h & \h0.33422198\h & \h0.32645717\h & \h0.32055289\h & \h0.22307455\h & \h1.84717041\h &                &                & \\
$f_{B}$\h   & \h0.33822862\h & \h0.33257324\h & \h0.32793859\h & \h0.23252162\h & \h1.79396823\h & \h1.74831843\h &                & \\
$f_{B^0}$\h & \h0.34428780\h & \h0.33980076\h & \h0.33582501\h & \h0.24080281\h & \h1.75101736\h & \h1.71137428\h & \h1.67932607\h & \\
$f_{B_s}$\h & \h0.42932416\h & \h0.43383002\h & \h0.43678947\h & \h0.33167323\h & \h1.55465419\h & \h1.55315110\h & \h1.55188280\h & \h1.80804153 \\
\hline\hline
\end{tabular}
\end{table}
\end{turnpage}

\clearpage
\bibliographystyle{apsrev4-1}
\bibliography{References.bib}

\end{document}